\documentclass[12pt,a4paper,hyper]{JHEP3}
\usepackage[utf8x]{inputenc}
\usepackage{amsmath}
\usepackage{amsfonts}
\usepackage{microtype}
\usepackage{amssymb,amsfonts}
\usepackage{graphicx}
\usepackage{multirow}

\title{\center {Quantum entropy of supersymmetric black holes}\footnote{This article is based on the thesis of the author.}}
\preprint{}
\author{Jo\~ao Gomes$^{1}$\\

\it $^1${Laboratoire de Physique Th\'eorique et Hautes Energies (LPTHE)\\
\it{Universit\'e Pierre et Marie Curie-Paris 6; CNRS UMR 7589}\\
\it{Tour 13-14, 5$^{\grave{e}me}$ \'etage, Boite 126, 4 Place Jussieu} \\
\it {75252 Paris Cedex 05, France}}\\

}

\abstract{
We review recent progress concerning the quantum entropy of a large class of
supersymmetric black holes in string theory both from the microscopic and macroscopic sides. On the microscopic field theory side, we present new results concerning the  counting of black hole microstates for  charge vectors with nontrivial arithmetic duality invariants.  On the macroscopic gravitational side, we present a novel application of localization techniques to a supergravity functional integral to compute the
quantum entropy of these black holes. Localization leads to an enormous simplification of a path integral of string theory in $AdS_2$ by reducing it to a finite dimensional integral. The localizing solutions are labeled by $n_v+1$ parameters, with $n_v$ the number of vector multiplets in the theory of $\mathcal{N}=2$ supergravity. As an example we show, for four dimensional large black holes which preserve four supersymmetries in toroidally compactified IIB string theory, that the macroscopic degeneracy precisely agrees with all the terms in an exact Rademacher expansion of the microscopic answer except for Kloosterman sums which in principle can be computed. Generalizing previous work, these finite charge contributions to the leading Bekenstein-Hawking entropy can also be viewed as an instance of ``exact holography'' in the context of $AdS_2/CFT_1$ correspondence.  
}

\newenvironment{myenumerate}{
\begin{enumerate}
   \setlength{\itemsep}{1pt}
   \setlength{\parskip}{0pt}
   \setlength{\parsep}{0pt}}{\end{enumerate}}
\newenvironment{myitemize}{
\begin{itemize}
   \setlength{\itemsep}{1pt}
   \setlength{\parskip}{0pt}
   \setlength{\parsep}{0pt}}{\end{itemize}}

\newcommand{\C}{{\mathbb C}}
\def\={\;  = \;}
\def\inn{\,\in\,}

\renewcommand{\Im}{\mbox{Im}}

\newcommand{\mbbz}{\mathbb{Z}}
\newcommand{\mbbr}{\mathbb{R}}
\newcommand{\no}{\nonumber}
\newcommand{\cA}{{\cal A}}

\newcommand{\IR}{\mathbb{R}}
\newcommand{\IZ}{\mathbb{Z}}

\newcommand{\Tr}{\mbox{Tr}}

\def\half{{\frac12}}
\def\e{\epsilon}
\def\t{\tau}
\def\CN{\mathcal{N}}
\def\CV{\mathcal{V}}
\def\CL{\mathcal{L}}
\def\CS{\mathcal{S}}
\def\CO{\mathcal{O}}

\def\CF{\mathcal{F}}
\def\be{\begin{equation}}
\def\ee{\end{equation}}
\def\ve{\varepsilon}
\def\ndt{\noindent}
\def\a{\alpha}

\def\g{\gamma}
\def\wh{\widehat}
\def\p{\partial}
\def\bea{\begin{eqnarray}}
\def\eea{\end{eqnarray}}
\def\m{\mu}

\newcommand{\sech}{\mathrm{sech} \,}
\def\wt{\widetilde}
\def\slash#1{\rlap{\hbox{$\mskip 1 mu /$}}#1}      % good slash for
\def\Slash#1{\rlap{\hbox{$\mskip 3 mu /$}}#1}      % " upper

\newcommand{\bem}{\begin{pmatrix}}
\newcommand{\eem}{\end{pmatrix}}

\def\vth{\vartheta}

\def\r{\rho}

\def\s{\sigma}
\def\g{\gamma}
\def\t{\tau}
\def\a{\alpha}

\def\m{\mu}
\def\n{\nu}
\def\e{\epsilon}

\def\h{\eta}
\def\l{{\lambda}}

\def\D{\Delta}

\def\CF{{\cal F}}
\def\CS{{\cal S}}
\def\CL{{\cal L}}

\def\CN{{\cal N}}
\def\CO{{\cal O}}

\def\half{{\frac12}}

\def\CN{{\cal N}}

\def\bea{\begin{eqnarray}}
\def\eea{\end{eqnarray}}
\def\be{\begin{equation}}
\def\ee{\end{equation}}
\def\ba{\begin{align}}
\def\ea{\end{align}}
\def\bse{\begin{subequations}}
\def\ese{\end{subequations}}
\def\1F1{{}_1\!F_1}
\def\2F0{{}_2\!F_0}

\def\ve{\varepsilon}
\def\v{\varphi}

\def\a{\alpha}

\def\h3{$\textrm{H}_3^+$}
\def\IR{{\mathbb R}}
\def\IZ{{\mathbb Z}}

\newcommand{\beq}{\begin{equation}}
\newcommand{\eeq}{\end{equation}}
\newcommand{\ber}{\begin{eqnarray}}
\newcommand{\eer}{\end{eqnarray}}

\def\be{\begin{eqnarray}}
\def\ee{\end{eqnarray}}

\newcommand{\cO}{{\cal O}}

\def\p{\partial}
\def\wh{\widehat}

\def\mod{{\rm mod}}

\def\CN{{\cal N}}

\def\CF{{\cal F}}

\def\CL{{\cal L}}
\def\CV{{\cal V}}
\def\CO{{\cal O}}

\def\CS{{\cal S}}

\def\CV{{\cal V }}

\def\CS{{\cal S }}

\def\Tr{{\rm Tr}}

\font\manual=manfnt
\def\dbend{\lower3.5pt\hbox{\manual\char127}}

\def\bar{\overline}
\def\CS{{\cal S}}
\def\CN{{\cal N}}

\def\rt2{\sqrt{2}}
\def\irt2{{1\over\sqrt{2}}}

\def\wt{\widetilde}
\def\ndt{\noindent}
\def\s{\sigma}

\def\a{\alpha}

\def\g{\gamma}

\def\mod{{\rm mod}}

\def\Tr{{\rm Tr}}

\begin{document}

%\tableofcontents 

\section{Introduction}
Einstein's general theory of relativity predicts that a sufficiently massive object can deform  spacetime in such a way that it creates a region from where not even light can escape. This solution is called a black hole. The boundary of such a region is a null hypersurface called event horizon. This a surface of infinite redshift which then motivates the name black hole.

As pointed out first by Bekenstein \cite{Bekenstein:1973ur} and then by Hawking \cite{Hawking:1974sw} a black hole must carry entropy so that the second law of thermodynamics is not violated. The classic thought experiment is to throw a bucket of warm water inside the horizon. Since the entropy of the Universe cannot decrease the black hole must have entropy. It is well known that, in general theory of relativity, the black hole entropy is proportional to the area of the horizon in contrast with ordinary matter systems where it is proportional to the accessible volume,
\begin{equation}\label{Bekenstein-Hawking entropy}
 S_{BH}=\frac{A}{4G_N\hbar}.
\end{equation}Here $A$ is the area of the horizon and $G_N$ is the Newton's constant. Consistency with statistical mechanics naturally lead us to the following question: can we describe a black hole as an ensemble of quantum states in such way that we can relate the entropy $S_{BH}$ to the logarithm of the number of accessible states?
\begin{equation}\label{boltzmann equation}
 S_{BH}=\ln \Omega_{micro}
\end{equation}

To answer this question we need a theory of quantum gravity. String theory is the leading candidate for such a theory. Although we are still far from a description of the real world in terms of strings, this theory is able to incorporate gravity in a consistent way with other forces and it leads to the discovery of branes from where the holographic correspondence \cite{Maldacena:1997re} was born. String theory gives us a systematic procedure to compute corrections to Einstein's theory of gravity which can be important to understand finite size effects in quantum gravity.

The salient results covered by this article are:
\begin{itemize}

\item\emph{\textbf{Finite charge corrections to Bekenstein-Hawking entropy}}

The main focus is the computation of finite charge corrections to the leading Bekenstein-Hawking entropy. Formula (\ref{Bekenstein-Hawking entropy})  is valid for an action with only the Einstein-Hilbert term. Since in string theory both the $\alpha'$ and string-loop corrections depend on the phase \footnote{By phase we mean the compactification of string theory.} of the theory, finite size corrections to the area law can give us information about the microscopic details of the phase.  

 To implement the effect of the higher derivative corrections we need to use the Wald formalism \cite{Wald:1993nt,Iyer:1994ys}. The entropy is then given by a surface integral over the horizon geometry. To compute the Wald entropy we need first to find the black hole solution by solving the gravity equations and then perform the surface integral which is not an easy task. However, for extremal black holes, the near horizon geometry has enhanced symmetries which can be used to simplify the computation of the Wald entropy. The near horizon geometry $AdS_2\times S^2$ has $SO(2,1)\times SO(3)$ symmetries and is separated from infinity by an infinite throat. The moduli of the theory get attracted at the horizon and their value only depends on the charges. This is called attractor mechanism \cite{Ferrara:1996dd}. Combining both the symmetries of the near horizon geometry and the attracor mechanism Sen gives a simple prescription to compute the Wald entropy \cite{Sen:2005wa}. This method, called entropy function, resumes all the computation of Wald to a minimization problem and does not require solving the Einstein's equations. The entropy function is proportional to the Lagrangian computed on the attractor background and the minimization parameters are the attractor values of the different fields. The black hole entropy is then given by the minimum of that function. 

Unfortunately this formalism is not completely adequate in the full quantum theory. There can be non-local and/or non-analytic terms in the action coming from the integration of massless fields. In this case the Wald formalism can  not be applied since it requires a local and gauge invariant action. Moreover it is in our interest to compute not just perturbative but also non-perturbative  corrections to the Wald entropy as suggested by the microscopic answers. Thus even defining the proper notion of quantum entropy presents important conceptual problems.  In an attempt to solve these issues Sen proposes a different formalism called quantum entropy \cite{Sen:2008yk,Sen:2008vm}. The idea is based on $AdS_2/CFT_1$ correspondence and gives a quantum version of the entropy. In summary, we are instructed to compute a path integral of string theory on $AdS_2$ with a Wilson line insertion at the boundary. The holographic correspondence then relates this observable to the degeneracy of the black hole. In contrast with higher dimensional cases, in $AdS_2$ the electric fields are non-normalizable modes and therefore they have to be fixed while performing the path integral. This is equivalent to fixing the charges instead of the chemical potentials which means that we are working in the microcanonical ensemble. For large charges the path integral is peaked at the classical attractor saddle point and the computation reduces to that of the entropy function.  Since the equations of motion are no longer implied we can compute both perturbative and non-perturbative contributions in a systematic way.

\item \emph{\textbf{Supersymmetric Localization}}
  
For supersymmetric theories we can hope to use supersymmetric localization to compute exactly a path integral \cite{Schwarz:1995dg,Pestun:2007rz,Banerjee:2009af}. In a few words, localization means deforming the original physical action by a Q-exact term of the form $tQV$, where $Q$ is the action of some supersymmetry. If both the deformation and the observable, we are interested in computing, are supersymmetric by themselves, then it can be shown that the path integral does not depend on $t$. This is very practical because we can choose a parameter $t$ where the computation is more convenient. In the limit $t\rightarrow\infty$ the deformation $tQV$ dominates over the original physical action and the semiclassical approximation becomes exact since, in this case, $t$ plays the role of $\hbar^{-1}$. Application of this technique in a path integral requires the supersymmetry $Q$ to be realized off-shell. Fortunately for us, there is an off-shell formulation of supergravity even though only eight SUSYs are realized \cite{deWit:1979ug,deWit:1980tn,deWit:1984px,deRoo:1980mm}. When this formalism is applied to supergravity on $AdS_2\times S^2$ the path integral localizes to a subspace where the scalar fields can be excited above their attractor values at the cost of exciting the auxiliary fields \cite{Dabholkar:2010uh}. The solution is labeled by $n_V+1$ constants where $n_V$ is the number of vector multiplets in the theory. Using this technique we were able to reduce a very complicated path integral to a finite dimensional integral which resembles very much the formula proposed by Ooguri, Strominger and Vafa \cite{Ooguri:2004zv} but with some important differences. These differences include for example loop determinants, instantons or subleading orbifold saddle points. Once they are taken into account it should be possible to reproduce exactly the microscopic answers. 

For large black holes in $\mathcal{N}=8$ string theory the microscopic answer has a simple expression given in terms of a Jacobi form. The degeneracy can be written as a sum of Bessel functions in an exact expansion called Rademacher expansion. Using localization techniques we were able to reproduce \emph{all} these terms for arbitrary values of the charges except for Kloosterman sums that can in principle be computed \cite{Dabholkar:2011ec}. In this analysis we do a careful treatment of the measure on the localization locus which reveals crucial for the exact matching.  

 The big goal is to establish an exact equality between a degeneracy computed from the microscopic degrees of freedom and the quantum entropy computed from gravity. This obviously implies two big tasks,
\begin{itemize}
 \item The first is to compute the expectation value of a Wilson line in $AdS_2$ by performing a path integral over the horizon string fields. The localization technique is extremely useful in this case.
 
 \item The second is to compute precisely the microscopic degeneracy using some weak coupling description in the same spirit of Strominger and Vafa \cite{Strominger:1996sh}.
\end{itemize}

 For large charges both tasks simplify. In this regime we can use a Cardy formula to compute the microscopic degeneracy. On the gravity side large charge means large horizon radius and therefore we can neglect higher derivative corrections. The entropy area law suffices in this case.
 Performing both tasks exactly is equivalent as establishing an exact $AdS_2/CFT_1$ holography.

\item \emph{\textbf{Microscopic counting}}

The success of Strominger and Vafa black hole inspired many other works. Results in microscopic BPS counting flourished. For quarter BPS black holes in $\mathcal{N}=4$ string theory the results are particularly interesting. The microscopic partition function is given in terms of the fourier coefficients of a Siegel modular form, which is a very rich object from the mathematical point of view. Part of this review is devoted to the analysis of the quarter-BPS dyon spectrum in these theories and to the construction of the corresponding partition functions. Previous works \cite{Dabholkar:2004yr,Shih:2005uc,Shih:2005he,Gaiotto:2005hc,Dabholkar:2005dt,David:2006yn,Maldacena:1999bp,David:2002wn,Banerjee:2008pu,Dabholkar:2008zy} concern the spectrum of dyons which obey a particular primitivity constrain on the charges. As first noted in \cite{Dabholkar:2007vk}, $I=\textrm{gcd}(\mathbf{q}\wedge \mathbf{p})$ is the only discrete invariant relevant in this problem, where $(\mathbf{q},\mathbf{p})$ denotes the dyonic electric and magnetic charges vectors respectively. Consider the charge lattice $\Lambda$ where both the electric $\mathbf{q}$ and magnetic $\mathbf{p}$ charge vectors live. These charge vectors generate a two dimensional lattice inside $\Lambda$. The invariant $I$ basically counts the number of unit cells of $\Lambda$ inside a cell bounded by $\mathbf{q}$ and $\mathbf{p}$. A primitive dyon corresponds to a unit cell. When the primitivity condition is relaxed additional difficulties arise in the microscopic counting mainly due to the analysis of multi-particle bound states at threshold \cite{Vafa:1995bm}. Without loss of generality we consider the case when the electric charge vector $\mathbf{q}$ is a multiple $I$ of a primitive vector while $\mathbf{p}$ is primitive. In type IIB frame this implies studying a system of D-branes weakly interacting with a $I$ KK-monopoles. We study the low energy theory and propose a two dimensional supersymmetric sigma model \cite{Dabholkar:2008zy}. A modified elliptic genus then gives an index which is consistent with previous constructions \cite{Banerjee:2008pv,Banerjee:2008pu} and passes many physical tests. In brief, the index found is given in terms of the fourier coefficients of the primitive answer and carries a non-trivial dependence on the divisors of $I$. In \cite{Dabholkar:2009kd} we propose a non-trivial check of the counting formula. We map a particular set of states to perturbative momentum-winding states of IIA string theory where the counting can be easily done and agreement is found for any value of $I$.

\item \emph{\textbf{Cardy limit and $\mathbf{AdS_3/CFT_2}$ correspondence}}

In the last section we focus on a different approach based mostly on $AdS_3$ rather than $AdS_2$. Instead of computing the entropy valid for any charge we consider the simpler case when only one of the charges is very large keeping the other charges arbitrarily finite. The result is exact in the limit considered and is able to probe details of the phase we are working on \cite{Dabholkar:2010rm}. The main result is: \emph{for black holes which preserve at least four supercharges the asymptotic growth of the index has a Cardy like formula with an effective central charge that is given by a linear combination of the coefficients of the Chern-Simons terms computed at asymptotic infinity}. Whenever a black hole has a factor $AdS_2\times S^1$ in the near horizon geometry we can view it as an extremal BTZ black hole living in $AdS_3$ space in the limit when the circle $S^1$ has a very large radius. The momentum along the circle corresponds to the angular momentum $J$ of the BTZ black hole. Then the extremal condition $M=J$ implies that we are counting states of large mass and therefore we can use a Cardy formula. In this case holography is extremely powerful since it relates the central charges, which are anomaly coefficients in the CFT, to the coefficients of the Chern-Simons terms living in the bulk of $AdS_3$ \cite{Witten:1998qj,Kraus:2005zm,Kraus:2005vz}. Note that the entropy formulas obtained are exact in the limit considered, that is, when only one of the charges is taken to be very large while keeping the remaining finite. During the analysis we found convenient to consider a macroscopic index that captures all the degrees of freedom from the horizon till asymptotic infinity. In the process we need to take into account additional contributions from external modes to the bulk of $AdS_3$. 

\end{itemize}

The review is organized as follows. In section \S2 we give exact results on the microscopic counting of both primitive and non-primitive dyons. In section \S3 we explain the quantum entropy formalism based on the $AdS_2/CFT_1$ correspondence and its relation with the microscopic index. In section \S4 we use localization of supergravity on $AdS_2\times S^2$ to reduce a very complicated path integral to a finite dimensional integral. We end discussing its relation to the OSV proposal. In section \S5 we apply our results from localization in the problem of large black holes in $\mathcal{N}=8$ string theory. Since the microscopic answer is known exactly we conclude by comparing both the macroscopic and microscopic answers which agree exactly for any finite charge. In the last section we study the index in the particular charge limit where only one of the charges is taken to be very large. In this regime of charges the $AdS_3$ point of view becomes more useful.

\section{Microscopic counting}\label{microsc}
 
In string theory the Newton's constant $G_N$ is proportional to the square of
string coupling $g_s$. As a consequence the gravitational attraction,
proportional to $G_N M$, with $M$ the mass of the object, can be made
arbitrarily small by
decreasing $g_s$. In particular, for fundamental strings and D-branes $G_N\,M$
goes as $g_s^2$ and $g_s$ respectively while for the KK monopole or NS5-brane
they are
of order one. In this regime of very weak string coupling we can turn off
gravity and ``dissolve`` the black hole. The space becomes flat and these
objects weakly interact. In this regime we can count the microscopic BPS states by quantizing the low energy theory of the system.

The first successful example in matching the microscopic degeneracy with the Bekenstein-Hawking entropy  is the Vafa and Strominger five dimensional black hole \cite{Strominger:1996sh}. They consider a
system of D1 and D5-branes wrapping cycles of $K3\times S^1$ in type IIB string
theory along with momenta through the circle. Effectively we see a five
dimensional black hole carrying electric and magnetic charges. The low energy theory of the branes is a two dimensional supersymmetric conformal field theory. In the limit of large charges we can use a Cardy formula to compute the entropy of BPS states while on the  black hole regime the entropy has the area law. Both answers perfectly agree. 

For a large class of supersymmetric black holes it is known that the number of
BPS states is constant over regions of the moduli space separated by codimension
one walls where the states are marginally stable against decay
\cite{Denef:2000nb,Sen:2007vb,Sen:2008ht,Sen:2008ta,Denef:2002ru,Denef:2007vg}.
The constancy of the degeneracy follows from the non-renormalization of the mass
of a state that saturates the Bogomol'nyi–Prasad–Sommerfeld bound, that is, of a
BPS state. In other words the mass $M$ equals the central charge $Z(Q)$ which is
perturbatively not renormalized and therefore these BPS states sit in multiplets
of shorter dimension.  Due to this property, we can work in a region of the
moduli space where string theory is weakly coupled, count the number of BPS 
states and then extrapolate this result to strong coupling, in the
black hole regime. In the limit of large charges, or thermodynamic limit, the
curvature of the horizon becomes small and the entropy is given by the Beckenstein-Hawking area law. 

To count BPS states we use an index. This has the property of
being invariant under continuous deformations of the theory. This is exactly
what we mean by the constancy of the number of BPS states over the moduli space.
In particular we use a helicity trace index or spacetime index. As a matter of
fact, the index counts the number of bosonic minus fermionic states and
therefore it can be zero or even negative. This is puzzling because ultimately we want to compare it with the exponential of the Beckenstein-Hawking entropy which is a strictly positive quantity. The usual
understanding is that the number of states that get paired up is subleading in
the large charge limit. Later we will see that the correct thing to do is to
compare this microscopic index with an index constructed from the black hole
solution. This issue will be analysed in section \S6 where we make a clear
distinction between index and degeneracy.  

Since the index is invariant under U-duality it becomes important to classify duality orbits and corresponding charge invariants. For dyons in $\mathcal{N}=4$ string theory with electric and magnetic charge vectors $Q$ and $P$ we can construct many duality invariants out of the charges. Apart from the continuous T-duality invariants $Q^2$, $P^2$ and $Q.P$ there is one discrete U-duality invariant $I=\textrm{gcd}(Q\wedge P)$ which is particularly important in this problem. Very basically it encodes a primitivity condition in the dyon charge vector. A primitive dyon is one for which $I=1$. Previous works in $\mathcal{N}=4$ string theory concern the spectrum of primitive dyons \cite{Dijkgraaf:1996it,Gaiotto:2005gf,Shih:2005he,Shih:2005uc,Gaiotto:2005hc,David:2006yn,Dabholkar:2006xa,David:2006ud}. The main focus of this section is the counting of quarter-BPS states when the primitivity condition is relaxed. We propose a two dimensional supersymmetric sigma model whose index captures the spectrum of non-primitive dyons \cite{Dabholkar:2008zy}. The resulting index is consistent with many physical tests including a perturbative test \cite{Dabholkar:2009kd} and is in agreement with the answer proposed in \cite{Banerjee:2008pv,Banerjee:2008pu}.

% An important feature of extremal black holes is the attractor mechanism. The
% fact that the theory is supersymmetric at asymptotic infinity implies that no
% potential can be generated for the moduli fields. As a consequence the vaccuum
% is parametrized by a set of constant moduli. Interestingly, at the horizon the
% values of these moduli get attracted to a fixed value which depends only on the
% charges. In other words the horizon data forgets completely about the moduli at
% infinity and therefore the entropy is determined solely in terms of the charges.
% The fact that the entropy does not depend on the string coupling at the horizon,
% one of the moduli of the theory, is the first indication that it may be possible to compute it in a very different regime. 

This section is organized as follows. In section \S2.1 we consider the low energy theory
of Heterotic string on $T^6$ and give general properties of quarter BPS dyons.
In section \S2.2 we focus on U-duality and classification of orbits via charge invariants. In particular we identify an important U-duality invariant $I$ on which the counting depends non trivially. Further in section \S2.3 we analyse the role of invariant $I$ in the microscopic counting, explaining the construction of the partition functions in the cases $I=1$ and $I>1$.

\subsection{Heterotic string on $T^6$: generalities}
We consider heterotic string theory compactified on a six-dimensional torus
$T^6$. This is a four-dimensional string theory with $\mathcal{N}=4$
supersymmetry or sixteen supercharges. It can have  a dual description as IIA or
IIB string theory compactified on $K3\times T^2$ .

The four-dimensional low energy theory contains the metric $g_{\mu\nu}$, the
axion-dilaton $\lambda=a+ie^{-2\phi}$ and six $U(1)$ gauge fields $A_{\mu}$
together with their susy partners sitting in the gravity multiplet. It contains
in addition 22 vector multiplets. Each of these contains a $U(1)$ gauge field
and six real scalars plus susy partners. The axion-dilaton together with the
132=22x6 scalars from the vectors parametrize the moduli space of the theory 
\begin{equation}
 \frac{SL(2,\mathbb{R})}{SL(2,\mathbb{Z})\times SO(2)}\times
\frac{SO(22,6;\mathbb{R})}{SO(22,6;\mathbb{Z})\times SO(22,\mathbb{R})\times
SO(6,\mathbb{R})}.
\end{equation}This theory has U-duality group
\begin{equation}
 G(\mathbb{Z})=SL(2,\mathbb{Z})\times O(22,6;\mathbb{Z})
\end{equation}where the first factor corresponds to electric-magnetic duality
and the second factor corresponds to T-duality.

The 28 gauge fields can carry electric and magnetic $Q,P$ charges which can be arranjed in the dyon charge vector
$$
 \Gamma^i =\left[\begin{array}{c} Q^i \\
			  P^i
                    \end{array}\right]
$$
The index 'i' stands for the vector representation of $SO(22,6;\mathbb{Z})$ and the
electric-magnetic duality acts on the pair $(Q,P)$ by an $SL(2,\mathbb{Z})$
transformation. This is also the S-duality symmetry of the four dimensional
theory that acts on the axion-dilaton. Both the dyon and the axion-dilaton
transform as
\begin{displaymath}
 \lambda\rightarrow \frac{a\lambda+b}{c\lambda+d},\;\;\;\;\\
\left[\begin{array}{c} Q \\
			  P
                    \end{array}\right]\rightarrow \left(\begin{array}{cc}
                                                      a & b \\
						      c & d
                                                    
\end{array}\right)\left[\begin{array}{c} Q \\
			  P
                    \end{array}\right]
\;\;\;\textrm{with} \left(\begin{array}{cc}
                                                      a & b \\
						      c & d
                                                     \end{array}\right) \in
SL(2,\mathbb{Z})
\end{displaymath}

The $\mathcal{N}=4$ superalgebra has central charges
$Z_1(\Gamma,\phi_{\infty})>Z_2(\Gamma,\phi_{\infty})$. A dyon with mass $M$ that
saturates the BPS bound $M=Z_1(\Gamma,\phi_{\infty})$ will preserve 1/4 of the
supersymmetries. Note the dependence on the moduli $\phi_{\infty}$ measured at
infinity. For certain values of $\phi_{\infty}$ the state can become marginally
stable against decay into 1/2-BPS states. These regions are codimension one and
are called walls of marginal stability \cite{Sen:2007vb} . As a consequence the index will jump. 

A 1/4-BPS dyon breaks 12 supercharges out of 16. The 12 fermion zero modes
associated with the broken susys make the Witten index $\textrm{Tr}(-1)^F$
vanish. To correctly account for the additional fermion zero modes we need to
use a modified index \cite{Maldacena:1999bp}. Also known as helicity trace index
or spacetime index, it is defined as
 \begin{equation}\label{indexB6}
 B_6(\Gamma,\phi_{\infty})=-\frac{1}{6!}\textrm{Tr}(-1)^{F}(2h)^6
\end{equation}where $h$ is the helicity quantum number and $F=2h$ is the fermion
number. The insertion of $(2h)^6$ in the usual Witten index has the effect of
rendering the trace over the fermion zero modes non-zero. 

Lets work in more detail the contribution of the fermion zero modes. Each pair
carries $h=\pm1/4$. To simplify the counting we compute first
$g(y)=-\frac{1}{6!}\textrm{Tr}(-1)^Fy^{2h}$ and the index $B_6$ becomes the sixth
derivative of $g(y)$ at $y=1$. Tracing  over the six complex fermion zero modes
we obtain $g(y)=\frac{1}{6!}(y^{1/2}-y^{-1/2})^6$ which, after differentiation,
gives the net result of $1$. In most of the cases we use the Witten index
$\textrm{Tr}'(-1)^F$ where the ' denotes that the trace over the fermion
zero modes has been carried out.  Moreover long supermultiplets carry additional
fermion zero modes so they won't be captured by $B_6$.

The index $B_6$ should be U-duality invariant. This translates to \footnote{Note this equation is only valid in a region of the moduli space. At special codimension one regions the index can jump. Phenomena also known as wall-crossing.}
\begin{equation}
 B_6(\Gamma,\phi_{\infty})=B_6(\Gamma',\phi_{\infty}')
\end{equation}where both $\Gamma$ and $\Gamma'$  and $\phi_{\infty}$ and
$\phi_{\infty}'$ are related by a $G(\mathbb{Z})$ transformation. If two dyons
belong to the same duality orbit, immediately we know that they have the same
index. In the problem of microstate counting it is important to identify duality
orbits through charge invariants.

Both the electric and magnetic charge vectors live in a $\Gamma^{22,6}$ Narain
lattice from which we can construct the continuous T-duality invariants
\begin{equation}
 Q^2=Q^TLQ,\;P^2=P^TLP,\;Q.P=Q^TLP
\end{equation}with $L$ the $SO(22,6,\mathbb{R})$ invariant metric. 

One important continuous U-duality invariant is the quartic Cremmer-Julia
invariant
\begin{equation}
 \Delta(\Gamma)=\textrm{det}(\Gamma\Gamma^T)=Q^2P^2-(Q.P)^2.
\end{equation}Later we will see that the entropy is proportional to
$\sqrt{\Delta}$. Because the U-duality group is discrete we can have more
interesting invariants. One of major importance in the characterization of
duality orbits is the arithmetic invariant \cite{Dabholkar:2007vk}
\begin{equation}
 I=\textrm{gcd}(Q\wedge P)=\textrm{gcd}(Q_iP_j-Q_jP_j).
\end{equation}This invariant will play an important role in the counting of
1/4-BPS dyons.

\subsection{Duality orbits and invariants}

As mentioned before, $\mathcal{N}=4$ string theory has U-duality symmetry
\begin{equation}
 G(\mbbz)=SL(2,\mbbz)\times SO(22,6;\mbbz) 
\end{equation}composed of S and T-duality symmetries. As a consequence the index
should be invariant under U-duality transformations of the charge vectors. 
 
 Under a rotation $\Omega \in SO(22,6;\mathbb{Z})$, the charge vectors transform
as
\begin{equation}
 Q\rightarrow \Omega Q,\,P\rightarrow \Omega P,
\end{equation}while the Lorentzian metric $L$ and the Narain lattice
$\Lambda$ are left invariant
\begin{equation}
 \Omega^T L \Omega=L,\,\Omega\Lambda=\Lambda.
\end{equation}As mentioned before we can construct the T-duality invariants
$Q^2$, $P^2$ and $Q.P$ which are left invariant under the continuous
$G(\mbbr)\supset G(\mbbz)$ U-duality group. Additional discrete invariants can
be constructed. These are necessary to completely characterize a T-duality
orbit.

Consider a dyon with primitive $(Q,P)$ charge vectors, that is, a dyon that
cannot fragment into "smaller" dyons. This means that the charge vector cannot
be written as multiple of a $(Q_0,P_0)$ vector but it doesn't imply that the
electric and magnetic charge vectors have to be individually primitive. We can
represent these charge vectors in a sublattice $\Lambda_0$ generated by
$e_1,e_2$ as
\begin{equation}
 Q=r_1 e_1,\;P=r_2(u_1 e_1+r_3e_2),\;r_1,r_2,r_3,u_1\in\mathbb{Z}^+
\end{equation}such that $\textrm{gcd}(r_1,r_2)=\textrm{gcd}(r_3,u_1)=1$ and
$1\leq u_1\leq r_3$. Recent work on the classification of $SO(22,6;\mathbb{Z})$
T-duality invariants \cite{Banerjee:2007sr} allows the identification of the set
of integers
\begin{equation}
 Q^2,\; P^2,\;Q.P,\; r_1,\;r_2,\;r_3\;\textrm{and}\;u_1
\end{equation}as the complete set of T-duality invariants. 

In these variables the discrete U-duality invariant $I$ becomes $r_1r_2r_3$.
This means that for a primitive dyon, that is, for a dyon with $I=1$,
$r_1=r_2=r_3=u_1=1$ and therefore the orbit becomes labelled by $Q^2$, $P^2$ and
$Q.P$ only. As a matter of fact the partition function for a primitive dyon
depends only on the continuous invariants. For non-primitive dyons it is
expected the index $B_6$ to have non trivial dependence on $I$ and the remaining
integers.

We can also explore the consequence of S-duality on these integers. It was shown
in \cite{Banerjee:2008ri} that the set $(r_1,r_2,r_3,u_1)$ can be brought to the
form $(I,1,1,1)$ by an $SL(2,\mathbb{Z})$ transformation. The charge vector
acquires a much simpler representation
\begin{equation}
 Q=I e'_1,\;P=e'_1+e'_2.
\end{equation}In this new "frame" the derivation of the dyon partition function
becomes easier since most of the invariants are trivial. Moreover
the set $(I,1,1,1)$ is left invariant under the action of a subgroup
$\Gamma^0(I)$ of $SL(2,\mathbb{Z})$ and therefore we expect the index $B_6$ to
exhibit this symmetry explicitly. The subgroup $\Gamma^0(I)$ is defined by
matrices
\begin{displaymath}
  \left(\begin{array}{cc}
                 a & \textrm{mod}(I) \\
		c & d
    \end{array}\right) \in SL(2,\mathbb{Z}) 
\end{displaymath}

\subsection{The dyon partition function}

The Siegel modular form $\Phi_{10}$ is for 1/4-BPS dyons as the ramanujan
function $\Delta=\eta^{24}$ is for 1/2-BPS states. The first is a modular form
of $Sp(2,\mathbb{Z})$, the modular group of genus two riemann surfaces, while
the second is the lower dimensional version, that is, for genus one surfaces.
Using this analogy and consistency with electric and magnetic duality, lead
Dijkgraaf, Verlinde and Verlinde \cite{Dijkgraaf:1996it} long time ago to
propose $\Phi_{10}^{-1}$ as the dyon partition function. This clue was
remarkable and many other works followed in its derivation
\cite{Dabholkar:2006bj,Gaiotto:2005gf,Shih:2005uc,Gaiotto:2005hc}.

In the work \cite{David:2006yn}, which we review next, the authors gave a
detailed derivation of the dyon partition function from first principles.
Nevertheless only primitive dyons were concerned. Later it was shown in
\cite{Dabholkar:2007vk} that the discrete invariant $I$ plays a non-trivial role
in the counting. In \cite{Banerjee:2008pu,Banerjee:2008pv} the authors consider
the case $I>1$ and propose a degeneracy formula based on duality symmetries and
consistency checks much like Dijkgraaf, Verlinde and Verlinde did. Following
this proposal, in \cite{Dabholkar:2008zy} we attempt to give a physical sigma
model interpretation of that result.

\subsubsection{Primitive dyons: $I=1$}\label{sec primitive case}

Also known as Igusa cusp form, $\Phi_{10}$ is the unique weight 10 form of
$Sp(2,\mbbz)$. It depends on three complex numbers which encode the modular
parameters of a genus two riemann surface. They can be packaged in a symmetric
two dimensional matrix
\begin{displaymath}
  \tau=\left(\begin{array}{cc}
                 \rho & v \\
		v & \sigma
    \end{array}\right)  
\end{displaymath}taking values in the Siegel upper half plane, defined as 
\begin{equation}
 \textrm{Im}(\rho)>0,\,\textrm{Im}(\sigma)>0,\,\textrm{Im}(\rho)\textrm{Im}
(\sigma)-\textrm{Im}(v)^2>0.
\end{equation}
 Under a transformation
\begin{displaymath}
  g=\left(\begin{array}{cc}
                 A & B \\
		C & D
    \end{array}\right) \in Sp(2,\mathbb{Z})  
\end{displaymath}with $A,B,C,D$ $2\times2$ matrices, the matrix $\tau$
transforms as
\begin{equation}
 \tau\rightarrow \tau'=(A\tau+B)(C\tau+D)^{-1}
\end{equation}in analogy with $Sp(1)$ modular transformations in a torus.
Correspondingly $\Phi_{10}$ shows the modular property 
\begin{equation}
 \Phi_{10}(\tau')=\textrm{det}(C\tau+D)^{10}\Phi_{10}(\tau).
\end{equation}The subgroup $SL(2,\mbbz)$ can be realized in $Sp(2,\mbbz)$ via
matrices of the form
\begin{displaymath}
  g=\left(\begin{array}{cc}
                 A^T & 0 \\
		0 & A^{-1}
    \end{array}\right) \textrm{with}\;A\in\,SL(2,\mathbb{Z})  
\end{displaymath}As can be easily checked  this transformation leaves
$\Phi_{10}$ invariant. As explained before, $SL(2,\mbbz)$ invariance of the
index concerns the set $(r_1,r_2,r_3,u_1)=(1,1,1,1)$, that is, of primitive
dyons.

 The index is extracted performing an inverse fourier transform of
$\Phi_{10}^{-1}$ 
\begin{equation}\label{dyonPfunction}
 B_6(\Gamma,\phi_{\infty})=(-1)^{Q.P+1}\int_{\mathcal{C}(\phi_{\infty})}
d^3\tau\frac{e^{-\pi i \Gamma^T\tau \Gamma }}{\Phi_{10}(\tau)}.
\end{equation}where the integration goes over a three dimensional torus
\begin{equation}\label{contour1}
 0\leq\textrm{Re}(\rho)\leq 1,\;0\leq\textrm{Re}(\sigma)\leq
1,\;0\leq\textrm{Re}(v)\leq 1
\end{equation}at fixed large values of the imaginary part of $\tau$
\begin{equation}\label{contour2}
 \textrm{Im}(\rho)\gg1,\;\textrm{Im}(\sigma)\gg1,\;\textrm{Im}(v)\gg1.
\end{equation}This defines the integration countour $\mathcal{C}$. Note the
dependence of the integration contour on the moduli space measured at infinity
$\phi_{\infty}$. Later we show that this dependence can lead to wall crossing.
As expected from the analysis of duality orbits of $I=1$ the index shows
dependence on only $Q^2$, $P^2$ and $Q.P$ via $\Gamma^T\tau
\Gamma=Q^2\rho+P^2\sigma+2Q.Pv$. 

\subsubsection{Derivation from physical grounds}

This section is based on \cite{David:2006yn} which we review in the following.

Without loosing generality we can restrict to a charge sub-lattice
$\Gamma^{2,2}\subset \Gamma^{22,6}$ corresponding to the reduction on a
particular two-torus $T^2=S^1\times \tilde{S}^1$. In this sector we have four electric and
four magnetic charges. The charge configuration is taken be of the form
\begin{displaymath}
 \Gamma=\left[\begin{array}{c}
               Q\\
		P
              \end{array}\right]=
\left[\begin{array}{cccc}
                 \tilde{n} & n & \tilde{w} & w \\
		 W & \tilde{W} & \tilde{K} & K
    \end{array}\right]_H.
\end{displaymath}where the indice $H$ denotes the heterotic frame. The charges
$n,\tilde{n}$ denote momentum on the circles $S^1$ and $\tilde{S}^1$
respectively while $w,\tilde{w}$ stand for winding charges on the respective
circles. The magnetic charges $W,\tilde{W}$ correspond to NS5-branes wrapped on
$S^1\times T^4$ and $\tilde{S}^1\times T^4$ respectively. Additionally we can
have Kaluza-Klein monopoles $K,\tilde{K}$ associated with the circles $S^1$ and
$\tilde{S}^1$ respectively. We endow the lattice $\Gamma^{2,2}$ with a metric
$L$ invariant under $SO(2,2,\mbbz)$,
\begin{displaymath}
 L=\left(\begin{array}{cc}
	         0_{2\times2} & \mathbf{1}_{2\times 2}\\
		\mathbf{1}_{2\times 2} & 0_{2\times2} \end{array}\right).
\end{displaymath}With this metric we construct the T-duality invariants
\begin{equation}
 Q^2=2(\tilde{n}\tilde{w}+nw),\;P^2=2(W\tilde{K}+\tilde{W}K),\;Q.P=\tilde{n}
\tilde{K}+nK+W\tilde{w}+\tilde{W}w
\end{equation}

In the presence of NS5-branes the microscopic theory is strongly coupled and
there's not much information we can extract. We avoid this problem by going
to the IIB frame and consider a system of D-branes coupled to KK monopoles where
a weakly coupled description is available. We perform the following chain of dualities. A string-string duality maps Heterotic
string on $T^4\times S^1\times \tilde{S}^1$ to IIA on $K3\times S^1\times
\tilde{S}^1$ which is further T-dualized to give IIB on the dual circle
$\hat{S}^1$ and finally we do a ten dimensional S-duality. Lets see more
carefully what is happening to the charges under this chain of transformations.
\begin{enumerate}
 \item \textbf{Six dimensional string-string duality, Het to IIA}: the momentum
and kaluza klein charges don't transform while the Poincaré electric-magnetic
duality of the six dimensional NS-NS B field takes winding charge to NS5-brane
charge and vice-versa.
\begin{displaymath} 
 \Gamma=\left[\begin{array}{c}
               Q\\
		P
              \end{array}\right]=
\left[\begin{array}{cccc}
                 \tilde{n} & n & \tilde{W} & W \\
		 w & \tilde{w} & \tilde{K} & K
    \end{array}\right]_{IIA}.
\end{displaymath}
\item \textbf{T-duality along $\tilde{S}^1$, IIA to IIB}: this duality maps IIA
on the circle $\tilde{S}^1$ to IIB on the dual circle $\hat{S}^1$. The momentum
and winding charges associated with this circle are exchanged. The same happens
for NS 5-branes and KK monopoles.
\begin{displaymath}
 \Gamma=\left[\begin{array}{c}
               Q\\
		P
              \end{array}\right]=
\left[\begin{array}{cccc}
                 \tilde{w} & n & \tilde{W} & \tilde{K} \\
		 w & \tilde{n} & W & K
    \end{array}\right]_{IIB}.
\end{displaymath}
\item \textbf{Ten dimensional S-duality, IIB to IIB}: this transformation maps
winding charges to D1-branes and NS 5-branes to D5-branes. Other charges remain
untouched.
\begin{displaymath}
 \Gamma=\left[\begin{array}{c}
               Q\\
		P
              \end{array}\right]=
\left[\begin{array}{cccc}
                 \tilde{Q}_1 & n & \tilde{Q}_5 & \tilde{K} \\
		 Q_1 & \tilde{n} & Q_5 & K 
    \end{array}\right]_{IIB}.
\end{displaymath}Here $Q_1$ and $\tilde{Q}_1$ represent charges associated with
D1-branes wrapping a circle $S^1$ and $\hat{S}^1$ respectively. Analogously
$Q_5$ and $\tilde{Q}_5$ represent D5-branes wrapping $K3\times S^1$ and
$K3\times\hat{S}^1$.

\end{enumerate} 

For simplicity we take a charge configuration of the form
\begin{displaymath}
 \Gamma=\left[\begin{array}{c}
               Q\\
		P
              \end{array}\right]=
\left[\begin{array}{cccc}
                 0 & n & 0 & \tilde{K} \\
		 Q_1 & J & Q_5 & 0
    \end{array}\right]_H.
\end{displaymath}which corresponds to a system of $Q_1$ D1-branes and $Q5$ D5-branes wrapping $S^1$ and $K3\times S^1$ respectively in the background of $\tilde{K}$ KK-monopoles, with momentum $n$ and $J$ along the circles $S^1$ and $\tilde{S}^1$. This configuration is also known as D1-D5-KK system. If we
impose primitivity on the charge vectors we get the following condition 
\begin{equation}
 I=\textrm{gcd}(Q\wedge
P)=\textrm{gcd}(Q_1n,Q_1\tilde{K},nQ_5,J\tilde{K},Q_5\tilde{K})=1
\end{equation}which can be satisfied imposing $\tilde{K}=1$ and
$\textrm{gcd}(Q_1,Q_5)=1$. The general case with arbitrary number of KK
monopoles will be studied later for non-primitive dyons. The condition
$\textrm{gcd}(Q_1,Q_5)$ is known to be a physical requirement for the existence
of D1-D5 bound states at threshold
\cite{Vafa:1995bm,Maldacena:1999bp,David:2002wn}.

In weak coupling limit both the D-branes and the KK-monopole are weakly interacting. We can see the D1-D5 brane system moving as a particle in the transverse four
dimensional Taub-Nut (TN) geometry which is the solution of Einstein's equations
in the presence of a Kaluza-Klein monopole. The ten dimensional geometry is
\begin{equation}
 ds^2=-dt^2+ds^2_{Taub-Nut}+ds^2_{K3\times S^1}
\end{equation}with the Taub-Nut metric given by
\begin{eqnarray}\label{Taub-Nut}
 ds^2_{Taub-Nut}=\left(1+\frac{R}{r}
\right)\left(dr^2+r^2d\theta^2+r^2\sin(\theta)^2d\phi^2\right)+R^2\left(1+\frac{
R}{r}\right)^{-1}(2d\psi+\cos\left(\theta\right)d\phi)^2 \nonumber \\
\end{eqnarray} The TN space has the particularity that near the origin $r=0$ it looks
like $\mathbb{R}^4$ while for large $r$ it asymptotes to $\mathbb{R}^3\times
\tilde{S}^1$. From the point of view of the observer at infinity
he sees a theory in four-dimensions. The TN geometry possesses in addition a
normalizable 2-form $\omega_{Taub-Nut}^{(2)}$.

The microscopic theory can be described by three weakly interacting parts. Each
of these can be realized as a two dimensional supersymmetric sigma model on
$\mathbb{R}^t\times S^1$ \cite{David:2006yn}. We denote the weakly interacting parts as
\begin{enumerate}
 \item \textbf{Higgs branch of D1-D5}: describes the moduli space of vacua of
the low energy theory of the D1-D5 brane system on $K3$. In the Higgs branch
\cite{Witten:1997yu} the (1,5) strings acquire vevs forcing the D1 and D5 branes
to sit on top of each other. In the IR, the low energy theory is described by a
two dimensional SCFT with sigma model the Hilbert scheme of $Q_1Q_5+1$ points on
$K3$ which is isomorphic to the symmetric product of $K3$ at the orbifold point
\begin{equation}
 \mathcal{M}_{D1-D5}=\textbf{Sym}^{Q_1Q_5+1}(K3)
\end{equation}
This is a (4,4) SCFT with R-symmetry $SU(2)_L\times SU(2)_R$. The R-symmetry
corresponds to rotations in the transverse space.
\item \textbf{Center of mass motion of the D1-D5 system}: it describes the
vector multiplet degrees of freedom of (1,1) and (5,5) strings. We can see the
D1-D5 as a particle moving in the TN space. From the motion on TN we have 4
scalars transforming under the vector representation of $SU(2)_L\times SU(2)_R$
and 4 left-moving and 4 right-moving fermions transforming in the fundamental of
$SU(2)_R$ and $SU(2)_L$ respectively. The Taub-Nut background breaks half the
susy's. This gives rise to a $(0,4)$ SCFT on $\mathbb{R}^4$. The sigma model has
target space
\begin{equation}
 \mathcal{M}_{CM}=\mathbb{R}^4
\end{equation}This target space contrasts with the curved TN. We note that the
index is a quantity that doens't depend on the parameters of the theory. We use
this property to compute the index for very large $R$ radius which is equivalent
as putting the D1-D5 at $r=0$, where TN looks like $\mbbr^4$. This comment fails
for the case of zero modes where we need to be more careful. 

\item \textbf{KK monopole closed string excitations}: this describes the low
energies excitations of closed strings in the Taub-Nut background. We have 3
massless scalars coming from the breaking of $\mathbb{R}^3$ translation.
Additionally the reduction of the Ramond-Ramond C 4-form on
$\omega_{Taub-Nut}^{(2)}\wedge \omega_{K3}^{(2)}$ gives 19 left-moving scalars
and 3 right-moving scalars. Additionally the NS-NS B field and the Ramond-Ramond
2-form give together 2 extra scalars. In total we have 24 left-moving and 8
right-moving scalars. The Taub-Nut preserves half susy's of IIB on $K3$. This
gives in addition 8 right-moving fermions. We denote the resulting sigma model
by $\sigma_{KK}$. 

\end{enumerate}

The analysis of the zero modes of the supersymmetric field theory on Taub-Nut
requires special care. The dynamics is of a $\mathcal{N}=4$ superparticle with 4
bosonic and 4 fermionic coordinates moving in the Taub-Nut space. So far we have
been working in a point in the moduli space where $S^1$ and $\tilde{S}^1$ are
orthogonal. A mixing between the circles can be achieved by a $d\psi+ady$
translation. As a consequence the tension of D1-D5 brane system $\sqrt{g_{yy}}$ generates a
potential
\begin{equation}
 V(r)=a^2R^2\left(1+\frac{R}{r}\right)^{-1}
\end{equation}which under supersymmetrization originates other fermionic terms.
Under this potential supersymmetric bound states can form and contribute to the
total index.

Concerning the fermionic zero modes resulting from the broken susy's, the analysis goes as
follows. IIB string theory on $K3\times\textrm{TN}$ preserves 8 left-moving
susy's on the $1+1$ world volume theory. The breaking of 8 supersymmetries gives
rise to 4 complex fermion zero modes. Additionally the D1-D5 breaks 4 of the
remaining 8 susy's contributing with 2 complex fermion zero modes. This gives a
total of 6 complex fermion zero modes as expected for a 1/4-BPS dyon. 

We now proceed to the construction of index. 

We use the index $B_6'=\textrm{Tr}(-1)^{2h}$ where tracing over fermion zero modes has been carried out. We find convenient to compute the generating function, also known as elliptic genus,
\begin{equation}
 \chi(q,\bar{q},y,\tilde{y};\mathcal{M})=\textrm{Tr}_{\textrm{R-R}}(-1)^{
2J_L-2J_R}q^{L_0} \bar{q}^{\bar{L}_0}y^{2J_L}\tilde{y}^{2J_R}
\end{equation}
with
\begin{equation}
 q=e^{2\pi i \rho},\;y=e^{2\pi i v},\;\tilde{y}=e^{2\pi i \tilde{v}},
\end{equation}
which corresponds to the partition function of the sigma model with
Ramond-Ramond boundary conditions. The generators $L_0$ and $\bar{L}_0$ are the
usual left and right Virasoro dilatation generators while $J_L$ and $J_R$
correspond to the Cartan generators of $SU(2)_L\times SU(2)_R$, the little group
in five dimensions. We contrast this with the little group  in four
dimensions which is $SO(3)$. Due to the particular fibration structure of the TN space
(\ref{Taub-Nut}) there is an interesting connection between five and four
dimensional black holes known as 4d-5d lift \cite{Gaiotto:2005gf}. While at the tip of
TN space the geometry looks like $\mathbb{R}^4$, at asymptotic infinity it looks
like $\mathbb{R}^3\times \tilde{S}^1$. If now we put the D1-D5 system at the tip
of the TN, the transverse space looks five dimensional. Therefore we can relate
the degrees of freedom of the five dimensional BMPV black hole
\cite{Breckenridge:1996is} to the D1-D5-KK four dimensional black hole. The rotation generator $J_L$ measured at $r=0$ is further identified
with $U(1)$ translations on the circle $\tilde{S}^1$ at asymptotic infinity.

Due to supersymmetry, the right movers are forced to stay in the ground state and therefore
the dependence of the function $\chi$ on $\bar{q}$ drops meaning that only BPS
states are being counted. We now show the different contributions to the
generating function:
\begin{enumerate}
 \item \textbf{Higgs branch of D1-D5}:
\begin{equation}
 \sum_{N=0}^{\infty}p^{N-1}\chi(q,y;\textbf{Sym}^N(K3))=\frac{1}{p}\prod_{n\geq1
,\,m\geq 0,\,l\in \mathbb{Z}}\frac{1}{(1-p^nq^my^l)^{c(4nm-l^2)}}
\end{equation}with $c(n)$ defined via the equation
\begin{equation}
 \chi(q,y;K3)=8\left[\frac{v_2(\tau,z)^2}{v_2(\tau,0)^2}+\frac{v_3(\tau,z)^2}{
v_3(\tau,0)^2}+\frac{v_4(\tau,z)^2}{v_4(\tau,0)^2}\right]=\sum_{n,\,j\in
\mathbb{Z}} c(4n-j^2)e^{2\pi i n +2\pi ijz}
\end{equation}
\item \textbf{CM contribution}:
\begin{equation}
 \chi(q,y;\mathbb{R}^4)=\frac{\prod_{n\geq
1}(1-q^n)^4}{\prod_{n\geq1}(1-q^ny)^2(1-q^ny^{-1})^2}
\end{equation}
\item \textbf{KK closed string excitations}:
\begin{equation}
 \chi(q,\sigma_{KK})=\textrm{Tr}(-1)^Fq^{L_0}\bar{q}^{\bar{L}_0}=\frac{1}{q}
\frac{1}{\prod_{n\geq1}(1-q^n)^{24}}
\end{equation}This is a four dimensional index. States don't carry charge
$\tilde{n}$ and dependence on $y$ drops. This index is the same as the 1/2-BPS
index that counts electric states in the heterotic string. In fact the system
KK-P can be mapped to Heterotic momentum-winding states using duality symmetry.

\item \textbf{$\mathcal{N}=4$ superparticle in Taub-Nut}:
\begin{equation}\label{TNsuperparticle}
 \textrm{Tr}(-1)^{F}y^{\tilde{J}}=\sum_{j\geq1} j e^{2\pi ijv}=\frac{e^{2\pi i
v}}{(1-e^{2\pi i v})^2}
\end{equation}where $\tilde{J}$ is the momentum charge on the circle
$\tilde{S}^1$. 
Note that the last expression can be expanded either in powers of $e^{2\pi iv}$
or $e^{-2\pi iv}$. This generates ambiguity when trying to extract the fourier
coefficient. We show later that this is related to wall-crossing.

\end{enumerate}
Putting all factors together we get
\begin{equation}
  -\frac{1}{pqy}\prod_{\substack{n,m\geq0,l\in \mathbb{Z}\\ l<0\,for\,k=l=0
                        }}\frac{1}{(1-p^nq^my^l)^{c(4nm-l^2)}}
\end{equation}with
\begin{equation}
 p=e^{2\pi i \sigma},\;q=e^{2\pi i \rho},\;y=e^{2\pi i v}
\end{equation}which is equal to $-\Phi_{10}^{-1}$.

The chemical potentials $\rho,\,\sigma$ and $v$ couple to $P^2=2Q_1Q_5$,
$Q^2=2N$ and $Q.P=\tilde{J}$ respectively. To extract the index $B_6$ we perform
an inverse fourier transform
\begin{equation}\label{partition function}
 B_6(Q^2,P^2,Q.P)=(-1)^{Q.P+1}\int_{\mathcal{C}} d\rho d\sigma dv \frac{e^{-i\pi
\rho Q^2-i\pi \sigma P^2 -2\pi i vQ.P }}{\Phi_{10}(\rho,\sigma,v)}
\end{equation}where the contour $\mathcal{C}$ is as given in
(\ref{contour1},\ref{contour2}). The additional factor $(-1)^{Q.P}$ is
reminiscent of going from five to four dimensions
\cite{Gaiotto:2005gf,Shih:2005he}. 

\subsubsection{Consistency checks}
In the limit of large charges $Q^2\gg1$, $P^2\gg1$ and $Q.P\gg1$ we can make an
asymptotic expansion of $B_6$ (\ref{partition function}). The leading term can then be compared with the
black hole entropy valid in the same limit. The idea is to deform the contour
$\mathcal{C}$ such that it passes near a pole whose residue contribution is much
leading than the left over integral
\cite{Dijkgraaf:1996it,David:2006yn,Banerjee:2008ky}. The Siegel form
$\Phi_{10}$ has second order zeros at
\begin{equation}
 n_2(\rho\sigma-v^2)+bv+n_1\sigma-m_1\rho+m_2=0,
\end{equation}with $n_1,n_2,m_1,m_2\in \mbbz$ and $b\in 2\mbbz+1$ obeying the
condition $n_1m_1+n_2m_2+b^2/4=1/4$. The residue at
$(n_1,n_2,m_1,m_2,b)=(1,0,0,0,1)$, modulo $SL(2,\mbbz)$ transformations, gives
the leading term  in the asymptotic expansion which is the correct
result for the entropy
\begin{equation}\label{BHentropy}
 B_6(Q^2,P^2,Q.P\gg1)\approx e^{\pi\sqrt{\Delta(Q,P)}}+
\mathcal{O}(e^{\pi\frac{\sqrt{\Delta}}{2}})
\end{equation}Subleading perturbative corrections to the microscopic answer can
be computed. In fact for sufficiently large charges we can approximate $B_6$ by 
\begin{equation}\label{entropy function from Siegel}
 B_6\approx K_0(-1)^{Q.P}\int
\frac{d^2\tau}{\tau_2^2}\left(26+\frac{\pi}{\tau_2}|Q+\tau
P|^2\right)e^{\frac{\pi}{2\tau_2}|Q+\tau
P|^2-24\ln\eta(\tau)-24\ln\eta(-\bar{\tau})-12\ln(\tau_2)}
\end{equation}which in the saddle point approximation reduces to
(\ref{BHentropy}). Subleading non-perturbative corrections are suggestive of
multi-center black hole contribution \cite{Banerjee:2008ky}.

The residues for $n_2=0$ are even more subleading. They encode phenomena
associated with wall-crossing.
Although the integrand in (\ref{partition function}) is manifestly $SL(2\mbbz)$
invariant the contour is not. After such transformation we may cross a pole in
deforming the contour to its original form. It happens that only a pole $n_2=0$
can be crossed in the deformation. Take for example the residue at
$(n_1,n_2,m_1,m_2,b)=(0,0,0,0,1)$ which corresponds to the pole $v=0$. Near this
pole the partition function behaves like
\begin{equation}
 \frac{1}{\Phi_{10}}\approx \frac{1}{v^2\eta^{24}(\rho)\eta^{24}(\sigma)}
\end{equation}
 In this case the index jumps by the amount
\begin{equation}\label{jumpindex}
 \Delta B_6=\textrm{Res}_{v=0}=(-1)^{Q.P}(Q.P)\int \frac{e^{-\pi i\rho
Q^2}}{\eta^{24}(\rho)}\int \frac{e^{-\pi i\sigma P^2}}{\eta^{24}(\sigma)}
\end{equation}In a different context, this can be easily recognized as the Denef's split attractor
formula for 1/2-BPS black holes in $\mathcal{N}=2$ supergravity
\cite{Denef:2000nb,Denef:2002ru,Denef:2007vg}
\begin{equation}
 \Delta \Omega=(-1)^{\langle \Gamma^1,\Gamma^2\rangle}\langle
\Gamma^1,\Gamma^2\rangle\Omega(\Gamma^1)\Omega(\Gamma^2)
\end{equation}with $\langle \Gamma^1,\Gamma^2\rangle=Q.P$ and $\Omega(\Gamma)$
is the index that counts 1/2-BPS states. In $\mathcal{N}=4$ string theory the
index that counts 1/2 BPS states is \cite{Dabholkar:1989jt}
\begin{equation}
 d_{1/2}(Q^2)=\int \frac{e^{-\pi i\rho Q^2}}{\eta^{24}(\rho)}
\end{equation}

From a microscopic point of view we can understand the jump in the index as the
decay of a 1/4-BPS dyon into its 1/2-BPS constituents, that is,
\begin{equation}
 (Q,P)\rightarrow (Q,0)+(0,P).
\end{equation}From the gravity point of view it corresponds to the appearance or
disappearance of a two center black hole \cite{Sen:2007pg}. In fact for large
charges the jump in the index can be interpret as coming from the contribution
of two centers which are very far from each other
\begin{equation}
 \ln(\Delta B_6)\approx 2\pi\sqrt{Q^2}+2\pi\sqrt{P^2}.
\end{equation}
Other poles with $n_2=0$ correspond to more complex decays which are basically
related by a $SL(2,\mbbz)$ transformation to the $v=0$ case. For more details we
refer the reader to \cite{Sen:2007vb}.

Physically the picture is the following. The $SL(2,\mbbz)$ transformation acts
not just on the charges but also on the axion-dilaton $\lambda_{\infty}$. Since the mass has a non-trivial dependence on $\lambda_{\infty}$ it will change as we move on the moduli space. When the mass of the quarter-BPS dyons equals the sum of the masses of the half-BPS dyons for $\lambda *$, that is, 
\begin{equation}
 m_{1/4-BPS}(Q,P)|_{\lambda *}=m_{1/2-BPS}(Q)+m_{1/2-BPS}(P)
\end{equation}it becomes marginally stable and can decay into its 1/2-BPS constituents. The regions in the moduli space where the dyon becomes marginally
stable are codimension one walls. Schematically we have the moduli space divided
into chambers $(X,X',X'',\ldots)$ separated by codimension one walls. The index
$B_6$ is piecewise constant in these chambers.

Consider the example of a dyon with $Q^2=P^2=-1$ and $Q.P=j>0$. We can easily
extract the index $B_6$ from (\ref{TNsuperparticle}). It gives
$B_6=(-1)^{j+1}j$. Under a S-duality transformation
\begin{displaymath}
 \left[\begin{array}{c}
        Q\\
	P
       \end{array}\right]
 \rightarrow
\left(\begin{array}{cc}
                 0 & 1\\
		 -1 & 0
    \end{array}\right) \left[\begin{array}{c}
        Q\\
	P
       \end{array}\right]
\end{displaymath}the T-duality invariants are mapped to $Q^2=P^2=-1$ and
$Q.P=-j<0$. This is equivalent to the change of contour $\textrm{Im}v\rightarrow
-\textrm{Im}v $. In deforming the countour to its original value we pick a
residue at $v=0$. In this case the jump is easy to compute and formula
\ref{jumpindex} gives $\Delta B_6=(-1)^{j+1}j$.  At the same time the
axion-dilaton gets transformed to $-1/\lambda$ and the dyon jumps from one
chamber to another separated by a wall at $\textrm{Re}(\lambda)=0$. In this new
chamber the index (\ref{TNsuperparticle}) only contains positive powers of
$e^{2\pi iv}$ which gives a zero index consistent with the predicted jump. 

In \cite{Cheng:2007ch} the authors propose a contour which captures only the
contribution from single center black holes. In this case the index becomes
moduli independent and therefore the dyon is free from decaying. The
prescription is the following
\begin{eqnarray}
 &&\textrm{Im}(\rho)=\Lambda\left(\frac{|\lambda|^2}{\lambda_2}+\frac{Q_R^2}{
\sqrt{\Delta_R}}\right)\\
&&\textrm{Im}(\sigma)=\Lambda\left(\frac{1}{\lambda_2}+\frac{P_R^2}{\sqrt{
\Delta_R}}\right)\\
&&\textrm{Im}(v)=-\Lambda\left(\frac{\lambda_1}{\lambda_2}+\frac{Q_R.P_R}{\sqrt{
\Delta_R}}\right)
\end{eqnarray}with $Q_R^2=Q^T(M+L)Q$, $P_R^2=P^T(M+L)P$, $Q_R.P_R=Q^T(M+L)P$ and
$\Delta_R=Q_R^2P_R^2-(Q_R.P_R)^2$. The matrix $M$ is a symmetric $28\times28$
matrix which encodes the 132 moduli of the theory and obeys the constraint
$M^TLM=L$, with $L$ the metric on $\Gamma^{6,22}$. The scalar $\lambda$ is the
axion-dilaton. The parameter $\Lambda$ is taken to be very large to
ensure the dyon doesn't leave this chamber.

\subsection{Non-primitive dyons: $I>1$} 

Derivation of the spectrum of non-primitive dyons from physical grounds is more
complex. As a matter of fact, it was noted long time ago that the counting of
non-primitive charge vectors, in the context of toroidally compactified IIB
string theory, was a difficult problem \cite{Maldacena:1999bp}. The case of the
D1-D5 system with $Q_1$ and $Q_5$ not coprime is a good example. Since the
system can split at no cost of energy, this signals the presence of
singularities in the moduli space of the low energy theory \cite{Vafa:1995bm}.

In the case of 1/4-BPS dyons with non-primitive charge vectors, similar
difficulties are encountered. Consider a charge configuration of the form
\begin{equation}\label{nonprimitive charge configuration}
 \Gamma=\left[\begin{array}{c}
               Q\\
		P
              \end{array}\right]=
\left[\begin{array}{cccc}
                 0 & nI & 0 & kI \\
		 Q_1 & J & Q_5 & 0
    \end{array}\right]_H.
\end{equation}with $(Q_1,Q_5)$ coprime. We choose charges such that
$\textrm{gcd}(Q\wedge P)=I>1$. In this case we have to consider a configuration multi KK-monopoles. If we were to repeat the analysis done in the
$I=1$ case, we would face the following difficulties
\begin{enumerate}
 \item Multi KK monopoles have collective coordinates which parametrize a
non-trivial moduli space. The study of bound states in this background is a very
difficult problem.
\item The multi KK geometry admits $I$ non-trivial 2-cycles. For each pair of KK
monopoles there is a 2-cycle that touchs both of them \cite{Sen:1997kz}. The
area of this 2-cycle is proportional to the distance between the two centers and
approches zero when the monopoles touch each other. A D3-brane wrapping such
cycle will give rise to tension less strings \cite{Witten:1995zh}. In the
counting we should consider a possible contribution from these strings.
\end{enumerate}

Aware of these problems, the authors in \cite{Banerjee:2008pu,Banerjee:2008pv}
proposed an index formula much as Dijkgraaf, Verlinde, Verlinde have made for
the case of primitive dyons. This formula is consistent with many properties
known for non-primitive dyons and corresponding black holes. The proposed index
has the form
\begin{equation}\label{Sen nonprimitive Pfunction}
 d(Q,P)=(-1)^{Q.P+1}\sum_{s|I} s^4 \int_{\mathcal{C}(s)} d^3\tau \frac{e^{-\pi i
\Gamma^T\tau \Gamma}}{\Phi_{10}(\rho,s^2\sigma,sv)}
\end{equation}with contour
\begin{equation}
 \mathcal{C}(s):\;0\leq \textrm{Re}\rho\leq 1,\;0\leq \textrm{Re}\sigma \leq
\frac{1}{s^2},\;0\leq \textrm{Re}v\leq \frac{1}{s}.
\end{equation}After a simple manipulation we can write it in a more convenient
form
\begin{equation}\label{non primitive degeneracy}
 d(Q,P)=\sum_{s|I}s\,d_1\left(\frac{Q^2}{s^2},P^2,\frac{Q.P}{s}\right)
\end{equation}where $d_1(a,b,c)$ denotes the fourier coefficient extracted from
the primitive answer (\ref{partition function}). The main driving principle for
the such construction is based on wall crossing for non-primitive decay. In the
case of a primitive decay\footnote{A primitive decay is one for which the products of the decay are primitive dyons.}, there is a one to one correspondence between the
decay and the pole in $\Phi_{10}^{-1}$. That is, take the most general primitive
decay
\begin{equation}
 (Q,P)\rightarrow (\alpha Q+\beta P,\gamma Q+\delta P)+(\delta Q-\beta P,-\gamma
Q+\alpha P)
\end{equation}with $\alpha \delta=\gamma\beta$ and $\alpha+\delta=1$. The set of
integers $(\alpha,\beta,\gamma,\delta)$ gives the location of the pole at $\rho \gamma -\sigma \beta+v(\alpha-\beta)=0$.

Once $I>1$ a marginally stable dyon can decay into products which are
non-primitive. This allows for a larger set of integers
$(\alpha,\beta,\gamma,\delta)$. In \cite{Banerjee:2008pv} they postulate that
such correspondence should remain even in the non-primitive case. This was
helpful in suggesting part of the pole structure of the partition function which
is indeed that of (\ref{Sen nonprimitive Pfunction}).

Take the example of a non-primitive decay
\begin{equation}
 (IQ_0,P)\rightarrow (IQ_0,0)+(0,P).
\end{equation}associated with the pole at $v=0$. The wall crossing formula
extracted from the residue of (\ref{Sen nonprimitive Pfunction}) gives a jump of the form
\begin{equation}
 \Delta d(Q,P)=(-1)^{Q.P}(Q.P)\sum_{s|I} d_{1/2}(Q^2/s^2)d_{1/2}(P^2)
\end{equation}

 Again for large charges, the term $s=1$ in (\ref{Sen nonprimitive Pfunction}) gives the leading contribution reproducing correctly the black hole entropy
\begin{equation}
 d(Q,P)\approx \sum_{s|I} e^{S_{BH}/s}.
\end{equation}with $S_{BH}=\pi\sqrt{\Delta}$. 

One additional requirement is the invariance
of the index under $\Gamma^0(I)\in SL(2,\mbbz)$. S-duality invariance demands
that $\Gamma'^T\tau\Gamma'=\Gamma^T\tau'\Gamma$ with $\tau'=(h^T)^{-1}\tau
h^{-1}$ and $h\in\Gamma^0(I)$. By embedding this subgroup in $Sp(2,\mbbz)$ via
\begin{displaymath}
  g=\left(\begin{array}{cc}
                 (h^T)^{-1} & 0 \\
		0 & h
    \end{array}\right) \textrm{with}\;h\in\,\Gamma^0(I)  
\end{displaymath}we can show that the integrand (\ref{Sen nonprimitive
Pfunction}) is left invariant due to
\begin{equation}
 \Phi_{10}(\rho',s^2\sigma',sv')=\Phi_{10}(\rho,s^2\sigma,sv).
\end{equation}There is yet another important check to this formula. At special
points in the moduli space of the heterotic string on $T^6$ we can have enhanced
gauge symmetry. Away from these points the symmetry is spontaneously broken and
the moduli fields play the role of the Higgs field. If their vevs are small the
symmetry breaking scale is small compared to the string scale. In this case the
theory contains states with masses much smaller then massive string states and
should be part of the dyon spectrum. In particular it should
include 1/4-BPS dyons of $\mathcal{N}=4$ SYM. 

Dyon charges in $SU(N)$ gauge theory are labelled by N-dimensional root vectors
labelled by a pair $(q,p)$. It can be shown that a primitive embedding of the
root lattice in the Narain lattice is possible. This means that 
\begin{equation}
 I=\textrm{gcd}(Q\wedge P)=\textrm{gcd}(q\wedge p).
\end{equation}Moreover the T-duality metric $L$ is the negative of the Cartan
metric which gives the following assignments
\begin{equation}
 q.q=-Q^2,\;p.p=-P^2,\;q.p=-Q.P
\end{equation}Additionally we have $q^2,p^2\geq0$ and $(q.p)^2\leq(q^2+p^2)/2$
because the Cartan metric is positive definite and therefore it implies $Q^2,P^2<0$ and $(Q.P)^2<-(Q^2+P^2)/2$. 

Counting BPS dyons in SYM is the problem analysed in \cite{Stern:2000ie}. The
authors computed an index $\mathcal{I}$ in the gauge theory for dyons with
torsion $r=\textrm{gcd}(q\wedge p)$ and found
\begin{equation}\label{SternYi}
 \mathcal{I}=r.
\end{equation}In terms of string theory dyons the conditions mentioned above
imply $Q^2/2=-I^2$, $P^2=-1$ and $Q.P=\pm I$. Neglecting issues of chamber
dependence, formula (\ref{non primitive degeneracy}) gives for these dyons
$d(Q,P)=I$ which agrees with (\ref{SternYi}). 

\subsubsection{Proposed sigma model from physical grounds}
 
The geometry associated with $I$ KK monopoles is the generalization of
(\ref{Taub-Nut}) to include multi centers 
\begin{equation}
 ds^2=V^{-1}(dx^4+\vec{\omega}.\vec{dx})^2+V\vec{dx}.\vec{dx}
\end{equation}where $x^4$ is a compact direction and $\vec{x}$ is the position
in $\mbbr^3$. The harmonic function $V$ and the connection $\vec{\omega}$ are
defined as
\begin{eqnarray}
 V=1+\sum_{s=1}^I V_s,\;\vec{\omega}=\sum_{s=1}^I\vec{\omega}_s\\
V_s=\frac{4r}{|\vec{x}-\vec{x}_s|},\;\vec{\nabla}\times \vec{\omega}_s=\nabla
V_s
\end{eqnarray}At asymptotic infinity, when $|\vec{x}|$ is very large, the
geometry looks like $\mbbr^3\times \tilde{S}^1$. The moduli $\vec{x}_s$ denote
the position of the each of the $I$ monopoles. If we zoom very close to one
center the geometry looks like $\mbbr^4$ given that $x^4$ has periodicity $16\pi
r$ to avoid a conical singularity.

In order to preserve supersymmetry we should consider the case when all the
monopoles sit on top of each other.

In this case the geometry looks like that of a single monopole with charge $I$
but with a conical singularity at the origin. The TN space becomes an
asymptotically locally euclidean space (ALE) $\mathbb{C}^2/\mbbz_I$. The subgroup
$\mbbz_I$ is embedded in $SU(2)_L$ of the tangent group $SU(2)_L\times SU(2)_R$
defined at the origin, preserving this way the same number of supercharges as a
single KK monopole \footnote{The holonomy group of TN is $SU(2)_R$, which means
that it breaks half of the background supersymmetries}. At asymptotic infinity
the radius of $\tilde{S}^1$ is measured in units of $1/I$ due to the orbifold.
This implies that an asymptotic observer will measure a total momentum charge
which is a multiple of $I$. This is consistent with the fact that $Q.P$ is a
multiple of $I$ (\ref{nonprimitive charge configuration}).

This instructs us to study the D1-D5 in transverse $\mathbb{C}^2/\mbbz_I$ which is the problem studied in \cite{Okuyama:2005gq}. The author uses the standard
approach of supersymmetric gauge field theory in ALE spaces
\cite{Douglas:1996sw}. We start by going to the covering space of
$\mathbb{C}^2/\mbbz_I$ which means enhancing the gauge group $U(N_1)\times
U(N_2)\times\ldots$ of the gauge theory in $\mathbb{C}^2$ to $U(IN_1)\times
U(IN_2)\times\ldots$. In our case the D1-D5 system in transverse $\mathbb{C}^2$
is a supersymmetric gauge theory with gauge group $U(Q_1)\times U(Q_5)$. A
careful analysis of the D-terms of the enhanced gauge theory reveals that the
moduli space of vacua factorizes with some additional identifications. This last
point is not clear in \cite{Okuyama:2005gq}. Denoting the moduli space of the
D1-D5 on $\mathbb{C}^2$ by $\mathcal{M}_1$. we propose
\begin{equation}
 \mathcal{M}_I=\textrm{Sym}^I(\mathcal{M}_1)/\mbbz_I.
\end{equation}The identification under permutations comes from gauging the
moduli space by the Weyl group $S^I\subset U(IQ_1)\times U(IQ_5)$ left unbroken
in the Higgs phase while the $\mbbz_I$ orbifold comes from the breaking of the
$SU(2)_L\times SU(2)_R$ R-symmetry of the parent $(4,4)$ theory to $U(1)_L\times
SU(2)_R$.

In other words we propose that the effective string that describes low energy
fluctuations of the D1-D5 system on $\mathbb{C}^2/\mbbz_I$ has a sigma model
with target space
\begin{equation}\label{simga model D1D5 on ALE}
 \sigma_I=\textrm{Sym}^I\left(\sigma(\mbbr)^4\times
\textrm{Sym}^{Q_1Q_5+1}(K3)\right)/\mbbz_I
\end{equation}where $\mbbz_I$ belongs to the $SU(2)_L$ R-symmetry of the parent
theory, that of D1-D5 on $\mathbb{C}^2$.  This space is singular in contrast
with $I=1$ case. Although we don't know how to resolve these singularities, the
index is well defined and can be computed. As for $I=1$ the effective string is
a $(0,4)$ SCFT.

Since we are interested in black holes which are not charged under $\mbbz_I$ we look for states of the untwisted sector, that is, states of the parent
$(4,4)$ theory invariant under $\mbbz_I$. This is equivalent to look for
states that carry $U(1)_L$ charge that is a multiple of $I$.

The sigma model carries two complex fermion zero modes, originally from the
$\mbbr^4$ factor, which have to be symmetrized along with the other states. To
correctly account for these we should use the appropriate helicity trace
\begin{equation}\label{B2index 3}
 B_2=-\frac{1}{2}\textrm{Tr}(-1)^{2J_L-2J_R}(2J_R)^2
\end{equation}in the same spirit of (\ref{indexB6}). Because the zero modes are
being symmetrized they have a non-trivial contribution to the index like in
\cite{Maldacena:1999bp}. The application of the theorem of symmetrized products
\cite{Dijkgraaf:1996xw} gives 
\begin{equation}\label{B2index}
 B_2(\sigma_I)=\sum_{s|I,s|n}
s\,\hat{c}\left(Q_1Q_5,\frac{nI}{s^2},\frac{lI}{s}\right)
\end{equation}with the coefficients $\hat{c}(a,b,c)$ defined via
\begin{equation}\label{hatc}
 B_2(\sigma_1)=\hat{c}(Q_1Q_5,n,l)
\end{equation}which is the answer for the D1-D5 on $\mathbb{C}^2$. The charge
$n$ is the momentum along the circle $S^1$ and $l$ is the angular momentum of
the black hole in five dimensions.

A derivation of (\ref{B2index}) goes as follows. 
The Hilbert space of a symmetrized product can be decomposed as a sum labelled
by partitions of $I$
\begin{equation}\label{hilbert D1D5 on C^2/ZI}
 \mathcal{H}_I=\sum_{\sum kN_k=I}\prod_k\otimes S^{N_k}(\mathcal{H}_{k}) 
\end{equation}with
\begin{equation}
 S^{N}(\mathcal{H})=\sum_{\textrm{permutations}\,\sigma}\,
\epsilon(\sigma)\prod_N\otimes\mathcal{H},
\end{equation}where $\epsilon=(-1)^{\sigma}$ for fermionic states and
$\epsilon=1$ for bosonic. The Hilbert space $\mathcal{H}_{k}$ denotes a multiple
wound string with size $k$ such that $\mathcal{H}_{1}$ is the effective string
in the case $I=1$.

Each of $\mathcal{H}_{k}$ carries two complex fermion zero modes. Now a state $|
h_{k}\rangle\in S^{N_k}(\mathcal{H}_{k})$ contributes to $J_R$ with $h_{k}$. The
operator $(2J_R)^2$ in (\ref{B2index 3}) becomes
$(2J_R)^2=\sum_k(2h_{k})^2+\sum_{k\neq l} (2h_{r_k}2h_{r_l})$ for a partition
$(N_k,k)$. Due to the presence of fermion zero modes, the trace of
$(-1)^{2J_R}(2J_R)^2$ over $\prod_k\otimes S^{N_k}(\mathcal{H}_{k})$ is zero
unless the partition obeys $kN_k=I$. In this case the index becomes 
\begin{equation}\label{B2 trace over symmetric states}
 B_2=\sum_{k\textrm{ with
}kr=I}\textrm{Tr}_{S^{r}(\mathcal{H}_{k})}(-1)^{2J-2h}(2h)^2. 
\end{equation}A state in $S^{r}(\mathcal{H}_{k})$ is given by symmetric or
antisymmetric wave functions $\sum_{\sigma}\epsilon(\sigma)\prod_i
|n_i,J_i,h_i\rangle$ if they are bosonic or fermionic accordingly. Each state
carries total momentum $\sum n_i=N$ and total angular momentum $\tilde{J}=\sum 2J_i$. Instead of computing directly the trace in (\ref{B2 trace over
symmetric states}), which is not trivial, we find it more convenient to compute
the partition function first
\begin{equation}
 f(\beta,q,y)_r=\textrm{Tr}_{S^{r}(\mathcal{H}_{k})}(-1)^{\sum 2J_i-\sum
2h_i}e^{\beta\sum 2h_i}q^{\sum n_i}y^{\sum 2J_i}.
\end{equation}and then extract $B_2$ by looking to a particular fourier
coefficient 
\begin{equation}\label{step1}
 B_2=\textrm{Coeff }q^Ny^{\tilde{J}}\textrm{ in }
\frac{\partial^2}{\partial\beta^2}f(\beta,q,y)|_{\beta=0}
\end{equation}
In $\mathcal{H}_k$ there is an orbifold action by $\mbbz_k$. The trace in the
untwisted sector gives
\begin{equation}
 \textrm{Tr}_{\mathcal{H}_k}(-1)^{ 2J-2h}e^{\beta 2h}q^{L_0}y^{2J}=\sum
c(km,\tilde{l},l)q^my^le^{\beta \tilde{l}}.
\end{equation}after projecting out states which are not invariant under
$\mbbz_k$. We are now in good position to apply the theorem (2.15) in
\cite{Dijkgraaf:1996xw} which tells how to trace over $S^{r}(\mathcal{H}_{k})$,
\begin{equation}
 \sum_{N\geq0}p^N\textrm{Tr}_{S^N(\mathcal{H}_k)}(-1)^{
2J-2h}e^{\beta2h}q^{L_0}y^{J}=\prod_{n,\tilde{l},l}\frac{1}{(1-pq^{n}y^le^{\beta
\tilde{l}})^{c(kn,\tilde{l},l)}}.
\end{equation}The function $f(\beta,q,y)$ is given by the coefficient of $p^r$
in the expression above. Even if this seems a hard task it is easier to perform
first the step (\ref{step1}) and then extract the $p^r$ coefficient, that is,
\begin{equation}
 \textrm{Coeff }p^r\textrm{
off}\;\frac{\partial^2}{\partial\beta^2}f(\beta,q,y)|_{\beta=0}
=\sum_{n,\tilde{l},l}
\tilde{l}^2c(nk,\tilde{l},l)rq^{nr}y^{lr}\label{2stderivatie}
\end{equation}where we have used the conditions 
\begin{eqnarray}
 &&\sum_{\tilde{l}}c(n,\tilde{l},l)=0\\
&&\sum_{\tilde{l}} \tilde{l}c(n,\tilde{l},l)=0
\end{eqnarray}that follow from the presence of two complex fermion zero modes in
$\mathcal{H}_1$.

The coefficient of $q^Ny^{\tilde{J}}$ in (\ref{2stderivatie}) gives the index
$B_2$
\begin{equation}
 B_2=\sum_{s|I}s\,\sum_{\tilde{l}}\tilde{l}^2c(NI/s^2,\tilde{J}/s,\tilde{l})
\end{equation}where $\hat{c}(n,l)=\sum_{\tilde{l}}\tilde{l}^2c(n,l,\tilde{l})$
as in (\ref{hatc}). 

In the limit of large charges the term $s=1$ in (\ref{B2index}) gives the leading contribution
\begin{equation}
 B_2(Q_1Q_5,n,l\gg1)\approx e^{2\pi\sqrt{Q_1Q_5nI-l^2I^2}}
\end{equation}which is in agreement with the Beckenstein-Hawking entropy of the
5d black hole in ALE space \cite{Gaiotto:2005gf}.

In the four dimensional case we have to consider in addition the closed string
excitations of multi KK monopole.

We separate the problem in two pieces. First we try to argue that the Hibert
space of KK-P states is the hilbert space of multiply wound strings using
IIB-Heterotic duality in four dimensions. Then using the 4d-5d lift together
with fermion zero modes and duality invariance, we suggest that the Hilbert
space of non-primitive dyons, denoted $\mathcal{H}_I$, is of the form
\begin{equation}
 \mathcal{H}_I=\sum_{\sum kN_k=I}\prod_kS^{N_k}(\mathcal{H}_{k})
\end{equation}where $\mathcal{H}$ is the Hilbert space of a primitive dyon much
like in the five dimensional case.

As mentioned at the beginning of this section, the study of bound states of
multi KK monopoles by quantizing the moduli space is a very difficult problem.
To circumvent this we map KK-P states in IIB to heterotic perturbative momentum
winding states after performing T and six dimensional string-string dualities. 

We want to study bound states of momentum $nI$ and winding $I$ at weak coupling.
Because both charges have a common factor $I$ the multiply wound string can
split without breaking supersymmetry when
\begin{equation}
 M(nI,I)=\sum_{\sum rk_r=I} r\, M(nk_r,k_r)
\end{equation}where $M(n,w)$ is the BPS mass of a string with momentum $n$ and
winding $w$. This suggests that the Hilbert space of a multiply wound string is
graded by partitions of $I$
\begin{equation}\label{hilbert multiply w strings}
 \mathcal{H}^{KK-P}_I=\sum_{\sum rk_r=I}\prod_r S^r(\mathcal{H}_{k_r})
\end{equation}where $\mathcal{H}_{k_r}$ corresponds to a $k_r$ multiply wound
string. Again the presence of fermion zero modes forces the partitions to obey
$rk_r=I$. Only in this case the index is non vanishing. A state in
$\mathcal{H}_{k_r}$ carries momentum $nI/r$ and winding $I/r$.

Naively, we would tensor (\ref{hilbert multiply w strings}) with (\ref{hilbert
D1D5 on C^2/ZI}) to obtain the full hilbert space of non-primitive dyons
\begin{equation}
 \mathcal{H}_{full}=\sum_{r|I}S^r(\mathcal{H}^1_{I/r})\,\otimes\sum_{k|I}
S^k(\mathcal{H}^2_{I/k})
\end{equation}but tracing the index $B_6$ (\ref{indexB6}) over this space would
give a non-duality invariant answer. Instead we propose that KK-P states should
be symmetrized along with D1-D5 states in such a way that
\begin{equation}
 \mathcal{H}_{full}=\mathcal{H}^0\otimes\sum_{r|I}S^r(\mathcal{H}^1_{I/r}
\otimes\mathcal{H}^2_{I/r})
\end{equation}
A state in $S^r(\mathcal{H}^1_{I/r}\otimes\mathcal{H}^2_{I/r})$ is of the form
$\sum_{\sigma}\epsilon(\sigma) \prod |n^1_i,h^1_i\rangle\otimes
|n^2_i,J_i,h^2_i\rangle$ with total momentum $\sum n^1_i+n^2_i=N$, angular
momentum $\sum 2J_i=\tilde{J}$ and total helicity $\sum2h^1_i+h^2_i$. A state in
$\mathcal{H}^0$ is made up of four fermionic zero mode states
$\prod_i^4\otimes|h_i=\pm1/4\rangle$ while a state in the symmetric product
carries only two fermion zero modes with $h^2_i=\pm1/4$. This distinction is
based on the fact that we shouldn't trace over the center of mass degrees of
freedom of the black hole. Since the black hole is free to move in the
transverse $\mbbr^3$, it will have bosonic zero modes along with the
corresponding fermionic partners.

Such construction gives a duality invariant answer. Consider first the trace
over the fermionic zero mode states in $\mathcal{H}^0$,
\begin{eqnarray}
 B_6&=&-\frac{1}{6!}\textrm{Tr}_{\mathcal{H}^0\otimes
\mathcal{H}'^I}(-1)^{2J-2(h^0+h^I)}(2h^0+2h^I)^6\\
&=&--\frac{1}{6!}\textrm{Tr}_{\mathcal{H}^0\otimes
\mathcal{H}'^I}(-1)^{2J-2(h^0+h^I)}\frac{6!}{4!2!}(2h^0)^4(2h^I)^2\\
&=&-\frac{1}{2}\textrm{Tr}_{\mathcal{H}'^I}(-1)^{2J-2h}(2h)^2
\end{eqnarray}where we have denoted
$S^r(\mathcal{H}^1_{I/r}\otimes\mathcal{H}^2_{I/r})$ by $\mathcal{H}'^I$. The
final trace has the form of the helicity trace $B_2$. 

Following the same steps used to compute $B_2$ in the five dimensional case, we
arrive at the final answer
\begin{equation}\label{B6 non primitive}
 B_6=\sum_{s|I}s\,c(Q_1Q_5,nI^2/s^2,JI/s)
\end{equation}with the coefficient $c(k,l,j)$ extracted from the primitive
answer.

 Further analysis of wall crossing phenomena in $\mathcal{N}=4$ string theory
based on multi center black hole splitting \cite{Sen:2008ht} suggests that only
two fermion zero modes have to be symmetrized. If we were to symmetrize $n$
complex fermion zero modes we would get a factor of $s^{n-1}$ in (\ref{B6 non
primitive}). A number $n$ different from two would contradict the results from
wall crossing and field theory dyon degeneracy. We don't have a physical
explanation of why this should be the case.

In terms of the effective string this hilbert space corresponds to a sigma model
with target space
\begin{equation}
 \textrm{Sym}^I(\sigma_{D1-D5-KK})
\end{equation}where $\sigma_{D1-D5-KK}$ is the sigma model we described in
section \S\ref{sec primitive case} for the primitive case.

\subsection{A Perturbative test of the dyon counting formula}\label{perturbative test}

The non-primitive answer (\ref{non primitive degeneracy}) is consistent with many physical tests. It reproduces the Beckenstein-Hawking entropy for large charges, it
correctly reproduces wall crossing phenomena and the degeneracy of $SU(N)$ field
theory dyons are correctly captured for small charges. 

Here we devise another microscopic test for the counting formula
\cite{Dabholkar:2009kd}. The strategy is to identify some states which are
non-perturbative in one frame but are perturbative in another. This strategy has
been used before in the case of half BPS states in various dualities. However in
$\mathcal{N}=4$ gauge theory, the quarter BPS states are necessarily
non-perturbative and cannot be mapped to any perturbative state. The reason is
that the only perturbative states in the gauge theory are the gauge bosons which
are half BPS. Interestingly this is not the case in string theory. A particular
set of quarter BPS states in $\mathcal{N}=4$ can be mapped to perturbative
states.

Consider the charge configuration
\begin{displaymath}
 \Gamma=\left[\begin{array}{c}
               Q\\
		P
              \end{array}\right]=
\left[\begin{array}{cccc}
                 0 & n & 0 & 0 \\
		 1 & 0 & 0 & 0
    \end{array}\right]_H.
\end{displaymath}It is easy to see that for these states the continuous
T-duality invariants all vanish, $Q^2=P^2=Q.P=0$. Nevertheless the invariant $I$
is non-trivial, $I=n$.

Under six-dimensional string-string duality, the heterotic NS5-brane is mapped
to type IIA
fundamental string, and the momenta are mapped to momenta. Thus, in the Type-II
frame,
our state corresponds to a perturbative type II fundamental string with winding
number one
with n units of momentum along the $S$ circle.    

The non-perturbative counting through formula (\ref{non primitive degeneracy})
gives
\begin{equation}
 B_6(Q^2=0,P^2=0,Q.P=0)=\sum_{s|I}\,s d_1(0,0,0)
\end{equation}The particular fourier coefficient $d_1(0,0,0)$ of
$\Phi_{10}^{-1}$ vanishes. Consequently the index $B_6$ vanishes for this charge
configuration independently of the invariant $I$,
\begin{equation}\label{B6 perturbative states}
 B_6(Q^2=0,P^2=0,Q.P=0,I)=0
\end{equation}

This charge configuration maps to a fundamental string with unit winding and $n$
units of momentum along $S^1$ in type IIA on $K3\times S^1\tilde{S}^1$. To
identify the spectrum we can use perturbative string theory.

For this propose we use light-cone gauge in Green-Schwarz formalism of string
theory. The world-sheet theory thus have a target manifold $\mbbr^2\times
T^2\times T^4/\mbbz_2$, where we denote $T^4/\mbbz_2$ the orbifold limit of
$K3$. 

We first compute the partition function
\begin{equation}
 Z(q,\bar{q},y)=\textrm{Tr}(-1)^{F}q^{L_0}\bar{q}^{\bar{L}_0}y^J
\end{equation}where $J$ is the $U(1)$ spin generator in the non-compact
directions. The index $B_6$ is extracted differentiating $Z(q,\bar{q},y)$ with
respect to $y$ six times and then setting $y=1$. Under this process only quarter
BPS states should be captured.

The partition function is
\begin{equation}\label{Z perturbative}
 Z(q,\bar{q},y)=(y^{1/2}-y^{-1/2})^4\prod_{n\geq1,j=\pm
1}\frac{(1-\bar{q}^ny^j)^2(1-q^ny^j)^2}{(1-\bar{q}^n)(1-q^n)^2(1-\bar{q}^ny^{2j}
)^2(1-q^ny^{2j})}\times \textrm{Tr}_{K3}(q,\bar{q},y)
\end{equation}with $\textrm{Tr}_{K3}(q,\bar{q},y)$ defined via
\begin{eqnarray}
 &&\textrm{Tr}_{K3}(q,\bar{q},y)=8\left[\frac{\vartheta_2(\tau,
v)^2\vartheta_2(\bar{\tau},v)^2}{\vartheta_2(\tau,0)^2\vartheta_2(\bar{\tau},
0)^2}+\frac{\vartheta_3(\tau,v)^2\vartheta_3(\bar{\tau},v)^2}{\vartheta_3(\tau,
0)^2\vartheta_3(\bar{\tau},0)^2}+\frac{\vartheta_4(\tau,v)^2\vartheta_4(\bar{
\tau},v)^2}{\vartheta_4(\tau,0)^2\vartheta_4(\bar{\tau},0)^2}\right]\no\\
\\
&&q=e^{2\pi i\tau},\bar{q}=e^{2\pi i\bar{\tau}},y=e^{2\pi i v}\no
\end{eqnarray}We refer the reader to \cite{Dabholkar:2009kd} for further
details.
The partition function (\ref{Z perturbative}) contains already a factor of
$(y^{1/2}-y^{-1/2})^4$ due to the presence of eight real fermion zero modes in
the Green-Schwarz formalism. For simplicity we remove first this factor and then
differentiate twice. Imposing the level matching condition $L_0-\bar{L}_0=I$ on
the left-moving BPS states we get
\begin{equation}
 d(I)=16\left[\sum_{s|I}s(3+(-1)^{s+1})-4\sum_{(2s+1)|I}\frac{I}{2s+1}\right]
\end{equation}which can be shown to vanish identically for any value of $I$.
This is in perfect agreement with the result  (\ref{B6 perturbative states}).
The same result was found in \cite{Gregori:1997hi-1} as a consequence of a theta
identity.

Note that for $n=0$, we actually have a half-BPS state which is dual to a
heterotic perturbative state. Since it breaks eight supersymmetries it carries
four complex fermion zero modes so we need to differentiate the partition
function four times. One correctly obtains a non-zero degeneracy which moreover
equals 24 consistent with the heterotic counting \cite{Kiritsis:1997hj}.

\section{$AdS_2/CFT_1$ correspondence and Quantum Entropy }

For a large class of supersymmetric black holes in string theory, the Beckenstein-Hawking entropy $S_{BH}$ finds perfect agreement with the logarithm of a microscopic index $B_{micro}$ in the limit of large charges,
\begin{equation}
 S_{BH}=\ln B_{micro}
\end{equation}

In this limit both the computations simplify. On the microscopic side we can use an asymptotic expansion of the index instead of computing it exactly, while on the gravity side, due to a large horizon, we can neglect higher derivative corrections and work only with two derivative terms in the full string action. It is of great interest to know if the agreement holds for finite charges and if that is the case how to compute finite charge corrections.

As explained in the previous section, it is possible to compute exactly the index $B_{micro}$ even for small charges for a large class of supersymmetric black holes in $\mathcal{N}=4$ string theory \cite{David:2006yn,Dabholkar:2008zy,Banerjee:2008pu,Jatkar:2005bh}. The use of an index instead of degeneracy is of great advantage. Because it captures only the BPS states, we can compare a microscopic computation with another performed when the black hole exists.

From the gravity side, the entropy function \cite{Sen:2005iz,Sen:2005wa} is a powerful way to compute the black hole entropy. Based on Wald's formalism it gives a useful prescription to compute finite charge corrections to the entropy. Instead of computing a complicated integral over the horizon, as demanded by Wald's formalism, it instructs us to minimize the lagrangian computed on the horizon solution. The entropy is then equal to the minima of that function which reduces the problem to solve some algebraic equations \cite{Sen:2007qy}.

Introduction of higher derivative/loop string corrections through the entropy function in a consistent way is problematic. In the full quantum theory we have to integrate over massless fields and as consequence non-local terms in the action can be generated. This is problematic since Wald's method requires a local, gauge and diffeomorphic invariant action.

To avoid this problem, Sen proposes a new framework based on $AdS_2/CFT_1$ correspondence  \cite{Sen:2008vm,Sen:2008yk}. The idea is to take the minimization process in the ``classical`` entropy function seriously by considering a path integral formulation. The Wald's entropy then corresponds to the classical saddle point approximation of this path integral. Indeed, via holographic correspondence, we can relate the degeneracy of states in the $CFT_1$ to a $AdS_2$ path integral of string theory with a Wilson line insertion at the boundary. This is very powerful in the sense that it gives a consistent framework to compute both perturbative and non-perturbative charge corrections to the entropy.

Additional care of the microscopic index is required if we want to match the microscopic answer with the horizon $AdS_2$ partition function. The reason is the following. The space-time index $B_6$ captures all the degrees of freedom coming both from the horizon and any other contribution sitting between the horizon and asymptotic infinity.  We call these additional degrees of freedom hair modes \cite{Banerjee:2009uk,Jatkar:2009yd}. Basically they correspond to deformations of the black hole solution with support outside of the horizon. On the microscopic side, they can correspond to the center of mass degrees of freedom of some brane system, for example.

This section is organized as follows. In section \S3.1 we develop the concept of Sen's quantum entropy function based on the $AdS_2/CFT_1$ correspondence. In section \S3.2 we explain the relation between index and degeneracy.

\subsection{Quantum entropy}

The entropy function which is based on the Wald formalism \cite{Wald:1993nt,Jacobson:1993vj,Iyer:1994ys,Jacobson:1994qe} relates the value of the classical string theory lagrangian, calculated in the near horizon geometry, to the black hole entropy \cite{Sen:2005iz,Sen:2005wa,Sen:2007qy}. So we expect the entropy to depend only on the horizon data. For extremal black holes this assumption is even better because the horizon region is separated from asymptotic infinity by an infinite throat. In this case we can expect the entropy to depend on the horizon data not just classically, via Wald's formalism, but also quantum mechanically. This is one of the pillars of Sen's quantum entropy function that we develop in the following.

The quantum entropy formalism \cite{Sen:2008vm,Sen:2008yk,Sen:2009vz} uses $AdS_2/CFT_1$ correspondence to give a quantum formulation of the black hole entropy. It states that
\begin{equation}\label{QEF}
 d_{hor}(q_I)=\left\langle e^{-iq_I\oint A^I}\right\rangle^{finite}_{AdS_2}
\end{equation}where $d_{hor}$ is the degeneracy associated with the horizon of the black hole. In other words, the degeneracy $d_{hor}(q)$ equals the expectation value of a Wilson line inserted at the boundary of euclidean $AdS_2$. The black hole entropy $S_{BH}$ at the quantum level is given by the logarithm of $d_{hor}$.
The symbol $\langle\;\rangle_{AdS_2}$ in (\ref{QEF}) denotes that we perform a path integral weighted by $e^{-A}$ where $A$ is the Euclidean string action.

We start by writing the near horizon field content of an extremal black hole after performing the Wick rotation $t=-i\theta$,
\begin{equation}
 ds^2=v\left[(r^2-1)d\theta^2+\frac{dr^2}{r^2-1}\right],\;\phi=\phi^*,\;F^I_{r\theta}=-ie^I
\end{equation}where $\phi^*$ are the attractor values of the scalars and $F^I_{r\theta}$ are the $U(1)$ gauge field strengths. The corresponding gauge field is $A^I_{\theta}=-ie^I(r-1)$ in the gauge $A_r=0$. Since the thermal circle $\theta$ is contractable in the $AdS_2$ geometry this forces the gauge field to vanish for $r=0$, otherwise it will be singular. The boundary stays at $r=\infty$.

Quantum mechanically, the $AdS_{2}$  functional integral is defined by summing over all field
configurations which asymptote to the these attractor values with the fall-off conditions \cite{Sen:2008yk, Sen:2008vm, Castro:2008ms}
\bea\label{asympcond}
d s^2 &=& v\left[ 
\left(r^2+\cO(1)\right) d\theta^2+ \frac{dr^2}{r^2+\cO(1)}  \right]\  .\\
\phi^{I} &= &\phi^{I}_{*} + \cO(1/r)\ ,\qquad
A^I = -i  \, e^{I} (r -\cO(1) ) d\theta\ .
\eea
All massive fields asymptote to zero because of their mass term.

 The path integral suffers from IR divergences due to the infinite volume of $AdS_2$. Therefore we introduce a cuttoff at $r=r_0$. In an expansion in the cutoff parameter, the regulated amplitude then has the form 
\begin{equation}\label{IR prescription}
 \langle \ldots\rangle_{AdS_2}=e^{r_0A+B+\mathcal{O}({r_0}^{-1})}
\end{equation} The prescription used to remove the IR infinities is to keep only the finite part $e^B$. Technically we introduce a boundary counter term $S_{bndy}$ to remove the contribution $r_0A$ and then take the limit $r_0\rightarrow \infty$. The reader could have wondered why there aren't $\ln(r_0)$ contributions. This has to do with the fact that in a perturbative expansion around the attractor background, say $\phi=\phi^*+\delta \phi$, with $\phi=\mathcal{O}(1/r)$, terms linear in $1/r$ vanish via the equations of motion. So no logarithmic term is generated.

In defining the path integral on $AdS_2$ we need to specify boundary conditions. Usual rules of $AdS/CFT$ correspondence \cite{Witten:1998qj} instruct us to fix the non-normalizable modes and integrate over the normalizable ones. Special care is needed in the two dimensional case of Anti-de Sitter space. In this case the gauge field $A$ has two solutions to the linearised Maxwell equations: $A_\theta=c+ar$, in the gauge $A_r=0$. In contrast with higher dimensional cases, here the electric field mode is the dominant one so we should fix it. This is equivalent to working in the microcanonical ensemble where we fix the charges instead of the chemical potentials. The insertion of the Wilson line has the precise effect of rendering the equations of motion valid near the boundary \cite{Sen:2009vz}. Consider a small variation $\delta A$ of the gauge field and look for the linearised equations of motion
\begin{eqnarray}
 &&\lim_{r\rightarrow\infty}-iq_I\int d\theta\delta A^I_{\theta}-\frac{\delta S}{\delta A^I_{\theta}}\delta A^I_{\theta}=0\no\\
 &&\lim_{r\rightarrow\infty}-iq_I\int d\theta \delta A^I_{\theta}-\int drd\theta \frac{\delta \mathcal{L}}{\delta F^I_{r\theta}}\delta F^I_{r\theta}=0\no\\
&&\lim_{r\rightarrow\infty}\left\{-iq_I\int d\theta\delta A^I_{\theta}-\int d\theta \frac{\delta \mathcal{L}}{\delta F^I_{r\theta}}\delta A_{\theta}\right\}+\textrm{E.O.M.}=0 \label{Maxwell EOM bdny1}
\end{eqnarray}Since $F_{r\theta}$ is non-zero at the boundary there is a non-trivial contribution to the linearised equations of motion at the boundary. If we hadn't introduced the Wilson loop, then the linearised equation (\ref{Maxwell EOM bdny1}) wouldn't be obeyed.

The other fields are fixed in the standard manner. For the metric field
\begin{equation}
 ds^2=v\left[(r^2-\mathcal{O}(1))d\theta^2+\frac{dr^2}{r^2-\mathcal{O}(1)}\right]
\end{equation}we allow the constant mode denoted by $\mathcal{O}(1)$ to fluctuate. On the other end for the scalar fields we fix the constant mode to the attractor value.

On the CFT side we should be computing
\begin{equation}
 \textrm{Tr}e^{-\beta H}
\end{equation}where $H$ is the hamiltonian that generates translations on the boundary and the parameter $\beta=T^{-1}$ is the inverse of the temperature. Holography relates the radius of the thermal circle to the temperature as $\beta=2\pi r_0$. If the spectrum of $H$ has a mass gap then only the ground states contribute to the trace when we take the zero temperature limit, or $r_0\rightarrow \infty$. This dual quantum mechanics should be understood as the infrared limit of the quantum mechanics describing the black hole after removing the hair modes and it has the particular and interesting property that its hamiltonian is zero. In other words the quantum entropy function is counts the number of ground states of the $CFT_1$ in a particular charge and angular momentum\footnote{If the black hole carries angular momentum then the horizon has isometry $SO(2,1)\times U(1)$ and the angular momentum is seen as a charge from the two dimensional point of view} sector. 

The initial intuition was that the quantum entropy function should reduce to the Wald entropy in limit of large charges. Say that we consider a perturbative solution around the attractor vacuum. In the limit of large charges we can carry out a saddle point approximation 
\begin{eqnarray}
 \langle e^{-iq_I\oint A^I}\rangle &\approx& e^{-2\pi q_Ie^I(r-1)+\int v dr d\theta\mathcal{L}(\phi^*,e^J)+\mathcal{O}(q^{-1})}\\
&\approx&e^{-2\pi q_Ie^I(r_0-1)+2\pi v (r_0-1)\mathcal{L}(\phi^*,e^J)}
\end{eqnarray}where $\mathcal{L}(\phi^*,e^J)$ is the lagrangian which is scalar, and $\phi^*$ denotes the attractor values of the moduli. The gauge field chemical potentials are fixed to constant values such that the gauge field is regular at the origin $r=1$ of $AdS_2$, that is, $A_{\theta}(r=1)=0$. Following the prescription (\ref{IR prescription}) we keep only the constant term, that is
\begin{equation}
 \langle e^{-iq_I\oint A^I}\rangle^{finite} \approx e^{2\pi q_Ie^I-2\pi v\mathcal{L}(\phi^*,e^J)}.
\end{equation}This is exactly the exponential of the entropy function computed at the attractor value of the fields. Note that we have used the action principle $e^{S_{eucl}}$ and not $e^{-S_{eucl}}$ as is often used. This is related to the euclidean continuation chosen. Under $t=-i\theta$, $e^{iS}$ becomes $e^{S_{eucl}}$ following \cite{Sen:2008vm,Sen:2009vz}. This is important because it is the renormalized action that should damp the path integral, which is what happens in this case. From now on we define the renormalized action $S_{ren}$ to include also the Wilson loop contribution
\be\label{Sren}
\CS_{\rm ren} :=    \CS_{\rm bulk}  + \CS_{\rm bdry} - i \, {q_i }  \int_{0}^{2\pi} A^i_{\theta}  \, d\theta \ 
\ee where $\CS_{\rm bdry}$ is a boundary counterterm which renders the action IR finite.
In addition we define the expectation value of the Wilson Loop in $AdS_2$ by
\be\label{qef}
W (q, p) = \left\langle \exp \big[-i \, q_i \int_{0}^{2 \pi}  A^i_{\theta} \, d \theta \big]  \right\rangle_{\rm{AdS}_2}^{\rm finite}\ .
\ee At the classical attractor saddle point,
\be\label{classlim0}
%Z_{AdS_{2}}(q)|_{\phi_0} \sim e^{2 \pi ( q_{i} e^{i} - v \CL)}  \equiv e^{\CS_0}\ ,
W(q, p) \sim  \exp[{2 \pi ( q_{i} e^{i} - v \CL)}]  \equiv \exp\left[{S_{Wald}(q, p)}\right] \ ,
\ee  

This is a very powerful method to compute finite charge corrections to the Wald entropy. An immediate application consists in extending the perturbative analysis beyond the classical approximation. This is the work of \cite{Banerjee:2010qc,Banerjee:2011jp,Sen:2011ba} where the authors compute logarithmic corrections coming from a one-loop determinant in various supersymmetric theories.

Another non-trivial aspect of this formalism is that it allows for the contribution of additional subleading $AdS_2$ orbifolds \cite{Banerjee:2009af,Sen:2009vz}. They seem to play a non-trivial role in explaining non-perturbative contributions to the entropy as expected from microscopics \cite{Banerjee:2008ky}.
   
\subsection{Index versus degeneracy}

The idea that only the horizon degrees of freedom are relevant for the black hole entropy leads automatically to the conclusion that two black holes with same near horizon geometry must have the same entropy. Nevertheless, the same is not true for the microscopic index. Two black holes can have the same entropy but carry a different index. This suggests that the horizon degeneracy should be combined with an exterior contribution to account for the difference.    

This idea is consistent with the fact that even if the black hole entropy doesn't depend on the asymptotic values of the moduli, the index computed from microscopics can jump once we vary the asymptotic values of the moduli \cite{Sen:2007vb}. It is known that for given a set of charges one can have single center as multi center black hole  solutions \cite{Denef:2000nb,Denef:2002ru,Denef:2007vg}. The microscopic index doesn't know whether a state corresponds to a single or multi center solution, so it should include in the counting all these possibilities. Now, a multi center solution can cease to exist once we cross a wall of marginal stability which causes the index to jump \cite{Denef:2002ru,Sen:2007pg}. This also means that the entropy of a single center black hole can never jump. This suggests that string theory in the near horizon should capture only the degrees of freedom of a single black hole.

This is all suggestive to rewrite the index as \cite{Banerjee:2009uk,Jatkar:2009yd}
\begin{equation}\label{Bmicro=dhorxdhair}
 B_{micro}(q)=\sum_n \,\prod_{\substack{i=1\\\sum_i q^i_1+q^i_2=q}}^n \,d_{hor}(q^i_1)d_{hair}(q^i_2)
\end{equation}where the $n^{th}$ term corresponds to the contribution of a $n$-centered black hole configuration, $d_{hor}$ is the degeneracy associated with the horizon degrees of freedom and $d_{hair}$ corresponds to an additional contribution coming from modes exterior to the horizon that we generically call hair modes. By setting the asymptotic values of the moduli to their attractor values we can guaranty that only the single center solution will contribute.

Consider the following puzzle which arises when trying to compare the index with the black hole entropy. Take the BMPV black hole \cite{Breckenridge:1996is}. Microscopically it corresponds to a system of D1-D5 branes wrapping $K3\times S^1$ and carrying momentum along $S^1$. It can also carry angular momentum without breaking supersymmetry. Now consider the D1-D5-KK system discussed in section \S2. The 4D-5D lift relates these two configurations. If we put the BMPV black hole at the tip of the Taub-Nut it becomes the D1-D5-KK black hole. This means that if we zoom close to the origin of the Taub-Nut, where the space looks flat, both black hole solutions will look the same. As a matter of fact, this is also equivalent to the near horizon limit. This means that both the BMPV and D1-D5-KK black holes have the same near horizon geometry and hence the same black hole entropy. Nevertheless, the index differs substantially from one configuration to the other \cite{David:2006yn,Maldacena:1999bp}. We expect that after removing the hair contribution both degeneracies will agree.

The strategy is to study normalizable deformations of the black hole solution and check whether they have or not support near the horizon. Those that vanish near the horizon correspond to hair modes. In \cite{Banerjee:2009uk} the authors analysed the deformations using linearised equations of motion. 

The first basic conclusion is that all the fermion zero modes are part of the hair degrees of freedom. This not surprising since the solution outside the horizon breaks supersymmetry and therefore the goldstino modes must have support outside of the horizon. Additionally, they found that for the BMPV black hole the center of mass modes of the D1-D5 system are part of the hair degrees of freedom. 

For the D1-D5-KK black hole they have also found that the center of mass degrees of freedom of the D1-D5 moving in the transverse Taub-Nut space are also part of the hair degrees of freedom. The additional contribution coming from closed string excitations of the KK solution is also part of the hair modes.

The conclusion is that after removing the hair contribution the microscopic horizon partitions for both of the black holes agree,
\begin{equation}\label{Z4d=Z5d}
 Z^{hor}_{5D}=Z^{hor}_{4D}
\end{equation}They also make the important observation that these new partition functions are free from poles which could induce jumps in the index. This is important if we want the black hole entropy to not have moduli dependence.

A more refined approach in \cite{Jatkar:2009yd} using non-linear equations arrives at the same conclusion (\ref{Z4d=Z5d}) but with some important differences. The analysis shows that the bosonic deformations corresponding to the center of mass degrees of freedom of the brane system have curvature singularities at future horizon. Hence they should be include as horizon modes.

In the following we show why, for a class of supersymmetric black holes, the index matches with the black hole degeneracy in the large charge limit.   

In section \S2 we defined the helicity trace index $B_6$, suitable to capture the
spectrum of quarter BPS dyons in $\mathcal{N}=4$ theory. It was defined as
\begin{equation}
 B_6=-\frac{1}{6!}\textrm{Tr}(-1)^{2h}(2h)^6
\end{equation}with $h$ the helicity quantum number in four dimensions. We
justified the inclusion of six powers of $(2h)$ in the trace to remove the
contribution of six complex fermion zero modes coming from the breaking of twelve supersymmetries, and this way rendering the index non vanishing.

In the same spirit  we can rewrite the index $B_6$ in terms of horizon and hair contributions as
\begin{eqnarray}\label{hor+hair index}
 B_6&=&-\frac{1}{6!}\textrm{Tr}(-1)^{2h_{hor}+2h_{hair}}(2h_{hor}+2h_{hair})^6\\
&=&-\frac{1}{6!}\textrm{Tr}(-1)^{2h_{hor}}\textrm{Tr}(-1)^{2h_{hair}}(2h_{hair}
)^6\\
&=&\sum_{q+\tilde{q}=Q}B_{hor}(q)B_{6\,hair}(\tilde{q})
\end{eqnarray}where we used the fact that only the term $(2h_{hair})^6$ survives in the taylor expansion of $(2h_{hor}+2h_{hair})^6$ since the fermion zero modes are part of the hair degrees of freedom. The index $B_{hor}$ for the horizon degrees of freedom is given by the Witten index
\begin{equation}
 B_{hor}=\textrm{Tr}(-1)^{2h_{hor}}
\end{equation} and the index $B_{6\,hair}$ for the hair modes was defined as
\begin{equation}
 B_{6\,hair}=-\frac{1}{6!}\textrm{Tr}(-1)^{2h_{hair}}
(2h_{hair})^6.
\end{equation}If all the hair degrees of freedom are fermionic
zero modes than $B_{6\,hair}=-1$ implying $B_6=\textrm{Tr}(-1)^F$.

For an extremal spherically symmetric black hole all states carry $h_{hor}=0$ implying the equality $\textrm{Tr}(-1)^{2h_{hor}}=\textrm{Tr}(1)$ \cite{Sen:2009vz}. This shows that index equals degeneracy for the horizon. Moreover, for a black hole that preserves at least four supercharges, closure of the supersymmetry algebra implies spherical symmetry. In other words, a $SU(2)$ subalgebra is necessary for the susy algebra to close and this factor can be identified with the rotation symmetry of the near horizon geometry. 
The index $B_{hor}$ can be computed using the quantum entropy function \cite{Sen:2008vm,Sen:2009vz} which in the limit of large charges reduces to the exponential of the wald entropy. This, together with the fact that the hair contribution to the index is usually negligible compared to the wald entropy, explains why for large charges index equals degeneracy
\begin{equation}
 B_6(Q,P\gg1)\approx B_{hor}(Q,P\gg1)\approx e^{S_{BH}(Q,P)}.
\end{equation}This also explains why for one-sixteenth BPS black holes in $AdS_5$ no microscopic index seems to have the right asymptotics consistent with black hole entropy \cite{Kinney:2005ej,Chong:2005hr,Gutowski:2004yv}. Since in this case the black holes preserves too little supersymmetry closure of the supersymmetry algebra does not imply rotational invariance.

\section{Quantum black holes and Localization}

The Bekenstein-Hawking entropy is in a sense a bit too universal in that it is  always given by a quarter of the horizon area. This is a consequence that, for very large distances, only the Einstein-Hilbert term in the action contributes. Finite charge corrections, on the other hand, can arise after introduction of higher derivative terms which are different in different phases of the theory. This dependence on the phase 
can yield  useful information about  different aspects of the short-distance theory. In this section we are interested in computing finite charge corrections by explicitly evaluating the quantum entropy function 
for supersymmetric black holes in a broad class of phases of string theory, namely vacua with $\CN=2$ 
supersymmetry in four dimensions \cite{Dabholkar:2010uh}.

In a theory with massless $n_{v} +1$ vector fields, a black hole is specified by a 
charge vector $(q_{I}, p^{I})$  with $I = 0, \ldots, n_{v}$. We would like to develop methods to systematically compute the quantum entropy for arbitrary finite values of the charges. As explained in section \S3 the quantum entropy function, via $AdS_2/CFT_1$ correspondence, gives a consistent  and powerful framework to compute perturbative and non-perturbative corrections. 

Via $AdS_2/CFT_1$ correspondence the microscopic degeneracy $d(q,p)$ is identified with the expectation value of the Wilson loop that we denote by $W(q, p)$. Evaluating the formal expression for $W(q, p)$ by doing the string field theory functional integral is of course highly nontrivial. To proceed further we  imagine first integrating out the infinite tower of massive string modes and massive Kaluza-Klein modes to obtain a \emph{local} Wilsonian effective action for the massless supergravity fields.  To compute the exact quantum entropy,  one has to then evaluate exactly this functional integral of a finite number of  massless fields with $AdS_{2}$ boundary conditions using the full Wilsonian effective action  keeping all higher derivative terms. This effective action can include in general not only perturbative corrections in $\alpha'$ but 
also  worldsheet instanton corrections. We can  regard the ultraviolet finite string theory as providing a finite, supersymmetric, and consistent cutoff at the string scale. The functional integral with such a finite cut-off and a Wilsonian effective action containing all higher order terms is  thus in principle free of ultraviolet divergences. This functional integral will be our starting point.

We are still left with the task of evaluating a complicated functional integral.
The near horizon geometry preserves eight superconformal symmetries and moreover the  action, measure, 
operator insertion, boundary conditions of the functional integral are all supersymmetric. 
This allows us  to apply localization techniques \cite{Banerjee:2009af} which simplifies the evaluation of the functional integral enormously. Localization requires identification 
of a fermionic symmetry of the theory that squares to a compact bosonic symmetry.  Using this symmetry, one can then localize the functional integral onto the `localizing submanifold' of  bosonic field configurations invariant under the fermionic symmetry.  We review  the  superconformal symmetries of the near horizon geometry and  relevant aspects of localization in  \S{\ref{Localization}. 

Since localization is employed at the level of the functional integral and not just at  the level of a classical action,  it is important to use an \textit{off-shell} formulation of supergravity. Off-shell formulations of supergravity are in general notoriously involved. At present a complete formulation of off-shell supergravity coupled to both vector and hyper multiplets is not known. To implement localization in a concrete manner,  we therefore first  consider in \S\ref{Solution} a simpler problem  of  computing this 
expectation value of the Wilson line  in a truncated model of   supergravity  coupled only to vector multiplets  with an action containing only  F-terms which are chiral integrals over superspace.  In particular we ignore possible D-terms and  hyper multiplets, which are discussed later in \S\ref{Connection}.  The action still contains an infinite number of higher derivative terms but all of F-type.
We denote the corresponding functional integral  for the expectation value of a Wilson line in this restricted 
theory on $AdS_{2}$ by $\widehat W(q, p)$.   
Computation of $\widehat W(q, p)$ is greatly simplified by the fact that,  for vector multiplets 
in $\CN=2$ supergravity,  there exists an elegant off-shell formulation developed  in 
\cite{deWit:1979ug, deWit:1984px, deWit:1980tn},  using the 
superconformal calculus. The spectrum consists of the Weyl multiplet that contains the graviton and the gravitini, $n_{v} + 1$ vector multiplets, and one compensating multiplet that eliminates unwanted degrees 
of freedom. We review this formalism in \S{\ref{off-shell}}.  

The main result of this section concerns the localization  of the functional integral for $\widehat W (q, p)$ which is  derived in \S{\ref{Solution}}. 

The organization of this section is as follows. We start by reviewing the technique of localization and the superconformal symmetries of the near horizon geometry. In section \S4.2 we review the superconformal construction of supergravity using F-terms. In section \S4.3 we apply localization and determine $\widehat W (q, p)$. We end commenting on limitations of this approach and also on possible contributions from D-terms and hypermultiplets.

\subsection{Superconformal symmetries and localization \label{Localization}}

We start with a  brief review in \S\ref{LocReview} of  the localization techniques
\cite{Duistermaat:1982xu,Witten:1988ze,Witten:1991mk,Witten:1991zz,Schwarz:1995dg,Zaboronsky:1996qn}
 to evaluate supersymmetric functional integrals. In \S\ref{Supersymmetries} we review the superconformal  symmetries  of the attractor geometry and how localization can be applied in the present context.

\subsubsection{A review of localization of supersymmetric functional integrals \label{LocReview}}

Consider a supermanifold $\mathcal{M}$ with an integration measure $d\mu$. Let $Q$ be an odd (fermionic) vector field on this manifold that satisfies the following two requirements:
\begin{enumerate}
\item $Q^{2} =H$ for some compact bosonic vector field $H$,
\item The measure is invariant under $Q$, in other words $div_{\mu} Q = 0$.
\end{enumerate}
The divergence of the fermionic vector field is the natural generalization of ordinary divergence, which satisfies in particular\footnote{For a bosonic vector field $V$ and for  a measure determined by a metric $g$,  this corresponds to the identity   $\int dx \sqrt{g} V^{m} \partial_{m}f = - \int dx \partial_{m}(\sqrt{g} V^{m}) f  = - \int dx \sqrt{g} (\nabla_{m} V^{m})  f$  when the boundary contributions vanish.}
 \begin{equation}
\int_{\mathcal{M}} d\mu \, Q (f) = - \int_{\mathcal{M}} d\mu (div_{\mu}Q)  \, f \, ,
\end{equation}
for any function $f$. Hence, the second property implies $\int_{\mathcal{M}} d\mu \, Q (f) = 0$ for any $f$. We would like to evaluate an integral of some $Q$-invariant function $h$ and a Q-invariant action $S$
\begin{equation}
I := \int_{\mathcal{M}} d\mu  \, h \, e^{\mathcal{- S}} .
\end{equation}
To evaluate this integral using localization, one first deforms the  integral to
\begin{equation}\label{I(t)}
I (\lambda)  = \int_{\mathcal{M}} d\mu  \, h \, e^{-\mathcal{S}  - \lambda QV} \ , 
\end{equation}
where $V$ is a fermionic, H-invariant function which means  $Q^{2} V = 0$, that is, $Q V$ is Q-exact. One has 
\begin{equation}\label{dI/dt=0}
\frac{d}{d\lambda}\int_{\mathcal{M}} d \mu   \, h  \, e^{- \mathcal{S} - \lambda QV} = \int_{\mathcal{M}} d \mu   \, h  \, QV \, e^{- \mathcal{S} - \lambda QV} = \int_{\mathcal{M}} d \mu   \, Q( h  \, e^{- \mathcal{S} - \lambda QV}) = 0 \ , 
\end{equation}
and hence $I(\lambda)$ is independent of $\lambda$. 
This implies that one can perform the integral $I(\lambda)$ for any value of $\lambda$ and in particular for $\lambda \rightarrow \infty$.  
In this limit, the functional integral  localizes onto the  critical points of the functional $S^{Q} := QV$ 
and the semiclassical approximation becomes exact. The localizing solutions in general have both bosonic and fermionic collective coordinates. 

One can choose 
\begin{equation}\label{locV}
V = (Q\Psi, \Psi)
\end{equation}
 where $\Psi$ are the fermionic coordinates with some positive definite inner product  defined on the fermions.
In this case, the bosonic part of  $S^{Q}$ can be written as a perfect square $(Q\Psi, Q\Psi)$, and hence critical points of $S^{Q}$ are the same as the critical points of $Q$. This also implies that $QV$ vanishes on the localizing manifold as required by the invariance of $I(\lambda)$ (\ref{dI/dt=0}). Let us denote this set of critical points of $Q$ by $\mathcal{M}_{Q}$. 
The reasoning above shows that the integral over the supermanifold $\mathcal{M}$ localizes to an integral over the submanifold $\mathcal{M}_{Q}$. 
In the large $\lambda$ limit, the integration for directions transverse can be performed exactly in the saddle point evaluation. One is then left with an integral over the submanifold $\mathcal{M}_{Q}$
\begin{equation}
I = \int_{\mathcal{M}_{Q}} d\mu_{Q} \, h \, e^{-\mathcal{S}} \textrm{sdet}(D_2)\, ,
\end{equation}
where $d\mu_{Q}$ is a measure induced on the submanifold by the original measure and $\textrm{sdet}(D_2)$ is the superdeterminant of transverse fluctuations. We denote $D_2$ the operator of quadratic fluctuations of the $QV$ action.

In our case in \S\ref{Solution}, $\mathcal{M}$ is the field space of off-shell supergravity, $\mathcal{S}$ is the off-shell supergravity action with appropriate boundary terms, $h$ is the supersymmetric Wilson line, $Q$ is a specific supercharge described in \S{\ref{Killing}} and \S{\ref{Solution}}, and $\Psi$ are all fermionic fields of the theory. 
We will find that the submanifold $\mathcal{M}_{Q}$ of localizing solutions is  a family of nontrivial   instantons as exact  solutions to the equations of motion that follow from extremization of $S^{Q}$ labeled by $n_{v} +1$ real parameters $\{ C^{I} \, ; \, I= 0, \ldots, n_{v}\}$.

\subsubsection{Superconformal symmetries of the near horizon geometry \label{Supersymmetries}}

In higher dimensional cases we normally take the near horizon limit of an extremal, that is, zero temperature brane configuration. This limit allows us to focus on energy fluctuations of the brane system which are small to the asymptotic observer but sufficiently large compared to the temperature of the system. For the case of extremal black holes we proceed a bit differently \cite{Sen:2011cn}. Since the black hole quantum mechanics has a mass gap separating the ground state from the first excited state the only low energy excitations are zero energy excitations. This means that the usual near horizon limit of an extremal black hole is not a sensible limit. Instead we proceed by taking the near horizon and extremal limits at the same time \cite{Sen:2008yk}.
 
The near-horizon geometry  of a supersymmetric black hole in four dimensions 
is $AdS_2\times S^2$. After Euclidean continuation,  the metric is
\begin{eqnarray}\label{ads2s2}
 ds^2=  v \left[(r^2-1)d\theta^2+\frac{dr^2}{r^2-1}\right] + v  \, \left[ d\psi^2 + \sin^2 (\psi) d \phi^2 \right]
  \,  . 
\end{eqnarray}
We have  taken the radius $v$  of the $AdS_{2}$ factor to be the same as the radius of the $S^{2}$ factor which is a consequence of supersymmetry. There are several other coordinates that are useful. Substituting $r = \cosh (\eta)$, the metric takes the form
\begin{equation}\label{metric2}
    ds^2 = v \left[ d\eta^2 + \sinh^2 (\eta) d\theta^2 \right] +  v\left[ d\psi^2 + \sin^2(\psi) d \phi^2 \right] \, . 
\end{equation}
One can also choose the stereographic coordinates 
\begin{equation}\label{coordinate-trans}
    w = \tanh (\frac{\eta}{2}) e^{i\theta} := \rho e^{i\theta}, \quad z = \tan (\frac{\psi}{2}) e^{i\phi} \ , 
\end{equation}
in which the metric takes the form
\begin{equation}\label{metric}
%    ds^2 = 4 R^{2}\frac{dw d\bar w}{(1- w \bar w)^2} + 4 R^{2} \frac{dz d\bar z}{(1 + z \bar z)^2} \, .
    ds^2 =v \frac{4 dw d\bar w}{(1- w \bar w)^2} + v  \frac{4 dz d\bar z}{(1 + z \bar z)^2} \, .
\end{equation}
Note that the interval for the coordinates are $ 1 \leq r  < \infty$ and $ 0 \leq \eta < \infty$, and $0 \leq \rho < 1$. In the $w$ coordinates, 
Euclidean $AdS_{2}$ can be readily recognized as the Poincar\'e disk with
$\rho$   as the radial coordinate of the disk and a boundary at $\rho =1$. 

The Weyl tensor 
for the metric \eqref{metric2} is zero and hence this metric is conformally flat. For later use it will useful to know this conformal transformation. To map we first map the Poincar\'e disk to the upper half plane by  the transformation
\begin{eqnarray}
  u = x  + i y ,\, \qquad u = i \frac{1 - i w}{1 + iw} \, .
 \end{eqnarray}
The metric \eqref{ads2s2} in the new coordinates becomes
\begin{equation}\label{AdS_2 conformal flat}
 ds^2=\frac{dx^2+dy^2+y^2d\Omega_2^2}{y^2} \ , 
\end{equation}
with $-\infty < x < +\infty$ and $0  \leq y < \infty$. 
{}From the above equation,  we  see that $AdS_2\times S^2$ is conformally flat. We also know that $\mathbb{R}^4$ is conformal to $S^4$ so it would be useful to compute the conformal factor relating $AdS_2\times S^2$ to $S^4$. In the $(\eta,\theta)$ coordinates we have the following conformal rescaling
\begin{equation}\label{conformal-trans}
 ds^2(AdS_2\times S^2)=\cosh^{2}(\eta) ds^2(S^4) \ .
\end{equation}
Note that the conformal factor diverges at the boundary.
Under a Weyl transformation 
\begin{equation}
  g_{\mu\nu} \rightarrow e^{2\Omega}g_{\mu\nu} , 
\end{equation}
a field with Weyl weight $a$ 
transforms as 
\begin{equation}\label{scaling}
\Phi\rightarrow e^{-a\Omega}\Phi  . 
\end{equation}
Hence, such a field in the conformal frame with $AdS_2\times S^2$ metric will be mapped  to the field 
in the conformal frame with $S^{4}$ metric by
\begin{equation}\label{Weylscaling}
 \Phi_{AdS_2\times S^2}=\frac{\Phi_{S^4}}{\cosh(\eta)^{a}}.
\end{equation}
This transformation will be useful later in \S{\ref{Solution}}.

The superconformal symmetry of the near horizon geometry is the semidirect product $SU(1, 1|2)\rtimes SU(2)'$.  The invariant subgroup
$SU(1, 1|2)$ will be of our main interest which contains the bosonic subgroup 
$SU(1, 1) \times SU(2)$. The first factor can be identified with the conformal 
symmetry of $AdS_2$ and is generated by $\{ L, L_\pm\}$. The second factor can
be identified with the rotational symmetry  of $S^2$ and is generated by $\{ J,
J_\pm\}$. The factor $SU(2)'$ originates from the R-symmetry of $\mathcal{N}=2$ supergravity in four dimensions.   The odd elements of the superalgebra are the superconformal symmetries
$G^{ia}_r$. The commutations relations are
\begin{eqnarray}\label{scalgebra}
 \left[L, L_{\pm }\right] &=& \pm L_{\pm } \ ,  \qquad\qquad  \left[L_{+} ,
L_{-}\right] = -2 L  \ ,   \\
 \left[J, J^{\pm}\right] &=& \pm J^{\pm} \ ,   \qquad\qquad \left[J^+, J^-\right] = 2
J  \ ,   \\
 \left[L, G^{ia}_{\pm}\right] &=& \pm \half G^{ia}_{\pm} \ ,   \quad\qquad 
\left[L_\pm, G^{ia}_{\mp} \right] = -i  G^{ia}_{\pm} \ ,   \\
 \left[J, G^{i\pm}_r\right] &=& \pm \half G^{i\pm}_{r} \ ,   \quad\qquad \left[J^\pm,
G^{i\mp}_{r} \right] =   G^{i\pm}_{r}  \ ,  \\
 \{ G_+^{i\pm}, G_{-}^{j\pm}\} &=& \pm 4  \e^{ij} J^\pm \ ,   \qquad \{ G_\pm^{i+},
G_{\pm}^{j-}\} = \mp 4 i \e^{ij} L_\pm \ ,   \\
 &&\{ G_\pm^{i+}, G_{\mp}^{j-}\} = 4 \e^{ij} (L \mp J) \ ; \quad 
 \epsilon^{+-} = - \epsilon{-+} = 1 \, .
\end{eqnarray}
Explicit expressions for the Killing spinors corresponding to these superconformal supersymmetries will be obtained  in \S\ref{Killing} and will be required for localization in \S\ref{Solution}.
 
It is easy to see from the algebra that the generator
$Q = G^{++}_{+} + G^{--}_{-}$ squares to $4(L-J)$. Since $L$ is the generator of rotations of the Poincar\'e disk 
and $J$ is the generator of rotations of $S^{2}$, the square $Q^{2}$ is the generator of a compact bosonic 
symmetry. This is the generator that we will use for localization.

\subsection{Off-shell formulation of the theory \label{off-shell}}

In this section, we review the off-shell formulation of supergravity due to 
\cite{deWit:1979ug, deWit:1984px, deWit:1980tn}. 
This formalism has several  attractive features.
\begin{itemize}
\item First, it allows the supersymmetry transformations 
to be realized in an off-shell manner which will be crucial for us to apply localization 
to the functional integral.
\item Second, one can also include within 
the formalism a class of curvature squared corrections to the theory that are 
encoded in the Weyl multiplet. This has made  it possible to study the  higher derivative 
corrections to supersymmetric black holes using the full power of supersymmetry for solving 
BPS equations in the classical theory.  
\item Third, in the off-shell formalism, the supersymmetry transformations are specified  once and for all and do not need to be modified as one modifies the action with higher derivative terms. This is analogous to the situation for diffeomorphisms where the transformation properties of the metric, for example, are specified once and for all and does not depend on the form the action. Since the localization action that we use is constructed using these supersymmetry transformations, the localizing solutions that we will obtain by minimizing this action will therefore be universal and not dependent on the form of the physical action. This is clearly greatly advantageous both at the technical and conceptual level. 
\end{itemize}

In this section we  rederive the classical properties of the black hole in this new language. 
This section is meant to set the stage and fix all the notations for the quantum calculation which 
we discuss in \S\ref{Solution}. It will therefore be concise; a detailed account of the off-shell 
formalism can be found,  for example, in the review \cite{Mohaupt:2000mj}. 

We  use the {\it conformal supergravity} approach to $\mathcal{N}=2$ off-shell supergravity in four dimensions
 developed using {\it superconformal multiplet calculus}. 
The main idea is to extend the symmetries of the $\mathcal{N}=2$ Poincar\'e supergravity
 to the $\CN=2$ superconformal algebra. This bigger algebra has dilatations, special 
conformal transformations, conformal $S$-supersymmetries, and  local $SU(2) \times U(1)$ symmetries 
as extra symmetries compared to the Poincar\'e group\footnote{Note that the extra superconformal 
symmetry of this formalism is a gauge symmetry, not to be confused with the physical superconformal 
algebra of the near-horizon geometry of extremal black holes discussed in \S\ref{Supersymmetries} which is generated by the Killing vectors and Killing spinors of the  background.} . The conformal supergravity is 
then constructed as a gauge theory of this extended symmetry group. 

Upon gauge fixing the extra superconformal symmetries, one gets the Poincar\'e supergravity. 
In this sense, they are both gauge equivalent. However, the multiplet structure of the superconformal 
supergravity is smaller and simpler than the Poincar\'e theory. The form of the supersymmetry transformation 
rules is also simpler in the superconformal formalism, and one has a systematic way of deriving 
invariant Lagrangians. Following this approach, one gets an off-shell formulation of supergravity 
coupled to vector multiplets.

In \S\ref{Multiplets}, we first list the multiplets of the superconformal theory that will enter the theories we consider. 
In appendix \S\ref{susyvar}, we summarize some relevant aspects of the superconformal 
multiplet calculus including the supersymmetry variations of the various multiplets listed below. 
In \S\ref{Superaction} we discuss the invariant action of our interest.

\subsubsection{Superconformal multiplets \label{Multiplets}}

Our \emph{conventions} are as follows. In the Minkowski theory, all  fermion fields below are 
represented by Majorana spinors. In the Euclidean theory, they will be symplectic-Weyl-Majorana \cite{Cortes:2003zd}. 
Greek indices $\mu,\nu,\dots$ indicate the curved spacetime, latin indices $a,b,\dots$ indicate 
the flat tangent space indices, and $i,j, \dots$ denote the $SU(2)$ index. The $SU(2)$ indices
are raised and lowered  by complex conjugation. 
$A^{-} \equiv \varepsilon_{ij}\,A^{ij}$ for any $SU(2)$ tensor $A^{ij}$. We will also use
the superscript $\pm$ to denote (anti) self-duality in spacetime, the conventions should be clear from 
the context. 
We  use the  covariant derivative $D_{a}$, which  is defined to be covariant with respect to all the 
superconformal transformations as well as gauge fields of the theory if present. We also use
the bosonic covariant derivative $\nabla_{a}$ is defined to be covariant with respect to all the 
bosonic transformations and the gauge fields, except the special conformal transformation.

We now summarize the field content of various multiplets.

\begin{enumerate}
\item {\it Weyl multiplet}: This is the gravity multiplet which contains all gauge fields arising from gauging the full superconformal symmetries. The field content is:
\be\label{Weylfields}
{\bf w} = \left( e_{\mu}^{a}, w_{\mu}^{ab}, \psi_{\mu}^{i}, \phi_{\mu}^{i}, b_{\mu}, f_{\mu}^{a}, A_{\mu}, \CV_{\mu \, j}^{\, i},  T_{ab}^{ij}, \chi^{i}, D \right) \, .
\ee
The fields $(e_{\mu}^{a}, w_{\mu}^{ab})$ are the gauge fields for translations (vielbien) and Lorentz transformations;
$\psi_{\mu}^{i}, \phi_{\mu}^{i}$ are the gauge fields for Q-supersymmetries and the  conformal S-supersymmetries; 
$(b_{\mu}, f_{\mu}^{a})$ are the gauge fields for dilatations and the special conformal transformations; and 
$(\CV_{\mu \, j}^{\, i}, A_{\mu})$ are the gauge fields for the $SU(2)$ and $U(1)$ R-symmetries. Imposition of the `conventional constraints' determines $w_{\mu}^{ab}, \phi_{\mu}^{i},  f_{\mu}^{a}$ in terms of other fields and one is left with $24+24$ independent degrees of freedom. 
The $SU(2)$ doublet of Majorana spinors $\chi^{i}$, the antisymmetric anti self-dual auxiliary field $T_{ab}^{ij}$ 
and the real scalar field $D$ are all auxiliary fields, some of which will play a non-trivial role later. 
This multiplet contains the gravitational degrees of freedom. 

\item {\it Vector multiplet}: The field content is
%\be\label{Vectorfields}
%{\bf X}^{I} = \left( X^{I}, \O_{i}^{I}, W_{\mu}^{I}, Y^{I}_{ij}  \right)
%\ee
\be\label{Vectorfields}
{\bf X}^{I} = \left( X^{I}, \Omega_{i}^{I}, A_{\mu}^{I}, Y^{I}_{ij}  \right)
\ee
with $8+8$ degrees of freedom. $X^{I}$ is a complex scalar, the gaugini $\Omega^{I}_{i}$ are an $SU(2)$ 
doublet of chiral fermions, $A^{I}_{\mu}$ is a vector field, and $Y^{I}_{ij}$ are an $SU(2)$ triplet of 
auxiliary scalars. 
This multiplet contains the gauge field degrees of freedom. 
%In the following, we shall sometimes refer to the vector field in the vector multiplet 
%by $A^{I}_{\mu}$ instead of $W^{I}_{\mu}$ -- this should be clear from the context. 

\item {\it Chiral multiplet}: The field content is
\be\label{Chiralfields}
{\bf \widehat{A}} = \left( \widehat{A},  \widehat{\Psi}_{i},  \widehat{B}_{ij},  \widehat{F}^{-}_{ab},  \widehat{\Lambda}_{i}, \widehat{C}  \right)
\ee
with $16+16$ components. $ \widehat{A},  \widehat{C}$ are complex scalars, $\widehat{B}_{ij}$ is a complex 
$SU(2)$ triplet, $\widehat{F}^{-}_{ab}$ is an antiselfdual Lorentz tensor, and $ \widehat{\Psi}_{i},  \widehat{\Lambda}_{i}$ 
are $SU(2)$ doublets of left-handed fermions. 
The action will also contain the conjugated right handed multiplet. 
One can impose a supersymmetric constraint on the chiral multiplet to get a reduced chiral 
multiplet with $8+8$ degrees of freedom. 

The covariant quantities of a vector multiplet 
are associated with a reduced chiral multiplet. 
The covariant quantities of the Weyl multiplet are also 
associated with a reduced chiral multiplet ${\bf W}^{ij}_{ab}$. Products of chiral multiplets 
are also chiral, and one thus gets a chiral multiplet 
${\bf \widehat{A}} = {\bf W}^{2}  = \ve_{ik} \ve_{jl} {\bf W}_{ab}^{ij} {\bf W}^{abkl}$. 
The lowest component of ${\bf \widehat{A}}$  is $\widehat A = (T^{ij}_{ab} \, \ve_{ij})^2$ and 
the highest component of ${\bf \widehat{A}}$ contains terms quadratic and linear in the curvature. 
The problem of building Lagrangians with terms quadratic in the curvature thus reduces to the 
simpler problem of coupling the chiral multiplet ${\bf \widehat{A}}$ to the superconformal theory.

\item {\it Compensating multiplet}: 
This multiplet will be used as a compensator to fix the extra gauge transformations.  
There are three types of compensators that have been used in the literature so far, a non-linear 
multiplet, a compensating hypermultiplet and a tensor multiplet. As an example, we 
discuss the non-linear multiplet \cite{Mohaupt:2000mj, Sahoo:2006rp}. 
Other multiplets have their relative advantages, in particular the compensating hypermultiplet 
is used extensively for the treatment of higher derivative terms \cite{LopesCardoso:2000qm}.\\
\ndt {\it Non-linear multiplet}:
\be\label{nonlinear}
\left( \Phi^{i}_{\,\a}, \lambda^{i}, M^{ij}, V_{a} \right)
\ee
where $\lambda^{i}$ is a spinor $SU(2)$ doublet, $M^{ij}$ is a complex antisymmetric matrix of Lorentz scalars, 
and $V_{a}$ is a real Lorentz vector. $\Phi^{i}_{\, \a}$ is an $SU(2)$ matrix of scalar fields with the $\a$ index 
transforming in the fundamental of a rigid $SU(2)$, it describes three real scalars.  
Naively, the multiplet has $9+8$ degrees of freedom, but there is a supersymmetric constraint 
on the vector $V_{a}$ which reduces the degrees of freedom to $8+8$: 
\be\label{susycons}
D^a V_a - 3 D - \half V^a V_a - \frac{1}{4} |M_{ij}|^2 + D^a \Phi^i_{\;\;\alpha}
D_a \Phi^{\alpha}_{\;\;i} + \mbox{fermions} = 0
\ee

\end{enumerate}

\subsubsection{Superconformal action \label{Superaction}}

The procedure to get invariant actions is as follows: one first finds an invariant Lagrangian for a 
chiral multiplet, this was solved in \cite{deRoo:1980mm}. The second step is to 
write down a scalar function, the {\it prepotential} $F(X^{I})$ of the vector multiplets which is a 
meromorphic  homogeneous function of weight 2. One then uses the 
chiral Lagrangian of the first step for the chiral multiplet $\bf F$. This gives the 
two derivative $\CN=2$ Poincar\'e supergravity after gauge fixing. 
To include coupling to curvature square terms, one extends the function $F$ to depend on the lowest component
$\widehat A$ of the chiral multiplet $ \widehat{\textbf{A}} = {\textbf W^{2}}$.
$F(X^{I}, \widehat{{A}})$ is holomorphic and homogeneous of degree two in all its variables. 
One then uses the chiral Lagrangian of the first step for the chiral multiplet~$\bf F$. 

We  use the following notations. The prepotential which is a meromorphic function of its arguments 
obeys the homogeneity condition:
\be \label{homogen}
F(\lambda X, \lambda^2 \wh A) = \lambda^2
F( X, \wh A)\, .
\ee
Its various derivatives are defined as:
\be \label{defFI}
F_I = \frac{\p F}{ \p X^I}, \quad F_{\wh A} = \frac{\p F}{\p
\wh A}, \quad F_{IJ} =\frac{\p^2 F}{ \p X^I \p X^J},
\quad F_{\wh A I} = \frac{\p^2 F}{ \p X^I \p \wh A},
\quad F_{\wh A \wh A} =
\frac{\p^2 F}{ \p \wh A^2}\, .
\ee

Following the above procedure, one gets a  invariant action for $I=1,2,\dots , N_{V}+1$ vectors coupled 
to conformal  supergravity. The bosonic part of the action is:
\bea\label{supconaction}
e^{-1} {\cal L} & = & i \Big[ \bar{F}_I X^I ( \frac{1}{6} R - D )
+  {\nabla}_{\mu} F_I {\nabla}^{\mu} \bar{X}^I  \nonumber \\
& &  \quad  + \frac{1}{4}   F_{IJ} ( F^{-I}_{ab} - \frac{1}{4}  \bar{X}^I T^{ij}_{ab} \, \ve_{ij})
 ( F^{-abJ} - \frac{1}{4}  \bar{X}^J T^{ij}_{ab} \, \ve_{ij}) \nonumber\\
&&- \frac{1}{8}  F_I ( F^{+I}_{ab} - \frac{1}{4}  X^I T_{abij} \, \ve^{ij}) \, T^{ij}_{ab} \, \ve_{ij}  - \frac{1}{8}   F_{IJ} Y^I_{ij} Y^{J ij}    - \frac{1}{32}  F \, (T_{abij} \, \ve^{ij})^{2} \nonumber \\
 & & \quad + \frac{1}{2}   F_{\widehat{A}} \widehat{C} - \frac{1}{8}   F_{\widehat{A} \widehat{A}}
(\ve^{ik} \ve^{jl} \widehat{B}_{ij} \widehat{B}_{kl} - 2 \widehat{F}^-_{ab} \widehat{F}^-_{ab}) 
+ \frac{1}{2}   \widehat{F}^{-ab} F_{\widehat{A}I} ( F^{-I}_{ab} - \frac{1}{4}  \bar{X}^I 
T^{ij}_{ab} \, \ve_{ij} )  \nonumber\\
 & &  \quad - \frac{1}{4}   \widehat{B}_{ij} F_{\widehat{A}I} Y^{Iij} \Big] + \rm{h.c.} 
\eea

To get to the $\CN=2$ Poincar\'e supergravity, one has to gauge fix the extra gauge transformations 
of the superconformal theory. To gauge fix the special conformal transformations, one sets the $K$-{\it gauge}:
\be\label{Kgauge}
b_{\mu} = 0 \ . 
\ee
To gauge fix the dilatations, one impose the $D$-{\it gauge}:
\be\label{Dgauge}
- i \big( X^{I} \bar{F}_{I}  - F_{I} \bar{X}^{I} \big) = 1 \ . 
\ee
To fix the chiral $U(1)$ symmetry, one fixes the $A$-{\it gauge}:
\be\label{Agauge}
X^{0} = \bar{X}^{0} \ . 
\ee
Due to these constraints on the scalars, the Poincar\'e supergravity has only $N_{V}$ independent scalars. 

In order to fix the $S$-supersymmetry, one imposes another gauge called the $S$-{\it gauge}. This constraint 
can be solved by eliminating one of the vector multiplet fermions. This gauge also breaks $Q$-supersymmetry, 
but a combination of the $S$ and $Q$ supersymmetries is preserved and corresponds to the physical 
supertransformations in the Poincar\'e theory. 

Finally, to fix the local $SU(2)$ symmetry, one imposes the $V$-{\it gauge}:
\be\label{Vgauge}
\Phi^{i}_{\, \a} = \delta^{i}_{\, \a}
\ee

At each step in the gauge fixing process, one has to be careful to respect the previous gauge choices, and this 
leads to compensating field dependent transformations in the rules for the various remaining transformations. 
This is one of the reasons the final theory is more complicated. Finally, one has to solve algebraic 
equations to get rid of the auxiliary fields $D$ and $\chi$. 
At the end of this procedure, one gets the $\CN=2$ Poincar\'e supergravity with a bosonic Lagrangian:
\bea\label{PoincaresugraLag}
8 \pi e^{-1} {\cal L} &=& ( - i (  X^I\bar{F}_I - F_I \bar{X}^I )) \cdot  ( - \frac{1}{2} R ) \nonumber \\
& & + \big[ i  {\nabla}_{\mu} F_I {\nabla}^{\mu} \bar{X}^I +  \frac{1}{4} i F_{IJ} ( F^{-I}_{ab} -  \frac{1}{4} \bar{X}^I T^{ij}_{ab} \, \ve_{ij}) 
  ( F^{-abJ} -  \frac{1}{4} \bar{X}^J T^{ij}_{ab} \, \ve_{ij})\nonumber\\
&&- \frac{1}{8} i F_I ( F^{+I}_{ab} - \frac{1}{4} X^I  T_{abij} \, \ve^{ij}) T^{ij}_{ab} \, \ve_{ij} - \frac{1}{8} i F_{IJ} Y^I_{ij} Y^{J ij}  -\frac{i}{32} F \, (T_{abij} \, \ve^{ij})^{2} \nonumber \\
 & & + \frac{1}{2} i F_{\widehat{A}} \widehat{C} - \frac{1}{8} i F_{\widehat{A} \widehat{A}}
(\ve^{ik} \ve^{jl} \widehat{B}_{ij} \widehat{B}_{kl} - 2 \widehat{F}^-_{ab} \widehat{F}^-_{ab}) 
+ \frac{1}{2} i \widehat{F}^{-ab} F_{\widehat{A}I} ( F^{-I}_{ab} - \frac{1}{4} \bar{X}^I T^{ij}_{ab} \, \ve_{ij})  \nonumber\\
 & &  - \frac{1}{4} i \widehat{B}_{ij} F_{\widehat{A}I} Y^{Iij} + \rm{h.c.} \big]  \nonumber\\
 & & -i (  X^I\bar{F}_I - F_I \bar{X}^I ) \cdot ({\nabla}^a V_a 
- \frac{1}{2} V^a V_a - \frac{1}{4} | M_{ij} |^2 + D^a \Phi^i_{\;\;\alpha} D_a \Phi^{\alpha}_{\;\;i}) \;. \\ 
\nonumber
\eea
Note that both the covariant derivatives defined above are used in this expression, they are related by 
\be\label{defDa}
D^a V_a = {\nabla}^a V_a - 2 f^a_a + \mbox{fermionic terms} \;.
\ee

%Which higher derivative corrections do we have finally -- Weyl tensor, write clearly. ZZZ

\subsection{Localization \label{Solution}}

We now turn to the  evaluation of  the supersymmetric black hole functional integral defined  in \S\ref{QEF} using the localization techniques discussed in 
\S\ref{Localization}.  We use the  formalism of \S\ref{off-shell} so that the supercharge used for localization is realized off-shell.

The on-shell equations of motion that follow from the above Lagrangian \eqref{PoincaresugraLag} 
admit a half-BPS black hole solution 
\cite{LopesCardoso:1998wt,LopesCardoso:1999cv,LopesCardoso:1999xn,LopesCardoso:2000qm}. 
The near horizon geometry is an $AdS_{2} \times S^{2}$ which admits eight 
conformal supersymmetries\footnote{As mentioned above, these conformal supersymmetries 
are not the conformal supersymmetries of the four-dimensional theory discussed in the last section, 
the latter are gauge symmetries in that formalism.}.  
The values of other fields are determined by the attractor mechanism 
\cite{Ferrara:1995ih,Ferrara:1995h,Strominger:1996kf} in terms of the charges consistent with the 
isometries.  The near-horizon $AdS_{2} \times S^{2}$ geometry with the attractor values of the 
other fields can also directly be derived from the BPS equations \cite{Cardoso:2006xz}.

We first review  this  on-shell solution in \S{\ref{Onshell}}. We then proceed to find the 
localizing instanton solution in  \S{\ref{Offshell}} and evaluate the renormalized action for this solution 
in \S{\ref{RenAction}}.  We will sometimes refer to the localizing solution as the off-shell solution since 
for this solution the scalar fields are excited away from the attractor values inside the $AdS_{2}$. In \S\ref{What} we  put together  these ingredients to reduce the functional integral of $\widehat W(q, p)$  to an ordinary integral on the localizing submanifold.

%\subsection{On-shell Attractor Geometry \label{Onshell}}
\subsubsection{On-shell attractor geometry \label{Onshell}}

Symmetries of $AdS_{2} \times S^{2}$ imply that various field in the near horizon region take the form
\bea \label{sol1}
&& ds^2=  v \left[-(r^2-1)dt^2+\frac{dr^2}{r^2-1}\right] + v  \, \left[ d\psi^2 + \sin^2 (\psi) d \phi^2 \right] \ , 
 \nonumber \\
&& F^I_{rt} = e^{I}_{*}, \quad F^I_{\psi\phi}=  p^I
\, \sin\psi, \quad 
X^I = X_{*}^I, \quad
T^-_{rt}= v \, w \ , \nonumber \\
&& D-\frac{1}{3} R = 0, 
%\quad \mathcal{A}_\mu = 0, \quad  
%\mathcal{V}^{i}_{~j\mu} = 0, \quad V_\mu = 0,
\quad M_{ij}=0, \quad \Phi^\alpha_i
=\delta^\alpha_i\, , \quad Y^I_{ij}=0 \ . 
\eea 
The values of the constants $(e^{I}_{*}, X^{I}_{*},v_{*})$ that appear in this solution are determined 
in terms of the charges $(q_{I}, p^{I})$ 
by the attractor equations which follow from the BPS conditions \cite{LopesCardoso:1998wt},
 or, equivalently using the entropy function formalism \cite{Sahoo:2006rp}:
\bea 
 && v  = \frac{16}{ \bar w w} \, , \quad  \hat A =  -4 \omega^{2} \ ,  \label{sol2a}  \\
 && e^I_{*} - i  p^I - \frac{1}{2} \bar X_{*}^I v w =  0 \ ,\label{sol2b}  \\ 
 && 4 i (\bar w^{-1}\bar F_I - w^{-1} F_I) =  \, q_I \ . \label{sol2c}  
\eea
Taking the real and imaginary parts of \eqref{sol2b} and substituting \eqref{sol2a} gives
\bea \label{scalarattval}
 4 (\bar w^{-1}\bar X_{*}^I + w^{-1} X_{*}^I)&=& e^I_{*} \, ,  \label{sol3a}
\\
4 i (\bar w^{-1}\bar X_{*}^I - w^{-1} X_{*}^I) &=&  p^I \, , \label{sol3b}
\eea
where $F_{I}$  should be thought of as functions of $X^{I}_{*}$.
This geometry preserves eight superconformal supersymmetries as reviewed \S{\ref{Killing}} which   extends the symmetries to the supergroup $SU(1, 1|2) \otimes SU(2)'$ discussed in \S{\ref{Supersymmetries}}. The field $w$ can be fixed by a gauge choice.
In the rest of the paper, we choose a gauge in which $w = \bar w = 4$ using the  local scaling symmetry of the Lagrangian and the $U(1)$ invariance. In this gauge, the radius $v$ of  both $AdS_{2}$ and $S^{2}$ 
equals one, this simplifies the discussion of Killing 
spinors\footnote{This is different from  the gauge used in the previous section and also from the gauge $\omega = 8$ which is commonly used. 
These gauge choices do not affect considerations in this paper, but a  better understanding of different gauge choices can be useful to simplify the analysis. We plan to return to this issue in  future. 
}.

\subsubsection{Localizing action and the localizing instantons \label{Offshell}}
%\subsection{Localizing action and the saddle-point solution \label{Offshell}}
%\subsection{Off-shell half-BPS solutions \label{Offshell}}

In order to use the technique of localization for our system, we need to pick a subalgebra of the 
full supersymmetry algebra discussed in  \S\ref{Supersymmetries}, whose bosonic generator is compact. 
We shall choose the subalgebra generated by the action of the supercharge 
\be\label{defQ1}
Q_{1} = G^{++}_{+} + G^{--}_{-} \ , 
\ee
which generates the compact $U(1)$ action:
\be\label{Q1anticom}
Q_{1}^{2} = 4 ( L - J ) \ . 
\ee
The  explicit form of the Killing spinors can be found in 
\S{\ref{Killing}}. The above choice of the supercharge corresponds to choosing the supersymmetry parameter $\zeta_{1}$ defined in \eqref{defzeta}.
In this section, we  use the notation $Q \equiv Q_{1}$, $\zeta \equiv \zeta_{1}$. 

The localizing Lagrangian is then defined by 
\begin{equation}\label{loclag}
\CL^{Q} := QV \quad {\rm with} \quad V := (Q \Psi, \Psi) \, ,
\end{equation}
where $\Psi$ refers to all fermions in the theory. The localizing action is then defined by
\begin{equation}
S^{Q} = \int d^{4} x \sqrt{ g} \, \CL^{Q} \, .
\end{equation}
The localization equations that follow from this action  are
\begin{equation}
Q \Psi = 0 \, . 
\end{equation}
These are the equations that we would like to solve.

We assume that the supergroup isometries of the near horizon geometry are not broken further by the Weyl multiplet fields. 
By construction, as long as these symmetries are maintained, the fermions of the Weyl multiplet do not transform under the action of  $Q$
 \eqref{Weylvar1} --\eqref{Weylvar3} in the $AdS_{2}$ attractor background. 
One can check that the fermions of the chiral multiplet and the non-linear multiplet also do not transform 
in this background. This prompts us to look for solutions where one still has the $AdS_{2}$ attractor geometry, but the scalars of the 
vector multiplets can move away from their attractor values\footnote{Solutions more general than our simplifying ansatz are in principle possible where the Weyl multiplet fields also vary inside the $AdS_{2}$ .}. As we will see there do exist nontrivial solutions where the vector multiplet fields get excited maintaining the symmetries of the attractor geometry.

The action of $Q$ on the fermionic field of the vector multiplet takes the form (\ref{susy variantions})
\begin{eqnarray}\label{Qvars}
 && Q \, \Omega^{Ii}_{+}=\frac{1}{2}(F_{\mu\nu}^{I-}-\frac{1}{4}\bar{X}^{I} \, T^{-}_{\mu\nu}) \, 
 \gamma^{\mu} \, \gamma^{\nu} \, \zeta^{i}_+ +2i \displaystyle{\not}\partial X^{I} \, \zeta^i_-+Y^{Ii}_j \, \zeta^j_+ \ , \\
 && Q \, \Omega^{Ii}_-=\frac{1}{2}(F_{\mu\nu}^{I+}-\frac{1}{4}X^{I} \, T^{+}_{\mu\nu}) \, 
 \gamma^{\mu} \, \gamma^{\nu} \, \zeta^{i}_- +2 i \displaystyle{\not}\partial \bar{X}^{I} \, \zeta^i_+ +Y^{Ii}_j \, \zeta^j_- \ .
\end{eqnarray}

Let us recall the attractor equations for the constant values of the various fields 
in terms of the  electric gauge field strengths $e^{I}$ and the magnetic charges $p^{I}$:
\begin{eqnarray} \label{attractor eqs again}
&& e_{*}^{I} - ip^{I} - 2\bar{X}_{*}^{I} = 0 \ ,  \qquad e_{*}^{I} + ip^{I} - 2 X^{I}_{*} = 0 \, \qquad Y^{I}_{ij*} = 0 \  .
\end{eqnarray}
We are interested in the off-shell solutions in which the vector multiplet scalars $X^{I}$ move away from their attractor values $X^{I}_{*}$. We therefore parametrize the off-shell $X^{I}$ fields as
\begin{equation}\label{offshellpar}
X^{I} := X_{*}^{I } + \Sigma^{I}\, , \qquad \bar X^{I} := \bar X_{*}^{I } + \bar \Sigma^{I},
\end{equation}
 so that $\Sigma^{I}$ and $\bar \Sigma^{I}$ are values the scalar fields away from the attractor values. We further write
\be\label{defHJ}
\Sigma ^{I} = H^{I} + iJ^{I} \ , \qquad \bar \Sigma^{I} = H^{I} - iJ^{I} \ .
\ee
Note that  $Y^{I}_{ij}=\epsilon_{ik}\epsilon_{jl}Y^{Ikl}$ are triplets under the  $SU(2)$ rotation. It will turn out that for the BPS equations that we solve, they all have to be aligned along the same direction in the $SU(2)$ space. Hence  we parametrize them as 
\begin{eqnarray}\label{offshellpar2}
  Y^{I1}_{\,\,\, 1}=-Y^{I2}_{\,\,\,2}=K^{I} \, ; \qquad 
 Y^{I1}_{\,\,\,2}=  Y^{I2}_{\,\,\,1} = 0   \, ,
\end{eqnarray}where we have defined $Y^i_j=\varepsilon_{jk}Y^{ik}$.
Similarly we  parametrize the gauge fields away from the attractor values as
\begin{equation}\label{offshellpar3}
F^{I}_{\mu\nu } = F^{I}_{\mu\nu * }  + f^{I}_{\mu\nu } \, .
\end{equation}
With this parametrization, we can add the two equations  \eqref{Qvars} and perform a Euclidean continuation to obtain
\be
Q \, \Omega^{Ii}=\frac{1}{2}  f_{ab}^{I} \, 
 \gamma^{a} \, \gamma^{b} \,  \zeta^{i} + 2i \displaystyle{\not}\partial H^{I} \, \zeta^{i} + 2 \displaystyle{\not}\partial J^{I}  \, \gamma_{5} \zeta^{i}  - 2 i H ^{I} \gamma^{0} \gamma^{1}\zeta^{i} - 2 J ^{I} \gamma^{2} \gamma^{3} \zeta^{i}+ Y^{Ii}_{j} \, \zeta^j \ .
\ee
for the Dirac spinors $\Omega^{Ii} = \Omega^{Ii}_{+} + \Omega^{Ii}_{-}$. Note that the $a, b$ are tangent space indices and all gamma matrices $\gamma^{a}$ above are constant matrices of Euclidean $\mathbb{R}^{4}$.

The inner product for spinors $\chi_{1}$ and $\chi_{2}$ in Euclidean space is simply 
\begin{equation}\label{inner}
(\chi_{1} \, , \chi_{2}) = \chi_{1}^{\dagger} \chi_{2} \, .
\end{equation}
With this inner product, the  localization Lagrangian \eqref{loclag} restricted to only the vector multiplet fermions is given by
\begin{equation} 
\CL^{Q} =  QV  : = Q(Q\Psi, \Psi)
\end{equation} 
with $V$ chosen as in \eqref{loclag} with $\Psi$ denoting the vector multiplet fermions. Note that $V$ is H-invariant because $\zeta$ is independent of the combination $\theta -\phi$ and $ H$ is the vector field that generates translations along $\theta -\phi$. 
The bosonic part of this Lagrangian is
\begin{equation}
\CL_{\rm bos}^{Q} \equiv QV \big|_{\text{bosonic}}= \sum_{I=0}^{n_{V}}  (Q \Omega^{I}, {Q\Omega^{I}}) \ .
\end{equation}
With our choice of the inner product \eqref{inner} this Lagrangian is manifestly positive definite. 

The choice of $Q$ is determined by the choice of the Killing spinor $\zeta$.  Substituting the explicit form of the Killing spinor $\zeta$ and the gamma matrices defined in \S{\ref{Killing}},   the bosonic Lagrangian $ \mathcal{L}_{\rm bos}^{Q}$ as a function of the fields $H, J, K, f$ can be evaluated after somewhat tedious algebra. 
We find that $\frac{1}{2}\CL^{Q}_{\rm bos}$ equals
\begin{eqnarray}\label{LQdef}
 &&\cosh(\eta)\big[ K -2 \sech(\eta)H)\big]^2\nonumber\\
 &+& 4\cosh(\eta)\big[H_{1}+H \tanh(\eta)\big]^2 
  + 4 \cosh(\eta) [H_{0}^{2}+ H_{2}^{2} + H_{3}^{2}]  \nonumber\\
  &+& 2 A \left[ f_{01}^{-} - J - \frac{1}{A} \left( \sin(\psi) J_{3} - \sinh (\eta) J_{1}\right) \right]^{2}   
   +  2 B \left[ f_{01}^{+} + J - \frac{1}{B} \left( \sin(\psi) J_{3} + \sinh (\eta) J_{1}\right) \right]^{2} \nonumber\\
 &+& 2 A \left[ f_{03}^{-}  + \frac{1}{A} \left( \sin(\psi) J_{1} + \sinh (\eta) J_{3}\right) \right]^{2} 
     +  2 B \left[ f_{03}^{+}  + \frac{1}{B} \left( \sin(\psi) J_{1} - \sinh (\eta) J_{3}\right) \right]^{2}\nonumber\\
      &+&  2 A \left[ f_{02}^{-}  + \frac{1}{A} \left( \sin(\psi) J_{0} + \sinh (\eta) J_{2}\right) \right]^{2}
      + 2 B \left[ f_{02}^{+}  - \frac{1}{B} \left( \sin(\psi) J_{0} + \sinh (\eta) J_{2}\right) \right]^{2} \nonumber \\
      &+& \frac{4 \cosh(\eta)}{A B}\left[ \sinh(\eta) J_{0} - \sin(\psi) J_{2} \right]^{2}
      +  \frac{4 \cosh(\eta) \sinh^{2}(\eta)}{A B} [J_{1}^{2} + J_{3}^{2} ]  \, ,
 \end{eqnarray}
where 
\begin{equation}
H^{I}_{a} := e_{a}^{\mu} \partial_{\mu} H^{I} \, , \quad  J_{a}^{I} := e_{a}^{\mu} \partial_{\mu} J^{I} \, ,
\end{equation}
and
\begin{equation}
A := \cosh(\eta) + \cos(\psi)       \, , \qquad B :=  \cosh(\eta) - \cos(\psi)   \, .
\end{equation}
It is understood that in \eqref{LQdef} all squares are summed over the index $I$. 
Recall that $a = 0, 1, 2, 3$ correspond to the directions along the coordinates $\theta, \eta, \phi, \psi$ respectively used for example in \eqref{metric2}.
Since $A$ and $B$ are positive,  $\CL^{Q}_{\rm bos}$ is a sum of positive squares. 

The minimization equations now follow by setting each of the squares in \eqref{LQdef} to zero. This leads to simple first order differential equations for various fields which have to be solved with boundary conditions  consistent with the definition of the original functional integral on  Euclidean $AdS_{2}$ space. Equations \eqref{offshellpar},  \eqref{offshellpar2}, \eqref{offshellpar3} imply that fields $\Sigma^{I}$ and $K^{I}$ and $f^{I}$  must vanish at the boundary. 

It is easy to see that  with these boundary conditions, $J^{I}$ and $f^{I}_{a b}$ must both vanish throughout space.  Setting the first line in \eqref{LQdef} to zero implies
\begin{equation}\label{Ksol1}
K^{I} = \frac{2 H^{I}}{\cosh (\eta)} 
\end{equation}
Setting the second line in \eqref{LQdef} to zero leads to differential equations that can be easily solved to obtain
\begin{eqnarray} 
\label{CHCKsol}
H^{I} = \frac{C^{I}}{\cosh(\eta)} \ . 
\end{eqnarray}
We  have thus succeeded in finding a family of exact  solutions to the localization equations
which respect the classical boundary conditions on $AdS_2$ and are smooth everywhere in the interior. 
In terms of the original variables defined in \eqref{offshellpar}, we have 
\begin{eqnarray} \label{HEKEsol}
&& X^{I}  =  X^{I}_{*} +  \frac{C^{I}}{\cosh(\eta)} \ , \qquad  \bar X^{I}  =  \bar X^{I}_{*} +  \frac{C^{I}}{\cosh(\eta)}\\
&& \qquad \qquad Y^{I1}_{1} = - Y^{I2}_{2} =  \frac{2C^{I}}{\cosh(\eta)^2} \,\,  .
\end{eqnarray}
Since the scalar  fields are now
excited away from their attractor values, they are no longer at the minimum  of the classical entropy function.  Even though scalar fields `climb up' the potential away from the minimum of the entropy function the solution remains Q-supersymmetric (in the Euclidean theory) because an auxiliary field gets 
excited appropriately to satisfy the Killing spinor equations.

It is worth pointing out that the solutions \eqref{CHCKsol} and \eqref{Ksol1} look much simpler if we use the conformal transformation \eqref{conformal-trans} in \S\ref{Supersymmetries} to map $AdS_{2} \times S^{2}$ to $S^{4}$. Since the scalar fields $X$ and the auxiliary fields $Y$  have Weyl weight 1 and 2 respectively,
and since the conformal factor is $\cosh(\eta)$,  the fields $\Sigma$ and $Y$  are simply constant   on $S^{4}$. 
This is very similar to the localizing solution found by Pestun \cite{Pestun:2007rz} in a very different context of computing the expectation value of Wilson line in super Yang-Mills theory on $S^{4}$. Of course, under this conformal transformation the attractor values also will transform and since they are constant on $AdS_{2} \times S^{2}$, they  will no longer be constant on $S^{4}$. It is therefore more natural to work in the $AdS_{2} \times S^{2}$ frame. In any case, for computing the quantum entropy, the $AdS_{2}$ boundary conditions play an important role as we will see in the next subsection.    As pointed out in \cite{Banerjee:2009af}, in this frame our computation  has close formal similarity with the gauge theory computation of `t Hooft-Wilson line in the formulation of \cite{Kapustin:2005py, Gomis:2009ir} which could be useful in the computation of one-loop determinants and the instanton contributions. Note  that we are using localization techniques to evaluate a bulk functional integral of supergravity whereas in \cite{Pestun:2007rz, Kapustin:2005py, Gomis:2009ir} it was used to evaluate a functional integral in the boundary gauge theory.

\subsubsection{Renormalized action for the localizing instantons \label{RenAction}}

To obtain the exact macroscopic quantum partition function we would like to evaluate the renormalized action restricted to the  submanifold $\mathcal{M}_{Q}$ in field space of localizing instantons. We will find that even though both the original action and the solution are rather complicated, the renormalized action  is a remarkably simple function   of the collective coordinates $\{C^{I}\}$ determined entirely by the prepotential. Recall that the renormalized action defined in  the last section takes the form
\be\label{Sren2}
\CS_{\rm ren} :=   \CS_{\rm bulk}  + \CS_{\rm bdry} + i \, \frac{q_I }{2}  \int_{0}^{2\pi} A^I_{\theta}  \, d\theta \ .
\ee
The charges used here are related to the ones used in \eqref{Sren} by  $q_{I} =  -2 	q_{i}$ to be consistent with the normalization of gauge fields used in the literature, for example, in the reviews \cite{Sen:2007qy, Mohaupt:2000mj}.

We proceed to evaluate the bulk action given as a four dimensional integral of the  the supergravity Lagrangian \eqref{PoincaresugraLag} over $AdS_{2} \times S^{2}$. 
We note first that since various auxiliary  fields vanish for the off-shell solution, 
the Lagrangian  \eqref{PoincaresugraLag} simplifies to (recall $\widehat A = (T^{ij}_{ab} \, \ve_{ij})^2$):
%\begin{eqnarray}
% 8\pi \, \mathcal{L} & = & -\frac{i}{2} \, (X^I\bar{F}_I-\bar{X}^IF_I) \, R
%  +\big[i \, \partial_{\mu}F_I \, \partial^{\mu}\bar{X}^I 
%  +\frac{i}{4} \, F_{IJ} \, (F^{-I}_{\mu\nu}-\frac{1}{4}\bar{X}^I \, T^{-}_{\mu\nu}) 
%    (F^{-J\mu\nu}-\frac{1}{4}\bar{X}^J \, T^{-\mu\nu}) \nonumber\\
%&& \quad +\frac{i}{8} \, \bar{F}_I \, (F^{-I}_{\mu\nu}-\frac{1}{4}\bar{X}^I \, T^{-}_{\mu\nu}) \, T^{-\mu\nu}
%   -\frac{i}{8} \, F_{IJ} \, Y^I_{ij} \, Y^{Jij} + \frac{i}{32} \, \bar{F} \, \hat{A} 
%   + \frac{i}{2} \, F_{\hat{A}} \, \hat{C} + {\rm h.c.} \big] \ . 
%\end{eqnarray}
\begin{eqnarray}
 8\pi \, \mathcal{L} & = & -\frac{i}{2} \, (X^I\bar{F}_I-\bar{X}^IF_I) \, R \nonumber\\
  &&+\big[i \, \partial_{\mu}F_I \, \partial^{\mu}\bar{X}^I 
  +\frac{i}{4} \, F_{IJ} \, (F^{-I}_{\mu\nu}-\frac{1}{4}\bar{X}^I \, T^{ij}_{\mu\nu} \ve_{ij}) 
    (F^{-J\mu\nu}-\frac{1}{4}\bar{X}^J \, T^{\mu\nu ij} \ve_{ij}) \nonumber\\
&& +\frac{i}{8} \, \bar{F}_I \, (F^{-I}_{\mu\nu}-\frac{1}{4}\bar{X}^I \, 
   T_{\mu\nu ij} \ve^{ij}) \,T^{ij}_{\mu\nu} \ve_{ij}
   -\frac{i}{8} \, F_{IJ} \, Y^I_{ij} \, Y^{Jij} + \frac{i}{32} \, \bar{F} \, \hat{A} 
   + \frac{i}{2} \, F_{\hat{A}} \, \hat{C} + {\rm h.c.} \big] \nonumber\\. 
\end{eqnarray}
Moreover,  for  $AdS_2\times S^2$ both the Ricci scalar $R$ and the Weyl tensor 
$C$ are zero. 
Substituting  $X^I = X^{I}_{*} + \Sigma^{I}$  and $\bar X^I =\bar X^{I}_{*} + \bar \Sigma^{I}$ from \eqref{offshellpar} and using the attractor equation \eqref{attractor eqs again}
in the form 
% fact that
%\begin{equation}
%F^{-I}_{\mu\nu}-\frac{1}{4}\bar{X_{*}}^IT^{-}_{\mu\nu} = 0 \ , 
%\end{equation}
\begin{equation}
F^{-I}_{\mu\nu}-\frac{1}{4}\bar{X_{*}}^I \, T^{ij}_{\mu\nu} \ve_{ij} = 0 \ , 
\end{equation}
we get 
\begin{eqnarray}
 8\pi \, \mathcal{L} = i \, F_{IJ} \, (\partial_{\eta}\bar{\Sigma}^I) (\partial_{\eta}\Sigma^J)
  -i \, F_{IJ} \bar{\Sigma}^I \, \Sigma^J +\frac{i}{4} \, F_{IJ} \, K^I  K^J 
  +2i \, \bar{F}_I \, \bar{\Sigma}^I- 2i \, \bar{F} + {\rm h.c.} \ . 
\end{eqnarray}
Substituting the solution \eqref{HEKEsol} into the above equation, we find that  the first three terms add up  to zero.
We are thus left with 
\begin{equation} \label{Lstep}
 8\pi\mathcal{L}=2i\bar{F}_I\bar{\Sigma}^I-2i\bar{F}+ h.  c. \, .
\end{equation}
Since we keep the classical values $X^{I}_{*}, \bar X^{I}_{*}$ fixed in this problem, 
differentiating with respect to $X^{I}$ is the same as differentiating with respect to $\Sigma^{I}$.
%so we can rewrite \eqref{Lstep} as 
%\begin{equation} \label{Lstep2}
% 8\pi\mathcal{L} = 2i \, \Sigma_{I}^{2} \, \p_{\Sigma_{I}} \left(\frac{\bar{F}}{\bar \Sigma_{I}} - \frac{F}{\Sigma_{I}} \right)
% 8\pi \, \mathcal{L} = 2i \,  \p_{\tau_{I}} \left( \tau_{I}  F -  \bar \tau_{I} \bar{F} \right) \ , \qquad \tau_{I} = \Sigma_{I}^{-1} \ . 
%\end{equation}
%For the solutions \eqref{HEKEsol} where $\tau_{I} = -i r/C_{I}$,
This can be explicitly evaluated to find
\begin{eqnarray} \label{Lstep3}
 8\pi \, \mathcal{L}  =  2 i \p_{r} \left(r (F - \bar F) \right) \ , \quad \textrm{with} \quad \Sigma^{I} = \frac{C^{I}}{r} \, .
\end{eqnarray}
%%
%\begin{eqnarray}
% 8\pi\mathcal{L} = - 2 i\frac{\cosh(\eta)}{\sinh(\eta)}\partial_{\eta}\bar{F}-2i\bar{F}+h.c.
% = -\frac{2i}{\sinh(\eta)}\partial_{\eta}\left[\cosh(\eta)(\bar{F}-F)\right].
%\end{eqnarray}
%
The $\mathcal{N}=2$ supergravity Euclidean action is
\begin{equation}
 \mathcal{S}_{\rm bulk} =  \int d^4x \sqrt{g} \,  \mathcal{L} \, .
\end{equation}
The off-shell fields do not depend on  the coordinates of the  $S^{2}$ and the angular variable $\theta$ of the $AdS_{2}$. These integrals can be done trivially and give an overall factor of $8\pi ^{2}$, so that 
\begin{eqnarray} \label{Sbulk}
 \CS_{\rm bulk}  & = &  {8 \pi^{ 2}} \int_0^{\eta_{0}} \CL \,  \sinh(\eta) \,  d\eta \  
 = 8\pi^{2} \int_1^{r_{0}} \CL \, dr \ , \nonumber\\
& = &    2 \pi i \int_1^{r_{0}} dr \partial_{r} \left(r (F - \bar F) \right) \ , \nonumber \\
& = &  2 \pi i r_{0} \Big [F\big(X_{*}^{I} + \frac{C^{I}}{r_{0}}\big) - \bar  F\big(X_{*}^{I} + \frac{C^{I}}{r_{0}}\big) \Big]  
 - 2 \pi i \Big[F(X_{*}^{I} + C^{I}) -  \bar F(X_{*}^{I} + C^{I}) \Big]. \nonumber \\ 
 \label{Sbulkeval} 
\end{eqnarray}
The first piece in \eqref{Sbulkeval} which is linear in $r_{0}$ can be rewritten as:
%\begin{eqnarray} \label{Sbulk1}
%&&2 \pi i r_{0} \Big(F\big(X_{*}^{I} + i \frac{C^{I}}{r_{0}}\big)  - \bar F\big(X_{*}^{I} + i \frac{C^{I}}{r_{0}}\big) \Big) \cr
%& & \qquad \qquad = - 2 \pi i r_{0} \big(F(X_{*}^{I}) -  \bar F(X_{*}^{I}) \big) - 2 \pi (F_{I}(X_{*}^{I}) - \bar F_{I}(X_{*}^{I})) \, C^{I} + \CO(1/r_{0})\cr
% & &  \qquad \qquad =  - 2 \pi i r_{0} \big(F(X_{*}^{I}) -  \bar F(X_{*}^{I}) \big) - 2 \pi i q_{I} C^{I} + \CO(1/r_{0}) \ , 
%\end{eqnarray}
\begin{eqnarray} \label{Sbulk1}
&& 2 \pi i r_{0} \Big(F\big(X_{*}^{I} + \frac{C^{I}}{r_{0}}\big)  - \bar F\big(X_{*}^{I} + \frac{C^{I}}{r_{0}}\big) \Big) =  \cr
& & \qquad \qquad =   2 \pi i r_{0} \big(F(X_{*}^{I}) -  
\bar{F}(X_{*}^{I}) \big) + 2 \pi i (F_{I}(X_{*}^{I}) - \bar{F}_{I}(X_{*}^{I})) \, C^{I} + \CO(1/r_{0})\nonumber\\
 & &  \qquad \qquad =    2 \pi i r_{0} \big(F(X_{*}^{I}) -  \bar F(X_{*}^{I}) \big) - 2 \pi q_{I} C^{I} + \CO(1/r_{0}) \, 
\end{eqnarray}
where we have used a Taylor expansion in the first line and the attractor equation 
\be
F_{I}(X_{*}^{I}) - \bar F_{I}(X_{*}^{I}) = i q_{I}
\ee
in the second. 

The Wilson line evaluates to 
\be\label{Wilsoneval}
  i \, \frac{q_I }{2}  \int_{0}^{2\pi} A^I_{\theta}  \, d\theta \  =  \pi q_{I} e^{I}_{*} (r_{0} -1) \ . 
\ee
Hence we choose 
\be\label{Sbdry}
%\CS_{\rm bdry} = \pi r_{0} \left( \frac{q_{I} \, e^{I}_{*}}{2} + 2 i \, \big(F(X_{*}^{I}) -  \bar F(X_{*}^{I}) \big)  \right) \ . 
\CS_{\rm bdry} =  - 2 \pi r_{0} \left(  \frac{q_{I} \, e^{I}_{*}}{2} + i \, \big(F(X_{*}^{I}) -  \bar F(X_{*}^{I}) \big)  \right) \ . 
\ee
so that $\CS_{ren} =  \CS_{\rm bulk} + \CS_{\rm bdry} + i \frac{q}{2} \oint A$ is finite. 

As explained in section \S3,  the main purpose 
of the boundary action is to cancel the divergence in the bare bulk action plus Wilson line
which grows linearly with the length of the boundary. 
%The boundary action is an integral of a local gauge invariant operator. 
In order to cancel this divergence, 
we use a boundary cosmological constant which 
%This cosmological constant  should not depend on the fields we are integrating over, and 
must be specified along with the other boundary data. 
Indeed we have found that  $\CS_{\rm bdry}$ which is a constant that grows linearly 
with the length of the boundary indeed only depends on the fixed charges and not on the 
fluctuating fields. 

In general, however, there could be a finite part of the boundary action which does depend 
on the fields that are integrated over. The full boundary action should be constrained by 
supersymmetry. We shall discuss the supersymmetry of the functional integral 
in appendix \S\ref{Supersymmetry}. The conclusion of the analysis in appendix \S\ref{Supersymmetry} 
is quite simple -- the  finite part of the boundary action in our problem actually vanishes 
due to supersymmetry, and therefore the above prescription for $S_{\rm ren}$ as a sum of terms \eqref{Sbulk}, 
\eqref{Sbdry} and \eqref{Wilsoneval} is already supersymmetric. In appendix  \S\ref{Supersymmetry},
we shall rewrite the above in a manner that is manifestly supersymmetric. This 
rewriting takes the form of a functional integral with a supersymmetric Wilson line 
\cite{Maldacena:1998im, Rey:1998ik}  with the bulk action as above \eqref{Sbulk}, and a boundary action which exactly cancels the boundary piece in  \eqref{Sbulkeval}.

We thus  obtain the following expression for the renormalized action:
\be\label{Srenfinal}
\CS_{\rm ren} =  - \pi \, q_I \, e^I_* - 2 \pi  q_{I} C^{I}  
 - 2 \pi i \big(F(X_{*}^{I} + C^{I}) -  \bar F(X_{*}^{I} + C^{I}) \big) \ , 
\ee
The notation $e^{I}_{*}$ refers to the classical values of the electric field strengths as a function of the 
charges $(q_{I}, p^{I})$.
Using the scalar attractor values  \eqref{scalarattval}, 
and the new variable 
\be\label{ephi}
\phi^I \equiv e_{*}^I+2 C^I \ ,
\ee
we can express the renormalized action in a remarkably simple form:
\begin{eqnarray}\label{Srenormalized}
 \mathcal{S}_{ren}(\phi, q, p) =  - \pi q_I\phi^I + \mathcal{F}(\phi, p)\, .
\end{eqnarray}
with
\begin{equation} \label{freeenergy2}
\mathcal{F}(\phi, p) = - 2\pi i \left[ F\Big(\frac{\phi^I+ip^I}{2} \Big) -
 \bar{F} \Big(\frac{\phi^I- ip^I}{2} \Big) \right] \, .
 \end{equation}
Note that the electric field remains fixed at the attractor value but $\phi^{I}$ can still fluctuate with  $C^{I}$ taking values over the  real line. We will discuss the significance of this fact in 
\S\ref{Connection}. Note also that the prepotential is evaluated at precisely for values of the scalar fields at the origin of $AdS_{2}$ and not at the boundary of $AdS_{2}$.
Thus the classical contribution to the localization integrand will be of the form
\begin{equation}
e^{S_{ren}} =  e^{- \pi  \phi^{I} q_{I} + \mathcal{F}(\phi, p)}
\end{equation}
There will be additional contributions to the integral which we discuss next.

\subsubsection{Evaluation of $\widehat W(q, p)$ \label{What}}

We have thus determined which field configuration to integrate over and the classical action for these configuration.
The full functional integral will require three additional ingredients.

\begin{myitemize}
\item The integration measure over the $\{C^{I} \}$ fields over 
the submanifold $\mathcal{M}_{Q}$ of critical points of $Q$ simply  descends from the measure $\mu$ of supergravity over the field space $\mathcal{M}$. We denote this measure by $[dC]_{\mu}$ which can be computed using standard methods of collective coordinate quantization.  
\item There will  be one-loop determinants of fluctuations around the localizing manifold which can be evaluated from the quadratic piece of the localizing action $S^{Q}$.  We denote  this determinant contribution by $Z_{det}$. It is in  principle a straightforward but technically involved computation. Very similar determinants have been analyzed in  detail  for gauge theory \cite{Pestun:2007rz}. In string theory, the one-loop determinants and the duality invariance  measure around the on-shell solution have been analyzed in \cite{Banerjee:2010qc} and in \cite{LopesCardoso:2006bg, Cardoso:2008fr} respectively.  Some aspects of these  computations both from gauge theory and from around the on-shell saddle point  could be adapted to study the measure and determinants around our off-shell instantons solutions \cite{Dabholkar:2011ec}.  
\item In addition, there will be a contribution from point  instantons and anti-instantons viewed as singular configurations that couple to the vector multiplet fields as long as they preserve the same supersymmetry.  In gauge theory computations \cite{Pestun:2007rz}, the instantons will  be localized at the center of $AdS_{2}$ and at the north pole of the $S^{2}$ whereas the anti-instantons will be be localized at the center of $AdS_{2}$ and at the south pole of the $S^{2}$. Since string theory contains gauge theory at low energies we expect a similar structure also in string theory.  We denote this generating function for the instantons  by $Z_{inst}$. The generating function for anti-instantons will be the complex conjugate of the generating function for instantons. We will thus get a factor of $|Z_{inst}|^{2}$ which will  depend on the details of the string compactification, the spectrum of wrapped brane-instantons, and  the duality frame under consideration.
In gauge theory this generating function is the equivariant instanton partition function computed by Nekrasov \cite{Nekrasov:2002qd}. Since the low energy limit of string theory will reduce to gauge theory on $AdS_{2} \times S^{2}$, it would be interesting to explore if there are generalization of the gauge theory results to string theory. 
 \end{myitemize}

Putting these ingredients together we can conclude that the functional integral will have the form
\begin{equation}\label{integral2}
 \widehat W(q, p) = \int_{\mathcal{M}_{Q}}  e^{-\pi  \phi^{I} q_{I}} e^{\mathcal{F}(\phi,  p)} |Z_{inst}|^{2} \, Z_{det} \, \, [dC]_{\mu}\,     .
\end{equation}Note that the Wilson line $\widehat{W}$ corresponds to localization on $AdS_2\times S^2$ while $W(q,p)$ can contain additional contributions coming from orbifolds.
We have thus successfully reduced the functional integral to ordinary integrals. The dominant piece of the answer given by $e^{-S_{ren}}$ we have already evaluated explicitly.

In specific string compactifications the undetermined factors $Z_{det}$ and $|Z_{inst}|^{2}$ can simplify. For example, with $\CN=4$ supersymmetry, in gauge theory  both $|Z_{instanton}|^{2}$ and $Z_{det}$ equal unity.  Similarly, it was found in  \cite{Banerjee:2010qc} that very similar determinant factors for vector multiplets equal unity $\CN=4$ theories. One expects that this simplification will extend to the factors appearing in \eqref{integral2} around the localizing solution in $\CN=4$ theories.

%\section{Quantum Entropy and the Topological String\label{Connection}}
\subsection{Quantum entropy and the topological string\label{Connection}}

We now turn  to the original problem of evaluating of $W(q, p)$.  There are several issues that have to be addressed to extend the supergravity computation to a full string computation.

\begin{myitemize}
\item First, the full action of string theory of course contains more fields in addition to vector multiplets, in particular the hyper multiplets. 
\item Second, even if we restrict our attention to vector multiplets, the action will in general contain not just the F-terms which are chiral superspace integrals but also the D-terms which are  nonchiral superspace integrals. 
\item Third, there can be additional contributions  from functional integral over  orbifolds of $AdS_{2}$ that are allowed in the full string theory but not visible in supergravity. 
\end{myitemize}
 We discuss these questions below.

\subsubsection{D-terms,  hyper-multiplets, and evaluation of $W(q, p)$\label{Eval}}

We have thus far considered only F-type terms for the action of the vector multiplets which are chiral integrals over $\mathcal{N} =2$ superspace of the form $\int d^{4}\theta$. 
The effective action of string theory will contain  in general  D-type terms which are nonchiral  integrals over $\mathcal{N} =2$ superspace of the form $\int d^{4}\theta d^{4}\bar \theta$.  It is not \textit{a priori} clear that these terms will not contribute to the functional integral. We would like to make  the following two observations in this connection.
\begin{myitemize}
\item Since our localizing action $S^{Q}$ follows from off-shell supersymmetry transformations,  it does not depend on what terms  are present in the physical action $S$.  Hence our localizing instanton solutions are universal and they will continue to exist even with the addition of the  D-terms. The question then reduces to evaluating  the D-terms on these solutions to obtain their contribution to the renormalized action.
\item It has recently been shown \cite{deWit:2010za} that a large class of D-type terms do not contribute to the Wald entropy.  This class of terms are constructed using  the `kinetic multiplet' $T$  obtained from a chiral multiplet  $\Phi$  of Weyl weight $0$  by $T = \bar D^{4} \bar \Phi$ which transforms like a chiral multiplet of Weyl weight $2$. One can construct now supersymmetry invariant terms in the action as  chiral integrals $\int d^{4}\theta$ with arbitrary polynomials involving the  kinetic multiplet and other chiral multiplets. Since  four antichiral derivatives have the same effect as the four antichiral integrals,  these terms  correspond to  D-terms with non chiral integrals $\int d^{4 }\theta d^{4} \bar \theta$  of terms involving the original field $\Phi$. The nonrenormalization theorem of  \cite{deWit:2010za}  shows that D-terms of this type do not contribute to the Wald entropy.
Since the renormalized action of the localizing instantons follows from the bulk action and has the same form as the entropy function, it should be possible to extend this nonrenormalization theorem to the renormalized action discussed in this paper. 
\end{myitemize}
These two points indicate that the D-terms, or at least a large subclass of them,  may in fact not contribute to the renormalized action. 

Adding hyper multiplets does not change the transformation rules of the vector multiplets. We therefore expect that  the localizing instantons that we have found here will continue to exist. There could be in principle additional localizing solutions where hyper multiplet fields are excited but this may not necessarily happen. It then only remains to check that the coupling of hyper multiplets and vector multiplets at high order cannot contribute to the renormalized action. Lacking an offshell formulation of couplings between hypers and vectors, we cannot at present address this question but perhaps something analogous to the nonrenormalization theorem discussed above can be extended to these terms as well.

In any case, these questions  can be systematically investigated in the context of our off-shell localizing instantons. If some of the D-terms do happen to contribute to the renormalized action, their contribution can  be taken into account by evaluating them on  the off-shell solutions. Similarly if there are new localizing instantons upon the inclusion of hypers, those too can be added as separate contribution to the final answer for  the functional integral.

If the hyper multiplets and D-terms can be ignored for reasons outlined above, one can conclude that  
$W_{0}(q, p)$ has the same form as $\widehat W(q, p)$ evaluated  in \S\ref{Solution} 
 \begin{equation}\label{integral3}
 W_{0} (q, p) = \int_{\mathcal{M}_{Q}} \, e^{- \pi  \phi^{I} q_{I}}  \, | Z_{top}(\phi, p ) |^{2}  \, 
 \, Z_{det}  \, [dC]_{\mu} \end{equation}
The contribution from the orbifolds of $AdS_{2}$ also has a very similar structure since the localizing instanton solution is still valid.

\section{Quantum entropy of large black holes in IIB on $T^6$ }
The $AdS_{2}/CFT_{1}$ correspondence thus provides  a simple and yet nontrivial example of holography. Note that   $d(q, p)$ is the statistical degeneracy of the ensemble of quantum microstates that correspond to the black hole which  in general  is a highly nontrivial function of the integer charges. If the black hole preserves at least four supersymmetries this degeneracy equals an index and hence can be computed reliably in several examples.  On the other hand, $W(q, p)$ is the generalization of the exponential of the Wald entropy of the black hole.
The  equality of $d(q, p)$ and $W(q, p)$ for arbitrary finite values of the charges\footnote{Rules of $AdS_{2}/CFT_{1}$ correspondence suggest that the natural ensemble to make this comparison is the microcanonical ensemble fixing all charges rather than chemical potentials  \cite{Sen:2008yk, Sen:2008vm}.} can thus be viewed as a  statistical interpretation of the \textit{exact} quantum entropy of  the black hole for finite charges, including all corrections--both perturbative as well as nonperturbative in $1/Q$ where $Q$ denotes a generic charge.  It would be a rather nontrivial check of the nonperturbative structure of string theory if $W(q, p)$ evaluated from a functional integral of string theory  can precisely reproduce the full functional dependence of this integer $d(q, p)$ on  the integral charges $(q, p)$.

In this section we apply the previous results in the concrete context of supersymmetric black holes  preserving four supersymmetries in $\mathcal{N}=8$ supersymmetric compactifications of string theory to four spacetime dimensions.  Since the structure of  the $\CN=8$ theory is particularly simple, it enables us to analytically perform the ordinary integrals that remain after localization and  evaluate $W(q, p)$ even after including nonpertubative effects.  
The resulting $W(q, p)$ matches in remarkable details with the  quantum degeneracies $d(q, p)$ of these black holes that are known independently \cite{Dabholkar:2011ec}.

\subsection{Microscopic Quantum Partition Function \label{Micro}}

Consider Type-II string compactified  on a 6-torus $T^{6}$. The resulting four-dimensional theory has $\CN=8$ supersymmetry with $28$ massless $U(1)$ gauge fields. A charged state is therefore characterized by $28$ electric and $28$ magnetic charges which combine into the $\bf 56$ representation of the U-duality group $E_{7, 7} (\mathbb{Z})$.  Under the  $SO(6, 6; \mathbb{Z})$ T-duality group, the $28$ gauge fields decompose as
\be
\textbf{28} =\textbf{12} + \textbf{16}
\ee
where the fields in the vector representation $\bf 12$  come from the NS-NS sector, while the fields in the spinor representation $\bf 16$  come from the R-R sector. We obtain an $\CN=4$ reduction of this theory by dropping four gravitini multiplets. Since each graivitini multiplet of $\CN=4$ contains four gauge fields, this amounts to dropping sixteen gauge fields which we take to be the R-R fields in the above decomposition.The  U-duality group of the reduced theory is
\be
SO(6, 6; \mathbb{Z}) \times SL(2, \mathbb{Z}) \,
\ee
where  $SL(2, \mathbb{Z})$ is the electric-magnetic S-duality group. 

\subsubsection{Charge Configuration \label{Config}}

We will be interested in one-eighth BPS dyonic states in this theory which perserve four of the thirty-two supersymmetries. 
To  simplify things, we consider the 6-torus to be the product  $T^4 \times S^{1} \times \wt S^{1}$ of a 4-torus and  two circles. Let  $n$ and $w$ be  the momentum and winding along the circle $S^{1}$, and  $K$ and $W$ be the corresponding Kaluza-Klein monopole and NS5-brane charges. Let $\wt n, \wt w, \wt K, \wt W$ be the corresponding charges associated with the circle $\wt S^{1}$. A general charge vector with these charges can be written as a doublet of $SL(2, \mathbb{Z})$
\begin{equation}\label{hetcharges}
    \Gamma = \left[
               \begin{array}{c}
                 Q \\
                 P \\
               \end{array}
             \right] =
             \left[
               \begin{array}{cccc}
                 {\wt n}& n  & {\wt w}& w \\
                 {\wt W}& W & {\wt K} & K \\
               \end{array}
             \right]_{B'},
\end{equation}
where the subscript $B'$ denotes a particular Type-IIB duality frame. The  T-duality invariants for this configuration are \cite{Giveon:1994fu}
\be
Q^{2} = 2(n w + \wt n \wt w) \, , \qquad P^{2} = 2(KW + \wt K \wt W) \, , \qquad Q \cdot P = n K + \wt n \wt K  + w W + \wt w \wt W \, ,
\ee
and the quartic U-duality invariant can be written as
\be
\Delta = Q^{2} P^{2} - (Q \cdot P )^{2} \, .
\ee
For our purposes it will suffice to excite only five charges 
\begin{equation}\label{charges}
    \Gamma =
             \left[
               \begin{array}{cccc}
                0& n  & 0& w \\
                 {\wt W}& W & {\wt K} & 0 \\
               \end{array}
             \right]_{B'}
\end{equation}
so that the T-duality invariants  are all nonzero. 
There are three other duality frames  that are of interest.

\begin{myitemize}
\item Frame $B$: In this frame the charge configuration becomes
%  \cite{deWit:ZZZ}
\begin{equation}\label{chargesB}
    \Gamma =
             \left[
               \begin{array}{cccc}
                0& n  & 0& \wt K \\
                 Q_{1}& \wt n & Q_{5} & 0 \\
               \end{array}
             \right]_{B} \, ,
\end{equation}
where $Q_{1}$ is the number of D1-branes wrapping $S^{1}$ and $Q_{5}$ is the number of D5-branes wrapping $T^{4}\times S^{1}$. This frame is particularly useful for the microscopic derivation of the degeneracies described in 
\S\ref{microcount}. 
With $\wt K =1$, the Kaluza-Klein monopole interpolates between $\IR^{3} \times \wt S^{1}$  at asymptotic infinity and $\IR^{4}$ at the center. The momentum $\wt n$ at infinity becomes angular momentum at the center. This allows for a 4d-5d lift  \cite{Gaiotto:2005gf, David:2006yn} to relate the degeneracies of the four-dimensional state  to those of five-dimentional D1-D5 system carrying momentum $n$ and angular momentum $\wt n$.
\item Frame $A$: In this  frame the charge configuration becomes 
\begin{equation}\label{chargesA}
    \Gamma =
             \left[
               \begin{array}{cccc}
                0& q_{0}  & 0 & -p^{1} \\
                 p^{2}& q_{2}& p^{3} & 0 \\
               \end{array}
             \right]_{A} \, ,
\end{equation}
where $q_{0}$ is the number of D0-branes, $q_{2}$ is the number of D2-branes wrapping $S^{1}\times \wt S^{1}$, $p^{1}$  is a D4-brane wrapping $T^{4}$, $p^{2}$ is a D4-brane wrapping $\Sigma_{67}\times S^{1}\times \wt S^{1}$ and $p^{3}$ is a D4-brane wrapping $\Sigma_{89}\times S^{1}\times \wt S^{1}$ where $\Sigma_{ij}$ is a 2-cycle in $T^{4}$ along the directions $ij$. We will use this frame for localization in \S\ref{Sugra} and \S\ref{Macro}.
\item Frame $B^{''}$: In this frame the charge configuration becomes
\begin{equation}\label{chargesBdp}
    \Gamma =
             \left[
               \begin{array}{cccc}
                0& n  & 0 & Q_{5} \\
                Q_{3}& Q_{1}& Q_{3} & 0 \\
               \end{array}
             \right]_{B''} \, ,
\end{equation}
where all D-branes wrap the circle $S^{1}$ and an appropriate cycle in the $T^{4}$. 
\end{myitemize}

We can choose a charge configuration which is even simpler: 
\begin{equation}\label{chargesF}
    \Gamma =
             \left[
               \begin{array}{cccc}
                0& n  & 0& 1 \\
                1 & \nu & 1 & 0 \\
               \end{array}
             \right]
\end{equation}
where $n$ is a positive integer and $\nu$ takes values $0$ or $1$. 
The U-duality invariant is 
\be \label{udualinvt}
\Delta  = 4n - \nu^{2} \, .
\ee
It is clear that $\nu = \Delta$ modulo $2$, and so these states are completely specified by $\Delta$.
The states preserve four of the thirty-two supersymmetries. We will henceforth denote the degeneracies of these one-eighth BPS-states with charges \eqref{chargesF} by
$d(\Delta)$ instead of  $d(q, p)$. 

We should emphasize that a large class of states with the same value of $\Delta$ can be mapped by U-duality to  the state \eqref{chargesF} considered here but that does not exhaust all states. 
Note that the invariant $\Delta$ is the unique quartic invariant of the continous duality group $E_{7, 7}(\mathbb{R})$ but in general there are additional arithmetic duality invariants of the arithmetic group $G(\IZ)$  that cannot be written as invariants of $G({\IR})$.   As a result,  not all states with the same value of $\Delta$ are related by duality.  Classification of arithmetic invariants of $G(\IZ)$  is a subtle number-theoretic problem. For example, for the $\CN=4$ compactification where the duality group $O(22, 6; \IZ) \times SL(2, \IZ)$, essentially the only relevant arithmetic invariant  is given by $I = \gcd (Q \wedge P)$; and the degeneracies are known for all values of $I$ 
\cite{Banerjee:2007sr, Banerjee:2008ri, Banerjee:2008pu, Dabholkar:2008zy}. 
To our knowledge a similar complete classification  of  $E_{7, 7}(\mathbb{Z})$ invariants is not known at present. This would be a problem if one wishes to use canonical or a mixed ensemble. For our purposes, since we will working in the microcanonical ensemble, it will suffice to know the degeneracies for the states in the duality orbit of  \eqref{chargesF}.

\subsubsection{Microscopic Counting \label{microcount}}

The degeneracies of the 1/8-BPS dyonic states in the type II string theory on a $T^{6}$
are given in terms of the Fourier coefficients of the following counting function 
\cite{Maldacena:1999bp, Shih:2005qf, Sen:2008ta}: 
\bea\label{ourmicro}
F(\tau, z) &=&  \frac{\vth_1^2(\t, z)}{\eta^6(\t)} \, .
\eea  
where $\vartheta_{1}$ is  the Jacobi theta  function and  $\eta$ is the Dedekind function. With $ q:= e^{2\pi i \tau}$ and $ y:= e^{2\pi i z}$, they have the product representations
\bea \nonumber
\vth_1(\t, z) &=& q^{\frac{1}{8}} ( y^{\half} - y^{-\half})\prod_{n=1}^{\infty}(1 - q^{n}) (1 -  y q^{n}) (1 - y^{-1}q^{n}) \, ,\\ 
\eta(\tau) &=& q^{\frac{1}{24}} \prod_{n=1}^{\infty}(1 - q^{n}) \, \, .
\eea

The derivation of the counting function is simplest in the $B$ frame \eqref{chargesB} where we have  a D1-D5 system  in the  background of a single Kaluza-Klein monopole. By the 4d-5d lift, 
the momentum $\nu$ can be interpreted as 5d angular momentum.  The 
counting problem essentially reduces to counting bound states in \emph{five} dimensions of a single D1-brane bound to a single D5-brane carrying  $n$ units of momentum and $\nu$ units of angular momentum. Since the D1-brane can move inside the $D5$ anywhere on the $T^{4}$, the moduli space of this motion  is $T^{4}$. The function $F$ is  the  generalized elliptic genus of the corresponding superconformal field theory with target space $T^{4}$. This is  evident from the product representation which can be seen as coming from four bosons and four fermions.

Analysis of the  Fourier coefficients of $F$ simplifies enormously by the fact that $F$ is a \emph{weak Jacobi form}. 
We recall below  a few relevant facts about Jacobi forms \cite{Eichler:1985ja}. 
\begin{myenumerate}
\item \emph{Definition:}
A Jacobi form of  weight $k$ and index $m$  is a  holomorphic function $\varphi(\tau, z)$ from $\mathbb{H} \times\C$ to $\C$ which 
is ``modular in $\tau$ and elliptic in $z $'' in the sense that it transforms under the modular group as
  \be\label{modtransform}  \varphi(\frac{a\t+b}{c\t+d}, \frac{z}{c\t+d}) \ = 
   (c\t+d)^k \, e^{\frac{2\pi i m c z^2}{c\t +d}} \, \varphi(\t,z)  \qquad \forall \quad
   \left(\begin{array}{cc} a&b\\ c&d \end{array} \right) \in SL(2; \mathbb{Z}) \ee
and under the translations of $z$ by $\mathbb{Z} \tau + \mathbb{Z}$ as
  \be\label{elliptic}  \varphi(\t, z+\lambda\tau+\mu)\= e^{-2\pi i m(\lambda^2 \t + 2 \lambda z)} \varphi(\t, z)
  \qquad \forall \quad \l,\,\m \in \mathbb{Z} \, , \ee
where $k$ is an integer and $m$ is a positive integer.

\item \emph{Fourier expansion:}
Equations \eqref{modtransform} include the periodicities $\varphi(\t+1,z) = \varphi(\t,z)$ and $\varphi(\t,z+1) = \varphi(\t,z)$, so  $\varphi$ has a Fourier expansion
  \be\label{fourierjacobi} \varphi(\t,z) \= \sum_{n, r} c(n, r)\,q^n\,y^r\,, \qquad\qquad
   (q :=e^{2\pi i \t}, \; y := e^{2 \pi i z}) \ . \ee
Equation \eqref{elliptic} is then equivalent to the periodicity property
  \be\label{cnrprop}  c(n, r) \= C_{r}(4 n m - r^2) \ ,
  \qquad \mbox{where} \; C_{r}(D) \; \mbox{depends only on} \; r \, \mod\, 2m \ . \ee
The function  is called a \emph{weak} Jacobi form if it satisfies the condition
  \be\label{weakjacobi} c(n, r) \= 0\qquad   \textrm{unless}  \qquad n \geq 0 \, .\ee

\item \emph{Theta expansion:}
The transformation property (\ref{elliptic}) implies a 
Fourier expansion of the form
  \be\label{jacobi-Fourier} \v(\t, z) \= \sum_{\ell\inn \IZ} \;q^{\ell^2/4m}\;h_\ell(\t) \; e^{2\pi i\ell z} \ee
    % \= \sum_{\ell\inn \IZ/2m\IZ} h_\ell(\t) \, \vartheta_{m,\ell}(\t, z) \ee
where $h_\ell(\tau)$ is periodic in $\ell$ with period $2m$.  In terms of the coefficients \eqref{cnrprop} we have
  \be\label{defhltau}  h_{\ell}(\t) \= \sum_{D} C_{\ell}(D) \,  q^{D/4m} \, \qquad \qquad (\ell \inn \IZ/2m \IZ)\;.  \ee
Because of the periodicity property of $h_{\ell}$, equation \eqref{jacobi-Fourier} can be rewritten in the form 
  \be\label{jacobi-theta} \v(\t,z) = \sum_{\ell\inn \IZ/2m\IZ} h_\ell(\t) \, \vartheta_{m,\ell}(\t, z)\,, \ee
where $\vartheta_{m,\ell}(\t,z)$ denotes the standard index $m$ theta function 
  \begin{eqnarray} \label{thetadef} \vartheta_{m,\ell}(\t, z) 
   \;:=\; \sum_{{\l\inn\IZ} \atop {\l\,=\,\ell\,(\mod\,2m)}} q^{\l^2/4m} \, y^\l \, \,
   \= \sum_{n \inn \mathbb{Z}} \,q^{m(n+ \ell/2m)^2} \,y^{\ell + 2mn}  \end{eqnarray}
This is the theta expansion of $\v$.  The vector $h := ( h_1, \ldots, h_{2m})$ transforms like a modular form of weight $k-\frac{1}{2}$ under $SL(2,\IZ)$.
\end{myenumerate}
With these definitions, $F(\t, z)$ is a weak Jacobi form of weight $-2$ and index $1$. The indexed degeneracies  for a state carrying $n$ units of momentum and $r$  units of angular momentum is then given by $c(n, r)$ in the Fourier expansion \eqref{fourierjacobi} of $F$. 
Using \eqref{cnrprop} 
for $m=1$,  we see that $c(n, r)$ depend only on $D = 4n-r^{2}$ and   $r$ mod $2$ which in this case  equals  $D$ mod $2$.  Hence, all information about the Fourier coefficients $c(n, r)$ of $F$ is contained in a single function  of $D$ alone which we denote by $C(D)$. Our task is thus reduced to determining $C(D)$  given \eqref{ourmicro}. 

To read off $C(D)$ more systematically we use the  theta expansion
%  \be\label{jacobi-theta} \varphi_{-2,1}(\t,z) = \sum_{\ell\in \IZ/2\IZ} h_\ell(\t) \, \vartheta_{m,\ell}(\t, z)\,, \ee
\be\label{jacobi-theta2}  
F(\t,z)  = h_0(\t) \, \vartheta_{1,0}(\t, z)\, +  h_1(\t) \, \vartheta_{1,1}(\t, z)\,  .  
\ee
The functions $h_{\ell}(\tau)$ in this case are given explicitly by:
\bea \label{h0h1defs1}
h_{0} (\t) & = & - \frac{\vth_{1,1}(\t,0)}{\eta^{6}(\t)} = -2  -12 q - 56 q^{2}- 208 q^{3}\dots \\
\label{h0h1defs2} 
h_{1} (\t) & = & \frac{\vth_{1,0}(\t,0)}{\eta^{6}(\t)} =   q^{-\frac{1}{4}} \bigl(1 + 8 q + 39 q^{2} + \dots \bigr) 
\eea
For even and odd $D$, the coefficients $C(D)$ can  be read off   from these expansions of  $h_{0}$ and $h_{1}$ respectively using \eqref{defhltau}. 

It is clear that  $D$ can be identified with the duality invariant $\Delta$ in \eqref{udualinvt}. 
The degeneracies are then given  in terms of $ C(D)$ by
\be\label{Ctod}
d(\Delta) =  (-1)^{\Delta +1} C (\Delta) \, .
\ee
The factor of $(-1)^{\Delta}$ arises because  the state in five dimensional spacetime is fermionic for odd $\D$ and contributes to the index with a minus sign. The overall minus sign arises in relating the 4d degeneracies to the 5d degeneracies using the 4d-5d lift \cite{Shih:2005qf, Sen:2008ta}.

\subsubsection{Index, Degeneracy, and Fermions}\label{Index}

The first few terms in the Fourier expansion of $F$ are given by
\be
F(\t, z) = \frac{(y-1)^2}{y} \, - \, 2\,\frac{(y-1)^4}{y^2}\, q \, + \,  \frac{(y-1)^4(y^2-8y+1)}{y^3}\, q^2 \, + \, \cdots \ , 
\ee
In Table \eqref{tablefcoeffs} we tabulate the coefficients $C(\Delta)$ for the first few values  of $\Delta$. 
\begin{table}[h]   \caption{\small{Some Fourier coefficients}}   \vspace{8pt}    \centering
   \begin{tabular}{c|cccccccccccccc}   \hline   
       $\Delta$ &  -1 & 0 &3& 4& 7 & 8 &11& 12& 15 \\
       \hline $C(\Delta)$ & 1 & $-2$ &8 &$-12$&39&$-56$&152&$-208$&513\\   
   \hline    \end{tabular}   \label{tablefcoeffs}   \end{table}
   
It is striking that the sign of $C(\Delta) $ is alternating. This implies from \eqref{Ctod} that the degeneracies $d(\Delta)$ are always positive.  This is, in fact, true not only for the first leading coefficients but  for all Fourier coefficients, 
as can be seen from the equations  
%\eqref{jacobi-theta2}, \eqref{h0h1defs1}, \eqref{h0h1defs2}. 
\eqref{jacobi-theta2}--\eqref{h0h1defs2}. 
Mathematically, the alternating sign of the Fourier coefficients  is a somewhat nontrivial property of the specific Jacobi form \eqref{ourmicro} under consideration \cite{BringmannMurthy:2012}.  Physically, the positivity of $d(\Delta)$ is even more surprising. After all, these are  \emph{indexed} degeneracies  corresponding to a spacetime helicity supertrace for  a complicated bound states of branes. There is no \emph{a priori} microscopic reason why these should be all positive. 

Holography gives a simple physical explanation of the positivity \cite{Sen:2009vz, Sen:2010mz}. 
The near-horizon $AdS_{2}$ geometry has an $SU(1, 1)$ symmetry. If the black hole geometry leaves at least four supersymmetries unbroken, then closure of the supersymmetry algebra requires that the near horizon symmetry must contain the supergroup $SU(1, 1|2)$. This implies that that such a supersymmetric horizon must have $SU(2)$ symmetry which can be identified with spatial rotations.  If $J$ is a Cartan generator of this $SU(2)$, then for  a classical black hole with spherical symmetry, this could mean (depending on the ensemble) that either $J$  is zero  
or the chemical potential conjugate to $J$ is zero. As explained earlier, the $AdS_{2}$ path integral naturally fixes the charges and not the chemical potentials and hence $J=0$.  Together, this implies
\be\label{ind-deg}
\Tr (1) = \Tr (-1)^{J} \, ,
\ee
that is, index equals degeneracy and must be positive. 
For a more detailed discussion see \cite{Dabholkar:2010rm}.

Note the the index equals degeneracy only for the horizon degrees of freedom, but usually one  does not compute  the index of the horizon degrees of freedom directly. It is easier to compute the index  of the asymptotic states as a spacetime helicity supertrace which receives contribution also from the degrees of freedom external to the horizon. It is crucial that the  contribution of these external modes is removed from the helicity supertrace before checking the equality \eqref{ind-deg}.
Typically, modes localized outside the horizon come from fluctuations of supergravity fields and can carry NS-NS charges such as the momentum but not  D-brane charges \cite{Banerjee:2009uk,Jatkar:2009yd}.
In a given frame such as the $A$ frame where all charges come from D-branes, one expects that the Fourier coefficients of  $F(\t, z)$ will give the  degeneracies of only the horizon degrees of freedom. 

For  the Wilson line expectation value \eqref{qef} the equality \eqref{ind-deg} implies   that the functional integral with periodic boundary conditions for  the fermions must equal the functional integral with antiperiodic boundary conditions. This is possible for the following reason. All fermionic fields have nonzero $J$ and  couple to the Kaluza-Klein gauge field coming from the dimensional reduction on the $S^{2}$. As discussed above, the 
microcanonical boundary conditions for the functional integral instructs us to 
integrate over all the fluctuations of the constant mode. By a change of variables in the functional 
integral, one can change the origin of the constant mode of the gauge field, and therefore the 
periodic and antiperiodic boundary conditions for the fermionic fields are equivalent.

%The functional integral with the microcanonical boundary conditions \eqref{asympcond} for this gauge field  includes arbitrary fluctuations of  the constant mode. Integrating over these fluctuations makes the periodic and antiperiodic boundary conditions for the fermionic fields equivalent.

\subsubsection{Rademacher Expansion}

One can make very good estimates of  Fourier coefficients of a modular form using  an  expansion due to Hardy and Ramanujan. The leading term  of this expansion gives the Cardy formula. A generalization  due to 
Rademacher  \cite{Rademacher:1964ra} in fact  gives  an {\it exact} expansion for these coefficients in terms of the coefficients of the polar terms {\it i.e.} terms  with $ D<0$.

One can apply these methods to the Fourier coefficients of  the vector valued modular form $\{ h_{l} \}$ $(l = 0, \ldots 2m-1)$   of negative weight $-w$ to obtain  \cite{Dijkgraaf:2000fq, Manschot:2007ha}  a Rademacher expansion  for the coefficients  $C_{\ell}(D)$ \eqref{defhltau}
 \bea\label{radi} \nonumber
 C_\ell (D) &= & (2\pi)^{2-w} \sum_{c=1}^\infty 
  c^{w-2} \sum_{\wt\ell \inn \IZ/2m \IZ} \, \sum_{\wt D < 0} \, 
C_{\wt\ell}(\wt D) \,  
K(D,\ell,\wt D,\wt\ell;c) \, \left| \frac{\wt D}{4m} \right|^{1-w} \, \wt I_{1-w}
 \biggl[ {\pi\over c} \sqrt{| \wt D|  D}
\biggr] \, ,
\eea
%
% \bea\label{radi} 
% C_\nu(n) &= & (2\pi)^{1-w} \sum_{c=1}^\infty\sum_{\mu =0}^{1}
%  c^{w-2} \, K\ell(n,\nu,m,\mu;c)  \times \\
%&& \quad\sum_{m -\mu^{2}/4 < 0} C_\mu(4m -\mu^{2}) \vert m -\mu^{2}/4 \vert^{1-w} \, \wt I_{1-w}
% \biggl[ {4\pi\over c} \sqrt{\vert m -\mu^{2}/4 \vert(n -\nu^{2}/4)}
%\biggr] \, ,
%\eea
%
where 
\begin{equation}\label{intrep}
 \wt{I}_{\rho}(z)=\frac{1}{2\pi
i}\int_{\epsilon-i\infty}^{\epsilon+i\infty} \, \frac{d\s}{\s^{\r +1}}\exp [{\s+\frac{z^2}{4\s}}]
\, 
\end{equation}
is called the modified Bessel function of index $\r$. This is 
related to the standard Bessel function of the first kind $I_{\rho}(z)$ by
\be
\wt I_{\rho}(z) = \big(\frac{z}{2} \big)^{-\rho} I_{\rho}(z) \, .
\ee
The sum over ($\wt\ell$, $\wt D$) picks up a contribution $C_{\wt\ell}(\wt D)$ from every  
non-zero term $q^{\wt D}$ with $\wt D < 0$ in $h_{\wt\ell}(\tau)$ \eqref{defhltau}.
The coefficients $ K\ell(D,\ell,\wt D,\wt\ell;c)$ are
generalized Kloosterman sums. For $c > 1$ it is  defined as
\be
\label{kloos0}
K(D,\ell;\wt D,\wt\ell;c):=
e^{-\pi i w/2}\sum_{{-c \leq d< 0}  \atop { (d,c)=1}}
e^{2\pi i \frac{d}{c} (D/4m)} \; M(\gamma_{c,d})^{-1}_{\ell\wt\ell} \; 
e^{2\pi i \frac{a}{c} (\wt D/4m)} \, , 
\ee
where
\be\label{gamform}
\gamma_{c,d} = \begin{pmatrix} a & (ad-1)/c \\ c & d \end{pmatrix}
\ee
is an element of $Sl(2,\IZ)$ and  $M(\gamma)$ is the  matrix representation  of $\gamma$ on the vector space spanned by the $\{ h_{l} \}$.  Note that it follows from \eqref{gamform} that $ad = 1\,  \mod\,  c$.

The Jacobi form $F(\t, z)$ has weight  $-2$ and index $m=1$, so its theta expansion gives a two-component vector $\{h_{0}, h_{1}\}$ of modular forms of weight $w = -5/2$. 
Since  there is  only a single polar term $(\wt\ell=1, \wt D=-1)$, the Rademacher expansion takes the form:
 \be\label{rademsp} 
 C(D) =   2{\pi} \, \big( \frac{\pi}{2} \big)^{7/2} \, \sum_{c=1}^\infty 
  c^{-9/2} \, K_{c}(D) \; \wt I_{7/2} \big(\frac{\pi \sqrt{D}}{c} \big)  \, , 
\ee
where the Kloosterman sum $K_{c} (D) $  is defined by
\bea
\label{kloos}
K_{c}(D) :=
e^{5\pi i /4}\sum_{-c \leq d< 0; \atop (d,c)=1}
e^{2\pi i \frac{d}{c} (D/4)} \; M(\gamma_{c,d})^{-1}_{\ell 1} \; 
e^{2\pi i \frac{a}{c} (-1/4)} \qquad  \,  (c >1) \,    \qquad  
\eea
with $\ell = D \, \mod \, 2$ and $ad = 1 \, \mod \, c$.

Under the $SL(2,\IZ)$ generators, the modular form $h_{\ell}(\t)$ transform as
\bea
h_{0} (\tau + 1)  =  h_{0} (\t) \, , \quad  \;  &&  \qquad h_{0} (-1/\tau)  =  \frac{1+i}{2} \, \t^{-5/2}  \big(h_{0} (\t) + h_{1}(\t) \big) \, ; \\ 
h_{1} (\tau + 1)  = -i \, h_{1} (\t) \, , &&  \qquad h_{1} (-1/\tau) =  \frac{1+i}{2} \, \t^{-5/2}  \big(h_{0} (\t) - h_{1}(\t) \big) \, .
\eea
From these transformations, we can read off the matrices $M(\g)$ for the generators  $S$  and $T$ 
\be
T=\bem 1 & 1 \\ 0 & 1 \eem \, ,  \qquad S = \bem 0 & 1 \\ -1 & 0 \eem 
\ee 
to be
\be\label{MTMS}
M(T) = \bem 1&0\\0&-i \eem \, , \qquad M(S) = \frac{e^{\pi i/4}}{\sqrt{2}} \bem 1&1\\1&-1 \eem \, . 
\ee
Using the expression for a general $SL(2, \IZ)$ matrix $\g$ in terms of the generators $S$ and $T$, and the 
representation \eqref{MTMS},  we can obtain the representation $M(\g)$.

We see from \eqref{rademsp} that the microscopic degeneracy is an infinite sum of the form
\be\label{dexp}
d(\Delta)  = \sum_{c=1}^{\infty}d_{c} (\Delta) \, .
\ee
where each term is given by
\be\label{dc}
d_{c} (\Delta)  =  (-1)^{ \Delta +1} \, 2\pi \big(\frac{\pi}{ \Delta}\big)^{7/2} \, 
 I_{\frac{7}{2}}\big(\frac{\pi \sqrt{\Delta}}{c}\big) \, \frac{1}{c^{9/2}}  K_{c} (\Delta) \, .
\ee
It is easy to check that 
\be\label{K1}
K_{1} = (-1)^{ \Delta +1} \frac{1}{\sqrt{2}} \, .
\ee 
We will see that the Wilson line from the macroscopic side also naturally has the same expansion 
\be\label{Wexp}
W(\Delta)  = \sum_{c=1}^{\infty}W_{c} (\Delta) \, ,
\ee
coming from $\IZ_{c}$ orbifolds of $AdS_{2}$. Our objective then is to compute each of these terms exactly using localization. We compute the leading term $W_{1}(\Delta)$ in \S\ref{Macro} and the subleading terms corresponding to $c>1$ in \S\ref{Nonpert}. 

\subsection{Localization of  Functional Integral in Supergravity\label{Sugra}}

Evaluating the formal functional integral \eqref{qef} over string fields for  $W(q, p)$ is of course highly nontrivial.  To proceed further, we first integrate out the infinite tower of massive string modes and massive Kaluza-Klein modes to obtain a \emph{local} Wilsonian effective action for the massless supergravity 
fields keeping all higher derivative terms. We can  regard the ultraviolet finite string theory as providing a supersymmetric and consistent cutoff at the string scale. Our task is then reduced to evaluating a functional integral in supergravity.  The near horizon geometry preserves eight superconformal symmetries and the  action, measure, 
operator insertion, boundary conditions of the functional integral \eqref{qef} are all supersymmetric\footnote{Supersymmetry of the Wilson line and the action is discussed in the appendix.}. The formal supersymmetry of the functional integral makes it possible to apply localization techniques \cite{Dabholkar:2010uh, Banerjee:2009af} to evaluate it. 

To apply localization to our system, we drop two gravitini multiplets to obtain a $\CN=2$ theory and also drop the hypermultiplets to consider a reduced theory. This is partially motivated by the fact that the hypermultiplets are flat directions of the classical entropy function and our black hole is not charged under the gauge fields that belong to the gravitini multiplets. This theory contains a supergravity multiplet coupled to eight vector multiplets with a duality group
\be
SO(6, 2; \mathbb{Z}) \times SL(2, \mathbb{Z}) \, .
\ee 
In the effective action for these fields we will further ignore the D-type terms. This is partially justified by the fact that the black hole horizon is supersymmetric and a large class of D-terms are known not to contribute to the Wald entropy as a consequence of this supersymmetry \cite{deWit:2010za}.
We will denote  the functional integral \eqref{qef} restricted to this reduced theory by $\widehat W(q, p)$ which is what we compute in the subsequent sections. We  find that  $\widehat W(q, p)$ itself agrees perfectly with \eqref{rademsp} for $d(q, p)$. This rather nontrivial agreement  can be regarded as  post-facto evidence that the reduced theory correctly captures the relevant physics.

\subsubsection{Functional Integral in $\mathcal{N}=2 $ Off-shell Supergravity \label{Functional}}
 
The renormalized action $S_{ren}(\phi)$ (\ref{Srenormalized}) has the same  functional form as the classical entropy function. In particular, its extrema $\phi = \phi_{*}$ correspond to the attractor values of the scalar fields and its value at the extremum
 $S_{ren}(\phi^{*})$ equals the Wald entropy for the local Lagrangian described with a prepotential $\CF$.  However, the  physics behind  the renormalized action  is completely different. Unlike the classical entropy function which is essentially a classical on-shell object, the renormalized action is a quantum object  obtained after a complicated holographic renormalization procedure using an off-shell localizing field configuration \eqref{HEKEsol}. Even though the scalar  fields in the localizing solution  asymptote to the attractor values at the boundary of the $AdS_{2}$, they 
have a nontrivial coordinate dependence  in the bulk  and they take the value $ X^{I}_{*} +  C^{I}$ at the center of $AdS_{2}$. In particular, they are excited away from their attractor values and  are no longer at the minimum  of $S_{ren}$.  Even though the scalar fields thus `climb up the potential' away from the minimum of the entropy function, the localizing solution remains Q-supersymmetric (in the Euclidean theory) because the  auxiliary fields $Y^{I}_{ij}$ get
excited appropriately to satisfy the Killing spinor equations. This is what enables us to integrate over $\phi$ for values  in field space far away from the on-shell values.

 The infinite dimensional functional integral \eqref{qef}  for the Wilson line in the reduced theory can thus be written as a finite integral (\ref{integral2})
 \begin{equation}\label{integral}
 \wh W (q, p) = \int_{\mathcal{M}_{Q}}   e^{ -\pi  \phi^{I} q_{I}}  \, e^{\mathcal{F}(\phi, p)}
 \,  \, |Z_{inst}|^{2} \, Z_{det}\,  [d\phi]_{\mu}
\end{equation}
The measure of integration $[d\phi]_{\mu}$ is computable from  the original measure $\mu$ of the functional integral of massless fields of string theory by standard collective coordinate methods. The factor $Z_{det}$ is the one-loop determinant of the quadratic fluctuation operator around the localizing instanton solution. Such one-loop determinant factors in closely related problems have been computed in \cite{Pestun:2007rz, Gomis:2009ir}. 
We have included $|Z_{inst}|^{2}$ to include possible contributions from brane instantons which is partially captured by the topological string for a class of branes. \\
Note that the exponential of the integrand is in the spirit of the conjecture by Ooguri, Strominger, and Vafa \cite{Ooguri:2004zv}. Our treatment differs from \cite{Ooguri:2004zv} in that the natural ensemble in our 
analysis is the microcanonical one. Moreover, we will be able determine the measure factor from first principles and the determine the subleading orbifolded localizing instantons that contribute to the functional integral. For earlier related work see \cite{Beasley:2006us, Denef:2007vg}.

To compute $\widehat W(q, p)$, it is necessary to evaluate all these factors explicitly, 
as well as to perform the finite dimensional integral over $\phi$. This is what we will do for our system in \S\ref{Macro}.  For the $\CN=2$ reduction of the $\CN=8$ theory that we consider, $n_{v}=7$ and the prepotential is given by
\be \label{ourprepot}
F(X) = -\frac{1}{2} \frac{X^{1}C_{ab}  X^{a} X^{b} }{X^{0}} \, , \qquad \qquad a, b = 2, \ldots, 7 \, .
\ee 
where $C_{ab}$ is the intersection matrix of the six 2-cycles of $T^{4}$. In  our normalization, it is given  by
\be
C_{ab} =  \left(\begin{array}{cc}0 & 1 \\1 & 0\end{array}\right) \otimes \textbf{1}_{3\times 3}
\ee
where $\textbf{1}_{3\times 3} $ is a $3 \times 3$ identity matrix. This prepotential describes the classical two-derivative supergravity action. Note that this does not depend the field $\hat A$ because there are no higher-derivative quantum corrections to the prepotential. 

\subsection{Integration measure}

The measure $[d\phi]_{\mu}$ is inherited from the standard measure on field space in the original functional integral. 
The collective coordinates $\{\phi^{I}\}$ of the localizing instanton solutions correspond to the values of the scalar fields $\{ X^{I} \}$  at the center of the $AdS_{2}$. The functional integration measure for the scalar fields is a pointwise product of  integration measure over the scalar manifold. The metric and hence the measure on the scalar manifold can be read off from the kinetic term of the scalar fields \cite{Mohaupt:2000mj, LopesCardoso:2000qm}.
The scalar kinetic action is
\begin{equation}\label{scalarKE}
8\pi  \mathcal{L}=  \sqrt{|g|} g^{\mu\nu}\left[ i(\partial_{\mu}F_I+i\mathcal{A}_{\mu}F_I)(\partial^{\mu}\bar{X}^I-i\mathcal{A}^{\mu}\bar{X}^I)+h.c. \right] \, , 
\end{equation}
where $\mathcal{A}_{\mu}$ is the gauge field for the $U(1)$ gauge symmetry of the off-shell supergravity 
theory. This field does not have a kinetic term and it is therefore determined by its equation of motion to be
%The equation of motion for $\mathcal{A}_{\mu}$ then follows from which we find
\begin{equation}
 \mathcal{A}_{\mu}^{*}=\frac{1}{2}\frac{\bar{F}_I\vec{\partial}_{\mu}X^I-\bar{X}^I\vec{\partial}_{\mu}F_I}{-i(\bar{F}_IX^I-F_I\bar{X}^I)} \, .
\end{equation}
The Lagrangian $ 8\pi  \mathcal{L}^*$ computed by substituting  $\mathcal{A}_{\mu}^{*}$ in \eqref{scalarKE} becomes
\begin{equation}\label{scalarKE1}
 - \sqrt{|g|} g^{\mu\nu}\left[N_{IJ}\partial_{\mu} X^I\partial_{\nu} \bar{X}^J - \frac{e^{-K}}{4}( K_{I}\partial_{\mu}X^I - \bar K_{ I}\partial_{\mu} \bar{X}^I) ( K_{I}\partial_{\nu} X^I -\bar K_{I}\partial_{\nu} \bar{X}^I) \right] \, ,
\end{equation}with
\begin{eqnarray}
\label{Ndef} N_{IJ}&:= & -i (F_{IJ} - \bar F_{IJ}) = 2\,\textrm{Im}(F_{IJ}) \, , \\
\label{Kdef} e^{-K}&:= & -i(X^I \bar{F}_I   - \bar{X}^I F_I) \, , \\
\label{Ldef} K_I &:= &  \frac{\partial K}{\partial X^{I}} =  i e^{K}\left( \bar F_I -  {F}_{IJ}\bar X^J \right)  .
\end{eqnarray}

The metric $g_{\mu \nu}$ is not the physical metric of Poincar\'e supergravity because it does not 
come with the canonical kinetic term.  It is related to the dilatation-invariant physical metric $G$ as 
\be\label{physmetric}
 G_{\mu\nu} =  e^{-K}g_{\mu\nu} \, ,
\ee
whose kinetic term is given by the standard Einstein-Hilbert action. We have 
\be\label{gGrel}
\sqrt{|g|} g^{\mu\nu} = e^{K} \sqrt{|G|} G^{\mu\nu} \, . 
\ee

It is natural to define the scalar functional integral measure using the physical metric $G_{\mu\nu}$. 
The measure can be determined by the metric induced by the inner product in field space:
\be
(\delta X, \delta X) = \int d^{4}x \,  \sqrt{|G|}\,  \delta X \, \delta X \, .
\ee
Substituting $X^{I} = (\phi^{I} + ip^{I} )/2$ in \eqref{scalarKE1}, and using  \eqref{physmetric}, \eqref{gGrel},
we obtain the induced metric on the localizing submanifold in the field space
\be\label{lineelem1}
d\Sigma^{2} =  M_{IJ} \, \delta \phi^{I} \delta \phi^{J} \, , 
\ee
with 
\be\label{Mmat1}
M_{IJ} = e^{K} \left[ N_{IJ} -\frac{e^{K}}{4} (K_{I} - \bar K_{I} )   (K_{J} - \bar K_{J} ) \right] \, .
\ee

It  is possible to write  the metric on the localizing manifold entirely
in terms of the K\"ahler potential\footnote{Upon gauge-fixing, on the space of projective coordinates  
$K_{IJ}$ becomes the K\"ahler metric.  We will refer to  $K$ as the K\"ahler potential even though we do not fix any gauge here.}
 $K$ \eqref{Kdef}. 
It is easy to check that
\begin{eqnarray} \label{LNiden}
N_{IJ}& = & \frac{\partial^{2} e^{-K}}{\partial X^{I} \bar X^{J}}=  
e^{-K}\left( \frac{\partial^{2} K}{\partial X^{I} \partial \bar X^{J}} -  \frac{\partial K}{\partial X^{I}}  \frac{\partial K}{\partial \bar X^{J}} \right) \, . 
\end{eqnarray}
Defining the metric $K_{IJ}$ in terms of the K\"ahler potential in the usual way 
\be\label{kahmet}
K_{I J} := \frac{\partial^{2 } K}{\partial X^{I} \partial \bar X^{J}} \, , 
\ee
and using \eqref{LNiden}, we can write the Lagrangian \eqref{scalarKE1} entirely in terms of 
the K\"ahler potential:
\begin{equation} \label{scalarKE2}
 8\pi  \mathcal{L}= - \sqrt{|g|} g^{\mu\nu}e^{-K}\left[ K_{IJ}\partial_{\mu} X^I\partial_{ \nu} \bar{X}^J - \frac{1}{4} \partial_{\mu}K \partial_{\nu} K \right] \, .
\end{equation}

Substituting  $X^{I} = (\phi^{I} + ip^{I} )/2$ in \eqref{scalarKE2}, we can rewrite the moduli space metric 
\eqref{lineelem1} as 
\be\label{Mmat2}
M_{IJ } = K_{IJ} -\frac{1}{4} \frac{\partial K}{\partial \phi^{I}} \frac{\partial K}{\partial \phi^{J}} \, . 
\ee
Since the metric $K_{IJ}$ given in  terms of the K\"ahler potential \eqref{kahmet}, this expresses the moduli space metric $M_{IJ}$ entirely in terms of the K\"ahler potential. 
The measure on the localizing manifold is simply the measure induced by this metric and is given by 
\be
\prod_{I =0}^{n_{v}} d\phi^{I} \sqrt{\det (M)} \, .
\ee

\subsection{Macroscopic Quantum Partition Function \label{Macro}}

The two-derivative action of $\CN=8$ is invariant under the continuous duality group $E_{7, 7}(\mathbb{R})$. We therefore expect to be able to write the macroscopic answer in terms of  $\Delta$  which  is the unique quartic invariant of  $E_{7, 7}(\mathbb{R})$. For this purpose, we will first write the renormalized action in new variables   so that it depends only on the invariant $\Delta$ and then work out the measure in the same variables to obtain a manifestly duality invariant expression for the Wilson line.

\subsubsection{Renormalized action and duality invariant variables \label{renaction} }

As discussed in \S\ref{Config} the electric and magnetic charge vectors  $Q$ and $P$ respectively  are related to the charges in the Type-IIA frame \eqref{chargesA} by
\bea
Q  =  ( q_{0}, -p^{1}; q_{a}) \, \qquad
P  = (q_{1}, p^{0}; p^{a} ) \quad .
\eea
The inner product is defined for example by
\be 
P \cdot P = 2  \, q^{1}  p^{0} +  p^{a} \, C_{ab} \, p^{b} \, ,
\ee
The charge configuration \eqref{chargesF} has only five nonzero charges $q_{0} = n $, $q_{1} = l $, $p^{1} = -w$, and $p^{2}$, $p^{3}$.  Hence, the three T-dualiy invariants  all have nonzero values given by
\be\label{tinvt}
Q^{2 } = 2 \, n  w \, , \quad P^{2} = 2 \, p^{2}  p^{3}\, , \quad Q \cdot P = w \, l  \, .
\ee

The natural variables to start with are the projective coordinates
\begin{equation}\label{para1}
S := X^{1}/X^{0} \, , \quad   \quad T^{a} := X^{a}/X^{0} \quad a = 2, \ldots, n_{v} \, ,
\end{equation}
with real and imaginary parts defined by
\be\label{para2}
S:=  a  + is \, , \quad T^{a} :=  t^{a} + i r^{a} \, .
\ee
For our localizing instanton solutions we obtain
\bea
a = \phi^{1}/\phi^{0} \, ,&\quad& s = -w/\phi^{0} \\
t^{a} = \phi^{a}/\phi^{0} \, , &\quad& r^{a} = p^{a}/\phi^{0}  \, . 
\eea
The renormalized action \eqref{Srenfinal} for this charge configuration and prepotential \eqref{ourprepot} is 
\be\label{Srenour}
S_{ren} = -\frac{\pi}{2\phi^{0}} \left[ -w (\phi^{2} - P^{2}) + 2 \, \phi^{1} (\phi \cdot P) \right]
-\pi  n  \phi^{0} - \pi  l  \phi^{1} \, ,
\ee
where $\phi^{2}= \phi^{a} \, C_{ab} \, \phi^{b}$ and $\phi \cdot P = \phi^{a} \, C_{ab} \, P^{b}$.
Using the parametrization \eqref{para1} and \eqref{para2} and the T-duality invariants \eqref{tinvt} it can be written as
\be\label{Sren1}
S_{ren} = \frac{\pi}{2} \left[P^{2} s + \frac{Q^{2}}{s} + \frac{2 \, Q\cdot P \, a}{s}\right] 
-\frac{\pi w^{2} \, t^{2}}{2s} + \frac{\pi aw \, t \cdot P}{s}  \, \, .
\ee

Our  next goal will be to define integration variables to write the action entirely in terms of the U-duality invariant  $\Delta$.
Since the action is quadratic in the $t^{a}$ variables, it is useful to complete the squares by defining
\be
\tau^{a} = \frac{w}{\sqrt{s}} \left( t^{a} - \frac{a \, p^{a}}{w}  \right)
\ee 
so that 
\be\label{Sren3}
S_{ren} = \frac{\pi}{2} \left[P^{2} s + \frac{Q^{2}}{s} + \frac{P^{2} \, a^{2}}{s} + \frac{2 \, Q\cdot P \, a}{s}\right] 
-\frac{\pi \, \tau^{2}}{2}  \, \, .
\ee
Note that the parenthesis  is a manifestly S-duality invariant combination which  is quadratic in the axion variable $a$. So we complete the square again by defining
\be\label{defbig}
\sigma = \frac{\pi P^{2 }s}{2} \, , \quad 
\alpha = \frac{1}{\sqrt{\sigma}} \left( P^{2}a  + Q\cdot P \right)
\ee
The renormalized action then becomes
\be
S_{ren} =  \left( \sigma + \frac{z^{2}}{4\sigma} \right) -\frac{\pi \, \tau^{2}}{2}  + \frac{\pi \, \alpha^{2}}{2}\, .
\ee
with 
\begin{equation}
z^{2}= \pi^{2} (Q^{2} P^{2} - (Q.P)^{2}) \, = \pi^{2} \Delta \,  . 
\end{equation}
The variables $(\s, \a, \tau^{a})$ can be regarded as the duality invariant variables. We now turn to the integration measure.

\subsubsection{Conformal compensator, Gauge-fixing, and Analytic Continuation \label{confcomp}}

The constants $C^{I}$ which characterize the localizing instanton solution \eqref{HEKEsol} are all real. Hence, the contour of integration for the variables $s$ and $t$  would appear to be along the real axis. The quadratic terms in $t$ in the  action  \eqref{Sren3} would lead to divergent Gaussian integrals. We will see below that this is nothing but the divergence of Euclidean quantum gravity arising from the integration over the conformal factor that has a wrong sign kinetic term. 

We recall that the scalar kinetic term \eqref{scalarKE2} can be written as 
\be\label{omegaaction}
-\sqrt{-g} g^{\mu\nu}  \left[ e^{-K}K_{IJ} \partial_{\m} X^{I} \partial_{\n} \bar X^{J} - \frac{1}{4}e^{-K}\partial_{\mu} K \partial_{\nu}K  \right]\, .
\ee 
The kinetic term  for the spacetime metric $g_{\mu\nu}$ is of the form\footnote{We  suppress  an overall 
factor of $1/8\pi$ that is irrelevant for the discussion here but is important for the normalization of 
the renormalized action in \S\ref{Macro}.}
\be
 -\frac{1}{6}\sqrt{-g}  e^{-K} R_{g}\, ,
\ee
We can thus identify $e^{-K/2}$ as a conformal compensator $\Omega$ which is often used to extend the gauge principle  to include scale invariance in addition to diffeomorphism invariance. The Einstein-Hilbert action is then replaced by 
\be
 \sqrt{-g}\left[ -\frac{1}{6}  \Omega^{2} \, R_{g}  - g^{\mu\nu }\, \partial_{\mu} \Omega \,\partial_{\nu}\Omega \right] \, ,
\ee
which is  now invariant under both diffeomorphisms and Weyl rescalings. As can be seen from \eqref{omegaaction},  the kinetic term for $\Omega$ has a wrong sign compared to a physical scalar, as is usual for the 
conformal compensator field.  
In D-gauge \cite{Mohaupt:2000mj}
$\Omega$ is gauge-fixed to a constant and one recovers the Einstein-Hilbert action. Our localizing solution is however in a different gauge in which the volume of $AdS_{2}$ in the metric $g$ is gauge- fixed and hence $\Omega$ is effectively a fluctuating field. 
This also explains why we have $n_{v} +1 $ scalar moduli $\{\phi^{I} \}$ even though there are only $n_{v}$  physical scalars. Essentially, our choice of gauge  enables us to borrow the conformal factor $\Omega$ as an additional scalar degree of freedom. The advantage  is  that the symplectic symmetry acts linearly on the fields $\{ \phi^{I} \}$. 

Since the  kinetic term for conformal compensator  $\Omega$ has a wrong sign, to make the Euclidean functional integral well defined, it is necessary to analytically continue the contour of integration in field space   \cite{Gibbons:1976ue}. 
For our prepotential \eqref{ourprepot}, the K\"ahler potential is given by
\be \label{KahlerpotST}
\exp[-K]  = 4 \, | X^{0}|^{2}\,  \Im (S) \, C_{ab} \, \Im (T^{a}) \, \Im (T^{b}) \, .
\ee
For $S$ and $T^{a}$ fixed, we see that $\Omega$ is proportional to $X^{0}$ up to a  phase that can gauge-fixed by using the additional  $U(1)$ gauge symmetry. Thus, the analytic continuation in the $\Omega$ space can be achieved by analytically continuing in the $X^{0}$ space. For the localizing solution,  $X^{0} =\phi^{0}$. 
Thus,  analytic continuation in $\Omega$ space can be achieved by analytically continuing  in  the  $\phi^{0}$ space. Correspondingly, we take the contour of integration of $\phi^{0}$  or equivalently of  $\s$  along the imaginary axis rather than along the real axis\footnote{In general there can subtleties  in such analytic continuation, see for example  \cite{Harlow:2011ny}. These  will  not be important  in the present context.}.

A familar example of such an analytic continuation is the functional integral for the worldsheet metric in first-quantized string theory. The conformal factor of the metric  is the Liouville mode which can be thought of as a conformal compensator. Critical bosonic string with $c=26$  can be regarded as a noncritical string theory with $c=25$ coupled to this Liouville mode. The Liouville mode  plays the role of  time coordinate in target space \cite{Das:1988ds} and has a wrong-sign kinetic term on the worldsheet. The corresponding functional integral then has to be defined by a similar analytic continuation \cite{Polchinski:1998rq}.

\subsubsection{Evaluation of the Localized Integral \label{Evalu}}

The localizing action $QV$ with abelian gauge fields is purely quadratic. Hence, the quadratic fluctuation operator around the localizing instantons does not depend on the collective coordinates $\{ C^{I}\}$. As a result, $Z_{det}$ is independent of $\{\phi^{I}\}$ and charges can be absorbed in the overall normalization constant.  Another simplification for the $\CN=8$ theory is that $|Z_{inst}|^{2} =1$ because the classical prepotential \eqref{ourprepot} that we have used is quantum exact. 

Thus, all that remains is to  compute the determinant of the matrix $M_{IJ}$  introduced in \eqref{Mmat1}. 
Since there are no terms that depend on $\hat A$ for our prepotential, it is homogenous of degree $2$ 
in the variables $X$. As a result, $F_{IJ}X^{J} = F_{I}$, and it follows from \eqref{Ldef} that
\be
K_{I} =   e^{K}N_{IJ} \bar X^{J} \, , \qquad  \bar K_{I} =  e^{K} N_{IJ}  X^{J} \, . 
\ee 
This allows us to write \eqref{Mmat1} as
\be \label{M1}
M_{IJ} = e^{K} \left( N_{IJ} + \frac{1}{4} e^{K} N_{IK} \, p^{K}  N_{JL} \, p^{L}\right) \, .
\ee
We have 
\be
\det (M) =  \exp \left[\frac{(n_{v} + 1 )}{2}K \right] \det (N) \det (1 + \Lambda) \, , 
\ee
where the matrix $\Lambda$ is defined by
\be
\Lambda^{I}_{J} = \frac{1}{4} e^{K} \, p^{I} N_{JL} \, p^{L}  \, . 
\ee
Some elements of this measure such as the  matrix $N_{IJ}$ were anticipated in the work of \cite{LopesCardoso:2006bg, Cardoso:2008fr, Cardoso:2010gc} based on considerations of sympletic invariance. Our derivation follows from the analysis of the induced metric on the localizing manifold and has additional terms depending on $K_{I}$ and $\exp (K)$ which are also sympletic invariant.  Unlike in the $\CN=4$ theory,  in the $\CN=8$ theory  the higher-derivative corrections are zero,  and do not provide  a useful guide for the  determination of nonholomorphic terms of the measure such as the powers of $\exp(K)$. 

It is easy to see that for our system $\textrm{Tr} (\Lambda^{n}) = \lambda^{n}$ where $\lambda$ is a numerical constant independent of charges. As a result, 
\be
\det ( 1+ \Lambda) = \exp (\textrm{Tr} \log ( 1 + \Lambda) ) = \exp ( \log ( 1 + \lambda) ) 
\ee
is a field-independent and charge-independent numerical constant. In what follows, we will ignore 
all such numerical constants in the evaluation of the measure and  determine the overall normalization 
of the functional integral in the end.  

Hence,  up to a constant,   $\det(M)$ is determined by  $\det (N)$ and $\exp (K)$.
For our prepotential,   evaluating on the  localizing instanton solution we obtain
\be
\exp[-K]  = 4 \, P^{2} s = 8 \s/\pi
\ee
which is manifestly duality invariant.
Similarly, 
\be
\det (N )  = \frac{s^{n_{v}-3}  \det (C_{ab})}{4 |X^{0}|^{4}} e^{-2K}  
= s^{n_{v}+3} \,  \big(\frac{P^{2}}{w^{2}} \big)^{2}
\ee
as can be checked using Mathematica. 
In terms of the duality invariant variables defined earlier, we see that  the measure is given by 
\be
\prod_{I=0}^{n_{v}} d\phi^{a} \,\sqrt{\det (N)} = \frac{1}{\sqrt{\s}} \, d\s \, d\a \, \prod_{2}^{n_{v}} d\tau^{a}
\ee
up to an overall constant that is independent of charges and fields. 
The total  measure is thus given by
\be\label{measure1}
\prod_{I=0}^{n_{v}} d\phi^{I} \, \sqrt{\det{(M)}} =  \frac{d\s}{\s^{\rho+ 1}} d\alpha \prod_{2}^{n_{v}} d\tau^{a}\, 
\ee
with $\rho = n_{v}/2$. 
Our total integral is hence manifestly duality invariant.

Performing the Gaussian integrals over $\a$ and $\tau^{a}$ we obtain
\be\label{intrep2}
\int \frac{d\s}{\s^{\rho + 1}} \exp \left( \sigma + \frac{z^{2}}{4\sigma} \, \right) 
\ee
 which  gives exactly the integral representation of  the  Bessel function $\wt I_{7/2}(z)$ for $n_{v} =7$.
The overall numerical  normalization needs to be fixed by hand but once it is fixed for one value of $\Delta$, one  obtains a nontrivial a function for all other values of $\Delta$ given by 
\be
W_{1}(\Delta) =  \sqrt{2} \, \pi \, \big(\frac{\pi}{\Delta} \big)^{7/2} \, I_{7/2}(\pi \sqrt{\Delta}) \, .
\ee
This macroscopic calculation thus precisely reproduces the first term with $c=1$ in \eqref{Wexp} and matches beautifully with  the first term in \eqref{dexp} from the Rademacher expansion  \eqref{rademsp} for  of the microscopic degeneracy $d(\Delta)$.

For large $z$, the Bessel function has an expansion
\begin{equation}\label{Cardy}
I_{\rho} (z) \sim  \frac{e^{z}}{\sqrt{2\pi z}} \left[ 1- \frac{(\mu -1)}{8z} + \frac{(\mu -1)(\mu -3^{2})}{2! (8z)^{3} }- \frac{(\mu -1)(\mu -3^{2}) (\mu -5^{2})}{3! (8z)^{5} } + \ldots \right] \, ,
\end{equation}
with $\mu = 4\rho^{2}$.  The exponential term $\exp (\pi \sqrt{\Delta})$ gives the Cardy formula and $\pi \sqrt{\Delta}$ can be identified with the Wald entropy of the black hole. Higher terms in the series give power-law suppressed finite size corrections to the Wald entropy.  This is however not  a convergent expansion but only an asymptotic expansion. This means that for any given $z$ only the first few terms of order some power of $z$ are useful for making an accurate estimate. Beyond a certain number of terms, including more terms actually makes the estimate worse rather than improve it.  For larger and larger $z$ one can include more or more terms and one obtains better and better approximation but it is never convergent.

It should be emphasized that  our computation of $W_{1}(\Delta)$ gives an exact integral representation \eqref{intrep2} of the Bessel function $ I_{7/2}(z)$  and not merely the asymptotic expansion \eqref{Cardy}.  This is made possible because localization gives an exact evaluation of the functional integrals and allows one to access large regions in the field space far away from the classical saddle point of the entropy function used to derive the Cardy formula.

It is instructive to compare  the integers $d(\Delta)$ with  the $W_{1}(\Delta)$  and the exponential of the Wald entropy \eqref{tablefcoeffs2} we tabulate these coefficients  for the first few values  of $\Delta$. 
\begin{table}[h]   \caption{\small{Comparison of the microscopic degeneracy $d(\Delta)$ with the 
functional integral $W_{1}(\Delta)$ and the exponential of the Wald entropy. 
The last three rows in the table equal each other 
asymptotically.}}   \vspace{8pt}    \centering
   \begin{tabular}{c|cccccccccccccc}   \hline   
       $\Delta$ &  -1 & 0 &3& 4& 7 & 8 &11& 12& 15 \\
       \hline \hline $d(\Delta)$ & 1 & $2$ &8&$12$&39&$56$&152&$208$&513\\  
       \hline $W_{1}(\Delta)$ & 1.040 & $1.855$ &7.972 &$12.201$&38.986&$55.721$&152.041&$208.455$&512.958\\  
        \hline $\exp(\pi\sqrt{\Delta})$ & -  & $1$ &230.765 &$535.492$&4071.93&$7228.35$&33506&$53252$&192401\\   
   \hline    \end{tabular}   \label{tablefcoeffs2}   \end{table}
   
Note that the area of the horizon goes as $4\pi \sqrt{\Delta}$ in Planck units. Already for $\Delta =12$ this  area would be much larger than one, and one might expect that  the Bekenstein-Hawking-Wald entropy would be a good approximation to the logarithm of the quantum degeneracy. However, we see from the table  that these two differ quite substantially.   Indeed, in this example, since there are no higher-derivative local  terms, Wald entropy equals the Bekenstein-Hawking entropy.  The discrepancy thus arises entirely from the quantum contributions from integrating over massless fields.   Localization  enables an exact evaluation of these quantum effects. The resulting  $W_{1}(\Delta)$  is in spectacular agreement with $d(\Delta)$ and in fact comes very close to the actual integer even for small values of $\Delta$.  

We see from the asymptotic expansion \eqref{Cardy} that the subleading logarithmic correction to the Bekenstein-Hawking entropy goes as $ -2 \log ({\Delta})$. This in agreement with the results in  \cite{Banerjee:2010qc, Sen:2011ba, Banerjee:2011jp} where the logarithmic correction was computed by evaluating  one-loop determinants of various massless fields around the classical background.  Using localization, this logarithmic correction follows essentially from the analysis of the induced measure on the localizing manifold without the need for any laborious evaluation of one-loop determinants. Moreover, since   localization accesses regions in field space very off-shell from the classical background  the entire series  of power-law suppressed terms in \eqref{Cardy} follow with equal ease.

\subsubsection{Nonperturbative Corrections, Orbifolds, and Localization\label{Nonpert}}

We have seen that localization
correctly reproduces the first term in the Rademacher expansion. This term already captures all power-law  and logarithmic corrections to the leading Bekenstein-Hawking-Wald entropy exactly to all orders.  
We turn next to the  computation of the higher terms in the Rademacher expansion \eqref{rademsp} with $c >1$. These terms are nonperturbative because they are exponentially suppressed with respect to the terms in  \eqref{Cardy}. 

It was proposed in \cite{Banerjee:2008ky,Sen:2009vz, Murthy:2009dq, Banerjee:2009af} that such non-perturbative corrections could arise from  $\mathbb{Z}_{c}$ orbifolds for all positive integers $c$ because such orbifolds respect the  same boundary conditions  \eqref{asympcond} on the fields.  In general, it is  difficult to justify keeping such subleading exponentials if the  power-law suppressed terms are evaluated only in  an asymptotic expansion. However,  localization gives an exact integral representation of the leading Bessel function in \S\ref{Evalu}.  The power-law suppressed contributions are computed exactly, and it is justified to  systematically take into account the exponentially suppressed contributions.

The  $\IZ_{c}$ orbifold configurations that contribute to the localization integral are obtained as follows. We mod out with a symmetry $R_{c}T_{c}$ which   combines a supersymmetric order $c$ twist $R_{c}$ on  $AdS_{2}\times S^{2}$ with an order $c$  shift $T_{c}$ along the $T^{6}$. The orbifold twist  is required to be supersymmetric  because  to preserve the $Q$ supercharge  used for localization, the orbifold action must commute with $L -J$ \cite{Banerjee:2009af}. At the center of  $AdS_{2}$ and at the poles of $S^{2}$ the twist looks like a generator of the supersymmetric $C^{2}/\IZ_{c}$ orbifold.
With an appropriate shift, this action is freely acting and can be used to get smooth solutions \cite{Murthy:2009dq}.  

To illustrate how this works together with localization let us first discuss the case when  $T_{c}(\delta) $ is a simple shift  of $2\pi \delta/c$ along the circle $S^{1}$. It  acts on the momentum modes by
\be\label{phase}
T_{c}(\delta) \, |{m}\rangle = e^{\frac{2\pi i \delta m}{c}} \, | m \rangle \, .
\ee
Let $\phi$ be the azimuthal angle along the $S^{2}$ and $ y$ be the coordinate of the circle $S^{1}$ with $2\pi$ periodicities.  We will denote the orbifolded coordinates with a tilde. The orbifold operaton $R_{c}T_{c}$ identifies points in $AdS_2\times S^2\times S^1$ with the  identification
\begin{equation}\label{identify}
 (\wt \theta, \wt \phi \, , \wt y)\equiv (\wt \theta+\frac{2\pi}{c} \,, \wt  \phi-\frac{2\pi}{c}\, , \wt y+\frac{2\pi \delta}{c})
 \end{equation}
The combined action $R_{c} T_{c} (\delta)$ means that as we go around the boundary of $AdS_{2}$ the momentum modes pick up a phase as in \eqref{phase}. This corresponds to turning on a Wilson line of the Kaluza-Klein gauge field
$\mathcal{A}$  that couples to the  momentum $n$ by modifying  the gauge field as
\be\label{orbA}
\cA =  -i e_{*} (\wt r-1) d \wt\theta  + \delta \, d \wt \theta\
\ee
The  metric on the orbifolded   $AdS_{2}$  factor  has the same form
\be
ds^{2} =  v_{* }\left[  (\wt r^{2} -1)  d\wt \theta^{2}  + \frac{d\wt r^{2}}{(\wt r^{2} -1)} \right] \qquad 1 \leq \wt r < \wt r_{0}; \, \qquad 0 \leq \wt \theta < \frac{2\pi }{c}
\ee 
as the original unorbifolded metric \eqref{metric} but the $\wt \theta$ variable now has a different periodicity and we have cutoff at $\wt r = \wt r_{0}$.   Thus, it is not immediately obvious that asymptotic conditions on the fields are the same  as for the unorbifolded theory. 
To see this, we change coordinates
\begin{equation}
\wt  \theta=\frac{ {\theta}}{c}\, ,\quad \wt \phi= {\phi}-\frac{ {\theta}}{c}\, ,\quad \wt y= {y}+\frac{ {\theta}}{c} \, , \quad \wt r = c r , 
\end{equation}
so that in the  new coordinates, the fields have the same asymptotics \eqref{asympcond}  as before:
\be\label{asymorb}
d s_2^2  \sim  v_{*}  \left[ r^2 d \theta^2 + \frac{d r^2}{r^2} \right] , \qquad 
\cA \sim  -i e_{*} r d \theta\ \, . 
\ee 
Moreover, the new coordinates have the same identification
\begin{equation}
  ( {\theta}, {\phi}, {y})\equiv ( {\theta}+2\pi, {\phi}, {y})\equiv ( {\theta}, {\phi}+2\pi, {y})\equiv ( {\theta}, {\phi}, {y}+2\pi)
\end{equation}
as in the unorbifolded theory.  Such orbifolded field configurations with the same asymptotic behavior will therefore contibute to the functional integral. 

The  orbifold action is freely acting if $\delta$ and $c$ are relatively prime.  Therefore, the localizing equations, which   are local differential equations, remain the same as before and one obtains the same localizing instantons  \eqref{HEKEsol} as before. 
To compute the renormalized action it is convenient to use the tilde coordinates. 
If we  put a cutoff at $r_{0}$, the  range of $r$ is $1/c \leq r \leq r_{0}$ and that of $\wt r$ is   $ 1 \leq \wt r \leq c r_{0}$ . 
 The physical action  is an integral of the same local Lagrangian density as the unorbifolded theory but now the ranges of integration are different.  Since the localizing instantons do not depend on the angular coordinates,  the nontrivial integration is over  the coordinate $\wt r$.  The $r_{0}$ dependent contribution from this integral is therefore $c$ times larger than before but the $r_{0}$ independent constant piece is the same as before.  On the other hand, from the angular integrations  one gets an overall factor of $1/c$ because the range of these coordinates is divided by $c$ by the identification \eqref{identify}.  Altogether, the renormalized action obtained by removing the $r_{0}$ dependent divergence is  smaller by a factor of $c$. Moreover, with the modified gauge field \eqref{orbA} the Wilson line contributes an additional  phase. 
In summary, instead of \eqref{Sren} we obtain 
\be\label{newsolact}
 \exp\left[ \frac{\CS_{ren}(\phi)}{c} +  \,\frac{2 \pi i\,  n \delta}{c} \right] \ ,
\ee
where  $\CS_{ren}$ is the unorbifolded renormalized action for the localizing instantons  given by \eqref{Srenour}. 

Since the phase above does not depend on $\phi$ we can integrate over $\phi$ as before  and then sum over all phases.  Thus $W_{c}$ factorizes  as
\be
W_{c} (\Delta) = A_{c} (\Delta) B_{c} (\Delta)
\ee
where $A_{c}$  comes from integration over $\phi$ and $B_{c}$  comes from the sum over phases. Since the renormalized action  is now  smaller by a factor of $c$, it is easy to see that the integral $A_{c}$  gives precisely the modified Bessel function but with an argument $z_{c } = z/c$ with possible powers of $c$ coming from the measure which we absorb for now in $B_{c}(\Delta)$.
The final answer thus has the form
\be
W_{c}(\Delta) =  \sqrt{2} \, \pi \, \big(\frac{\pi}{\Delta} \big)^{7/2} \,  
I_{\frac{7}{2}} \big(\frac{\pi \sqrt{\Delta}}{c} \big) \, B_{c}(\Delta) \, .
\ee
This is very close to the $c$-th term in the Rademacher expansion. To obtain agreement we would need to show
\be\label{desired}
B_{c } (\Delta) =  c^{-9/2}K_{c}(\Delta)  \, .
\ee
We see from  \eqref{kloos} that the Kloosterman sum is also a rather intricate sum over various $c$ and $\Delta$ dependent phases. This suggests that by summing over the phases for various allowed orbifolds   and properly  fixing their relative normalization with respect to the $c=1$ term,  it may be possible to compute $B_{c}(\Delta)$ to reproduce the desired expression \eqref{desired} in terms of the Kloosterman sum \cite{Dabholkar:2012}.

\section{Macroscopic index and $AdS_3/CFT_2$ correspondence}

In this section we use an intermediate approach based on $AdS_3$ rather than $AdS_2$ to compute finite charge corrections to the black hole entropy. The computation is exact in the limit when only one of the charges is taken to be very large keeping the other charges finite. The answer is particularly sensitive to the details of the phase we are working on and therefore we can learn about microscopic details of the theory. In the process we find easier to construct a macroscopic index which captures all the degrees of freedom from horizon till asymptotic infinity. In the limit considered it matches precisely with a microscopic computation. We state our main result as:\emph{ in the limit when only one of the charges is taken to be very large, the asymptotic growth of the macroscopic index which has the form of a Cardy formula is controlled by an effective central charge which is related to the coefficients of the Chern-Simons terms computed at asymptotic infinity.}

In section \S3, we explained that the index computed
at asymptotic infinity captures not only the horizon degrees of freedom but also
any exterior contribution that sits between $AdS_2$ and the asymptotic infinity. The total index is a
combination of an index for the horizon and other for the hair degrees of
freedom. A construction of such an index from the gravity side is the subject of this section.

The idea is to construct the index starting from the horizon and going gradually
to asymptotic infinity. As explained in the previous sections, to count the
horizon degrees of freedom, we should
perform a path integral over the string fields in $AdS_2$ with specific boundary
conditions. In general this is a very difficult problem even though we showed in the previous sections that localization allows to go very far. By including the contribution from the hair modes we
will find easier to compute the index instead of the degeneracy.
     
For extremal black holes with a $AdS_2 \times S^1$ factor in the near horizon
geometry, we can use the power of $AdS_3$ to study subleading corrections to the Beckenstein-Hawking entropy \cite{Dabholkar:2010rm}.

The near horizon geometry $AdS_2 \times S^1$ has isometry group $SL(2,\mbbr)\times U(1)$ while $AdS_3$ has $SO(2,2)=SL(2,\mbbr)_L\times SL(2,\mbbr)_R$. Once we make the radius of $S^1$ larger and larger we can
restore the $SO(2,2)$ isometry and view the solution as a BTZ black hole
living in $AdS_3$. Using the $AdS_3/CFT_2$ correspondence we compute
the entropy using a Cardy formula. The central charges of the dual
$CFT_2$ can be computed exactly by knowing the coefficients of the Chern-Simons
terms on $AdS_3$ \cite{Kraus:2005vz}.

This section is organized as follows. In section \S6.1 we explain the criteria
for which the index equals degeneracy in the large charge limit and how we can use the power of $AdS_3$  to compute the quantum corrected
entropy. In \S6.2 we consider a puzzle of M5-branes on $K3\times T^2$ and its relation to four dimensional black holes. In \S6.3 we consider the contribution of exterior modes to the asymptotics of the macroscopic index. In \S6.5 we derive the asymptotic behaviour of the index from microscopics. We end presenting various results for five dimensional black holes.

\subsection{From $AdS_2$ to $AdS_3$}

To determine $B_6$ by computing first $B_{hor}$ and then $B_{hair}$ as defined in formula (\ref{hor+hair index}) is a very difficult task. First it requires evaluating a string path integral on $AdS_2$ and second it requires determining the hair modes which is not an easy task \cite{Banerjee:2009uk,Jatkar:2009yd}. For these reasons we shall give an
alternative approach based on $AdS_3$ rather than on $AdS_2$. The main advantage, as we will see, resides on the fact that it is possible to determine exactly the central charges by just knowing the Chern-Simons terms in the bulk of $AdS_3$. Further we will see that the hair analyses simplifies once we combine them with the bulk contribution.

Consider a black hole whose near horizon geometry contains a factor
$AdS_2\times S^1$ with metric
\begin{equation}
 ds^2=v\left(-r^2dt^2+\frac{dr^2}{r^2}\right)+\left(dy-\frac{r}{R}
 dt\right)^2
\end{equation}The coordinate $y$ corresponds to the circle and $R$ is its radius. The fiber $r/Rdt$ corresponds, from the two dimensional point of view, to a charged electric gauge field. In particular if $v=1/R^2$ the space is locally $AdS_3$.

 The space $AdS_2\times S^1$ has isometry group $SL(2,\mbbr)\times U(1)$. The first factor correponds to $AdS_2$ isometries while the $U(1)$ corresponds to translations along the circle.  It differs from global $AdS_3$ due to a translation identification of $2\pi$ along the non-compact $y$ direction. If we make the radius $R$ very large we can restore the $SL(2,\mbbr)_L\times SL(2,\mbbr)_R$ of $AdS_3$. At the same time we take the
asymptotic value of $R$ to be very large keeping all the other moduli fixed to ensure that the space develops an intermediate $AdS_3$ factor. Now we can regard the black hole solution as an extremal BTZ black hole living in this intermediate $AdS_3$ space time \cite{deBoer:2008fk,Strominger:1998yg}. For an extremal BTZ black hole, in the limit of very large spin $J$, we can use a Cardy formula to compute the entropy 
\cite{Saida:1999ec,Kraus:2005vz,Kraus:2006wn},
\begin{equation}
 d(n\gg1)\approx e^{2\pi\sqrt{nc_L/6}}
\end{equation} From the $AdS_2$ point of view,
the spin $J$ is identified with the momentum $n$ along the circle $S^1$ which justifies the limit $n\rightarrow \infty$, that is the Cardy limit.

Since we expect the Cardy formula to hold in the full quantum theory we should take
this as the quantum generalization of the black hole entropy. The problem is reduced to the computation of the central charges. Since we don't know the
details of the dual $CFT_2$, we should be able to compute them from the bulk
theory after including quantum corrections. 

There are however some subtleties in this approach. The Cardy formula should
count the total degeneracy without caring if it is a single or multi center
solution in $AdS_3$. It should include not just the horizon degrees of freedom
but also any possible contribution of  additional modes exterior to $AdS_2$ but living inside
$AdS_3$. This should not be a problem because at the end we want to
compare the bulk answer with a microscopic index which already includes all those
contributions. Moreover we neglect the possibility of multiple $AdS_3$ throat
\cite{deBoer:2008fk} by working in an appropriate domain in the moduli space.

One additional problem we face is related to the fact that the $CFT_2$ does not capture all the degrees of freedom of the black hole. There could be additional modes living on the
boundary, like in the case of the $U(1)$ factor in $AdS_5$ \cite{Bilal:1999ty}, or between $AdS_3$
and asymptotic infinity. We shall call them exterior modes.

Since we take the asymptotic value of the radius to infinity while keeping its
attractor value large but fixed the physical momentum, measured at infinity, vanishes. This restores  the $1+1$ Lorentz symmetry of part of the solution lying
between $AdS_3$ and asymptotic infinity. Therefore we expect the dynamics of the 
exterior modes to be described by a two dimensional field theory. In addition we have to combine this contribution with the Cardy formula to recover the total index.

We repeat the computation (\ref{hor+hair index}) but now rewriting the index in
terms of the bulk $AdS_3$ and exterior degrees of freedom,
\begin{eqnarray}
 B_6&=&-\frac{1}{6!}\textrm{Tr}(-1)^{2h_{bulk}+2h_{exterior}}(2h_{bulk}+2h_{
exterior} )^6\\
&=&-\frac{1}{6!}\textrm{Tr}(-1)^{2h_{bulk}}\textrm{Tr}(-1)^{2h_{exterior}}(2h_{
exterior }
)^6\\
&=&\sum_{q+\tilde{q}=Q}B_{bulk}(q)B_{6\,exterior}(\tilde{q})
\end{eqnarray}where we defined $B_{bulk}=\textrm{Tr}(-1)^{2h_{bulk}}$ and $B_{6\,exterior}=-\frac{1}{6!}\textrm{Tr}(-1)^{2h_{exterior}}(2h_{exterior})^6$. We have assumed that the black hole when regarded as a solution in $AdS_3$ doesn't break any further supersymmetry. Under this
assumption the fermionic zero modes are all part of the exterior modes which 
implies that only the term $(2h_{exterior})^6$ will survive in the binomial
expansion of $(2h_{bulk}+2h_{exterior})^6$. 

We should note that in the BTZ near horizon geometry $AdS_2\times S^1$ the time circle is contractible in the full $AdS_3$ geometry while the $S^1$ circle is not. On the boundary theory we should compute a partition function with anti periodic conditions along the thermal circle and periodic
along $S^1$. In other words, the holographic correspodence instructs us to compute $\textrm{Tr}(1)$
with Ramond-Ramond boundary conditions instead of $\textrm{Tr}(-1)^F$ as presupposed by
$B_{bulk}$. Instead we will try to argue that the asymptotics of left moving excitations are not changed after the insertion of $(-1)^F$ in the trace. That is, for large momentum $n$ $B_{bulk}$ is still given by a
Cardy formula
\begin{equation}
 B_{bulk}\approx e^{\pi\sqrt{n\,c_L/6}}
\end{equation}where $c_L$ is the central charge of the left-Virasoro algebra of the $CFT_2$.

We find useful to construct the partition function
\begin{eqnarray}
 \hat{B}_{6}(\vec{q},\tau)&=&\sum_n B_{6}(\vec{q},n)e^{2\pi i n\tau}\\
&=&\hat{B}_{bulk}(\vec{q},\tau)\hat{B}_{exterior}(\tau)
\end{eqnarray}where $\vec{q}$ stands for a generic charge vector and $n$ is the
momentum along the circle $S^1$. We are assuming that the exterior modes don't carry
any other charge rather than momentum $n$.

The Cardy formula gives the asymptotic behavior of the degeneracy $d=\textrm{Tr}(1)$ for large $n$. The derivation is based on the fact that under a modular transformation we can relate small with large $\tau$ behavior. The partition function can then be related to its ground state energy via the formula
\begin{equation}
 \hat{d}(\tau)\approx e^{\frac{\pi i c_L}{12\tau}}
\end{equation}with $-c_L/24$ the ground state energy of the left-moving sector.
Now instead of computing $\hat{d}(\tau)$ we should compute
$\hat{B}_{bulk}(\tau)$ which requires an insertion $(-1)^{2h_{bulk}}$ in the
trace. Generically for a black hole that preserves at least four supercharges
the isommetries of $AdS_3$ together with supersymmetries give rise to a
$(0,4)$ superconformal algebra in the boundary theory. This includes an $SU(2)$
R-symmetry current whose global part can be identified with the rotation
symmetry group $SU(2)$ of the black hole. Namely, $h_{bulk}$ is identified with the zero mode of the $U(1)$ in the $SU(2)$ R-current algebra in the $SCFT_2$.
Since the twist $(-1)^{2h_{bulk}}$ by the zero mode of the right-moving current is not expected to change the ground state energy of the left-moving sector, we conclude that, for small $\tau$, $\hat{B}(\tau)$ still has the same behavior as $\hat{d}(\tau)$, that is,
\begin{equation}\label{B small tau}
 \hat{B}(\tau)\approx e^{\frac{\pi i c_L}{12\tau}},\;\;|\tau|\ll1
\end{equation}

The analysis for $\hat{B}_{exterior}$ is a bit different. We will try to argue in the next section that the asymptotic growth of the index $\hat{B}_{6\,exterior}$ is still given by a Cardy formula, but in this case is not possible to identify $c_L$ with the central charge of a dual conformal field theory. We will be able to show that, under certain assumptions,
$\hat{B}_{exterior}(\tau)$ has a behavior given by (\ref{B small tau}) 
\begin{equation}\label{small tau behaviour}
 \hat{B}_{exterior}\approx e^{\frac{\pi i c^{eff}_L}{12\tau}},\;\;|\tau|\ll1
\end{equation}with $c^{eff}_L$ a constant that effectively controls the asymptotic growth. This should not confused with the physical central charge.

A simple calculation shows that, in the large $n$ limit, the growth of the total index is given by a Cardy like formula 
\begin{eqnarray}
 B_{6}&=&\oint \hat{B}_{bulk}(\tau)\hat{B}_{exterior}(\tau)e^{-2\pi i n\tau}\\
      &\approx&e^{2\pi\sqrt{nc^{total}_L}/6},
\end{eqnarray}with $c^{total}_L=c_L+c^{eff}_L$. 

In  \cite{Kraus:2005vz} the authors show, using general properties of $AdS/CFT$
correspondence, that the left and right central charges $c_L$ and $c_R$ of the $CFT_2$ can be computed from the bulk lagrangian via the coefficients of the Chern-Simons terms. 

The gravitational Chern-Simons term constructed out of the gravivational $SO(2,1)$ $\Gamma$ connection 
\begin{equation}
 \Omega_3(\Gamma)=\Gamma\wedge d\Gamma+\frac{2}{3}\Gamma\wedge \Gamma\wedge \Gamma
\end{equation}induces a diffeomorphism anomaly which translates in the non-conservation of the boundary stress tensor. Holography relates the coefficient of $\Omega_3$ to $c_L-c_R$, that is, the difference between the left and right-Virasoro central charges. In addition there can be gauge Chern-Simons terms of the form
\begin{equation}
 \Omega_{3}(A)=A\wedge dA+\frac{2}{3}A\wedge A\wedge A
\end{equation}where $A$ is a non-abelian gauge field. Similarly, the lack of gauge invariance induces a non-conserved current in the boundary theory. In the case of the $SU(2)_L\times SU(2)_R$ R-symmetry, the coefficient of the Chern-Simons of each of the $SU(2)$ factors is related to the R-current anomalies $k_L$  and $k_R$ respectively. These results are exact in the sense that they include already all the higher
derivative corrections.

A ``classical'' approach to this problem consists in computing the central charges as a perturbative expansion in inverse powers of $l^2$, the size of
$AdS_3$. This is called c-extremization \cite{Kraus:2005vz} and requires, as the name suggests, extremizing the bulk action with respect to the parameter $l$. 

Consider the example of string theory on $AdS_3\times S^p$. We take the metric to be
\begin{equation}
 ds^2=l^2\left(d\eta^2+\sinh^2\eta d\Omega_2^2\right)+l_{S^p}^2d\Omega_p^2.
\end{equation}The values of $l$ and $l_{S^p}$ can be obtained by demanding the lagrangian computed on the background constant solution to be stationary under variation of the those parameters as implied by the equations of motion. In general the lagrangian $\mathcal{L}_{p+3}$ is a very complicated function of the size $l$ and the charges. Its dependence will come for example from non trivial fluxes on $S^p$ or higher derivative corrections. Due to the topological nature of the Chern-Simons terms these will not be relevant for c-extremization. 

Now, put the $CFT_2$ in the boundary sphere $S^2$ which has metric
\begin{equation}
 ds^2=e^{2w}d\Omega_2^2.
\end{equation}and consider a small variation $\delta w$ in the free energy $F$
defined via the partition function $Z=e^{-F}$,
\begin{equation}
 \delta F=\frac{1}{4\pi}\int d^2x \sqrt{g}T^{ij}\delta
g_{ij}=\frac{\delta w}{2\pi}\int d^2x \sqrt{g}T^{ij}g_{ij}=\frac{\delta
w}{2\pi}\int d^2x \sqrt{g}T^i_i
\end{equation}Using the trace anomaly equation
\begin{equation}
 T^i_i=-\frac{c}{12}R
\end{equation}with $c$ the central charge and $R$ the Ricci scalar, we find
\begin{equation}\label{free energy and central charge}
 \frac{\delta F}{\delta w}=-\frac{c}{3}R.
\end{equation}

To compute $F$ from the bulk lagrangian we take the vacuum background $AdS_3$ with some possible fluxes on $S^p$ corresponding to the charges. The lagrangian becomes a constant scalar that depends on $l_{S^p}$, $l$ and the other charges 
\begin{equation}\label{free energy+Sbndy}
 F=V_{S^2}V_{S^p}\mathcal{L}_{p+3} \int d\eta \sinh^2\eta +S_{bndy}
\end{equation}The counter-term $S_{bndy}$ is introduced to remove the infrared
divergence associated with the infinite volume of $AdS_3$. This is all analogous to the
discussion of the $AdS_2$ path integral in the context of the quantum entropy function. For a large cutoff $\eta_{max}$, equation (\ref{free energy+Sbndy}) becomes
\begin{equation}\label{free energy AdS3}
 F=V_{S^2}V_{S^p}\mathcal{L}_{p+3}(-\frac{1}{2}\eta_{max}+\frac{1}{2}e^{2\eta_{
max} } )+S_{bndy}
\end{equation}The counter term $S_{bndy}$ goes as $e^{2\eta_{max}}$. From here we also conclude that $\omega=\eta_{max}$. If we choose an appropriate boundary counter term we can eliminate the divergent term in (\ref{free energy AdS3}) proportional to $e^{2\eta_{max}}$. We are left with
\begin{equation}
 F=-\frac{1}{2}\eta_{max}V_{S^2}V_{S^p}\mathcal{L}_{p+3}
\end{equation}which is divergent since it still depends linearly on the cuttoff. However because we are only interested in variations of $F$ with respect to
$w$ or $\eta_{max}$ the final result is finite 
\begin{equation}\label{free energy linear}
 \frac{\delta F}{\delta w}=-\frac{1}{2}V_{S^2}V_{S^p}\mathcal{L}_{p+3}
\end{equation}

Using (\ref{free energy linear}) and (\ref{free energy and central
charge}) we compute the central charge
\begin{equation}\label{cextremization}
 c=\frac{3}{2}V_{S^2}V_{S^p}\mathcal{L}_{p+3}.
\end{equation}

This formalism is very similar to that of the entropy function \cite{Sen:2005wa,Sen:2007qy}. There the entropy was given by extremizing some entropy function proportional to the bulk
lagrangian. Here we extremize a bulk lagrangian to compute the central charge.

For the moment we considered the case when both the left and right central
charges are the same. Though this method is very powerful it requires knowing
all the terms in the action which are non zero in the background solution. 
In the following we show how to go beyond this perturbative approach by considering the Chern-Simons terms.

The presence of gravitational Chern-Simons terms renders the bulk action
diffeomorphic anomalous. This translates in the non conservation of the boundary
stress tensor. Similarly the presence of other gauge Chern-Simons terms implies
the non-conservation of the dual currents in the boundary theory
\cite{Witten:1998qj,Kraus:2005vz,Kraus:2005zm}.

The $AdS/CFT$ correspondence instructs to compute a path integral over string fields with specific boundary conditions \cite{Witten:1998qj}
\begin{equation}\label{Z_string}
 Z_{string}=e^{-S^{cl}_{string}-S^{1-loop}_{string}+\ldots}=e^{-\Gamma[g^0_{ij},
A_0 ] } 
\end{equation}where $A_0$ is the boundary value of the gauge field and $g^0_{ij}$ the boundary metric, just as an example. The effective action $\Gamma[g^0_{ij},A_0 ]$ includes both classical and quantum contributions. The $AdS/CFT$ dictionary then tells us to compute the boundary correlation functions of the dual operators via
\begin{equation}\label{variation Z_string}
 e^{-\Gamma[g^0,A]}|_{AdS_3}=\langle e^{-\int d^2x T^{ij}g^0_{ij}-\int d^2x \textrm{Tr}\,
A_{\mu}(x)J^{\mu}}(x)\rangle|_{CFT_2}
\end{equation}where $T^{ij}$ is the boundary stress tensor and $J^{\mu}$ is the dual current. The presence of a gauge Chern-Simons renders $\Gamma[A]$ not gauge invariant and consequently it induces an anomalous dual current. A gauge transformation with parameter $\Lambda$ gives
\begin{equation}
 \delta_{\Lambda} \Gamma[A]=-\int d^2x \textrm{Tr}\,\Lambda J^{\mu},_{\mu}(x)
\end{equation}where $\Lambda$ denotes a gauge transformation. Similarly a gravitational Chern-Simons will render the boundary stress tensor anomalously conserved. 

For four dimensional black holes which preserve at least four supercharges the near horizon geometry is guaranteed to be rotational symmetric as explained in section \S3. Apart from an $S^2$ factor the near horizon geometry is  $AdS_2\times S^1$. When we take the momentum and the asymptotic radius of $S^1$ to infinity we can view the black hole as an extremal BTZ living in $AdS_3$. In this case the bulk theory has $(0,4)$ supersymmetry in $AdS_3$. Generalizations to $(4,4)$ theories corresponding to string theory on $AdS_3\times S^3$ are straightforward. We will describe the case of five dimensional black holes by the end of this section.

In the case of $(0,4)$ supersymmetry, the supergravity action can be regarded as a gauge theory with supergroup $SU(1,1)\times SU(1,1|2)$. In terms of the superconnections $\Gamma_L$ and $\Gamma_R$, in the adjoint representation of $SU(1,1)$ and $SU(1,1|2)$ respectively, the action can be written as \cite{Sen:2007qy}
\begin{equation}
 S=a_L\int \textrm{Tr}(\Gamma_L\wedge d\Gamma_L+\frac{2}{3}\Gamma_L\wedge
\Gamma_L\wedge \Gamma_L)+a_R\int \textrm{Tr}(\Gamma_R\wedge
d\Gamma_R+\frac{2}{3}\Gamma_R\wedge
\Gamma_R\wedge \Gamma_R).
\end{equation} In terms of the bosonic fields  the action is
\begin{eqnarray}
 S&=&\int d^3x\sqrt{g}(R-2\Lambda\nonumber)\\
 &+& \frac{K}{2}\int\textrm{Tr}\left(\Gamma\wedge d
\Gamma+\frac{2}{3}\Gamma\wedge \Gamma\wedge \Gamma\right) +\frac{k_R}{4\pi}\int
\textrm{Tr}\left(A\wedge dA+\frac{2}{3}A\wedge A\wedge A\right) \nonumber\\
\end{eqnarray}where $\Lambda=-1/l^2$ is the cosmological constant, $A$ is the $SU(2)$ gauge field and $\Gamma$ is the $SO(2,1)$ tangent space connection.

Under a local Lorentz gauge transformation $\omega$ of $\Gamma$ ,
\begin{equation}
 \delta \Gamma=d\omega +[\Gamma,\omega]
\end{equation}the variation of the graviational Chern-Simons gives a boundary term which induces a variation on the action
\begin{equation}\label{var S}
 \delta S=\frac{K}{2}\int_{\partial AdS_3}\textrm{Tr}\,\omega d\Gamma.
\end{equation}

At the same time a variation is induced in the source that couples to the stress tensor $T^{ij}$ in the boundary,
\begin{equation}\label{bndy stress tensor}
 \delta S_{CFT}=\frac{1}{2}\int d^2x\sqrt{g}T^{ij}\delta g_{ij}
\end{equation}The local Lorentz gauge transformation $\omega$ induces a variation in the boundary metric. After some tedious calculation we can relate the difference between the left and right central charges of the CFT to the coefficient of the gravitational Chern-Simons term 
\begin{equation}\label{delta c}
 c_L-c_R=48\pi K
\end{equation}with $K/2$ the coefficient of the gravitational Chern-Simons.

A similar analysis can be done for the gauge Chern-Simons. In this case
the coefficient of the $SU(2)$ gauge Chern-Simons is related to the anomaly $k_R$ of the boundary $SU(2)$ R-current via
\begin{equation}
 k_R=4\pi \alpha
\end{equation}where $\alpha$ is the coefficient of the gauge Chern-Simons.
Closure of the $(0,4)$ superconformal symmetry in the boundary CFT relates the central charge $c_R$ of the right super-Virasoro algebra to the $SU(2)$ R-current anomaly via
$$
c_R=6k_R
$$ Armed with both the gravitational and gauge Chern-Simons coefficients $\beta$ and $\alpha$ we can compute $c_L$ 
\begin{equation}
 c_L=96\pi \beta+24\pi\alpha
\end{equation}

In the case of string theory on $AdS_3\times S^3$ the R-symmetry is
$SO(4)=SU(2)_L\times SU(2)_R$, the isometry group of $S^3$. In this case we
have an additional gauge Chern-Simons associated with the $SU(2)_L$ connection. In addition if the SCFT has $(4,4)$ superconformal symmetry we can identify the left central charge with the left-moving R-current anomaly through $c_L=6k_L$.

Since $c_L$ and $c_R$ are determined in terms of the Chern-Simons coefficients they cannot receive any higher derivative corrections. From equation (\ref{cextremization}) this may seem surprising because we would expect the central charges to depend on all the derivative corrections. 

In \cite{David:2007ak} the authors explain this puzzle. The first thing to note is that from the bulk point of view, $(0,4)$ supersymmetry prevents the addition of any higher derivative corrections except those that can be removed by a field redefinition. On the CFT side all the correlation functions of operators dual to the fields in the supergravity multiplet are determined completely in terms of $c_L$ and $c_R$,
the central charges of the left and right Virasoro algebras. Once
we determine $c_L$ and $c_R$ via the gravitational and gauge Chern-Simons
coefficients we can determine all correlation functions of the superconformal
currents and therefore the boundary S-matrix of the supergravity fields. Now, two different theories with same boundary S-matrix must be related by a field redefinition. In other words two actions with $(0,4)$ supersymmetry and the same Chern-Simons terms must be related by a field redefinition.

\subsection{Results from four dimensional black holes}

We consider the D1-D5-KK-P black hole studied in section \S2 (\ref{microsc}). Since we are interested in answering some puzzles related to M theory on $K3\times T^2$, raised in \cite{Maldacena:1997de,Lambert:2007is,LopesCardoso:1999ur},  we will proceed our analysis in the  M-theory frame.

In type IIB compactified on $K3\times S^1\times \tilde{S}^1$, this black hole carries the following set of charges: $Q_1$ D1-branes wrapping the circle $S^1$ \footnote{The number $Q_1$ is effectively the charge and not the number of D1-branes. Due to the present of a Chern-Simons term of the form $\int C^2\wedge R\wedge R$ in the D5-brane world volume theory, there is an induced $-N_5$ D1 charge, with $N_5$ the number of D5 branes wrapping
$K3$. Therefore the total D1 charge is $Q_1=N_1-N_5$, where $N_1$ is the number of D1 branes.
}, $Q_5$ D5-branes wrapping $K3\times S^1$, $\tilde{K}$ KK monopoles associated with the
circle $\tilde{S}^1$, $n$ units of momentum along $S^1$ and $J$ units of
momentum along $\tilde{S}^1$. First we use mirror symmetry on $K3$ to map
the D1-D5 to a D3-D3 system with $Q_1$ D3-branes wrapping a 3-cycle
$\Sigma_2\times S^1$ and $Q_5$ D3-branes wrapping a 3-cyle
$\tilde{\Sigma}_2\times S^1$, with $\Sigma_2$ and $\tilde{\Sigma}_2$ a pair of
dual 2-cycles of $K3$. We then make a T-duality along $\tilde{S}^1$ to map the
D3-branes to D4-branes and the KK monopoles to NS5-branes wrapped on $K3\times
S^1$. The momentum $J$ is mapped to winding. We note the dual circle by $\hat{S}^1$. So far we have $Q_1$ D4-branes on $\Sigma_2\times S^1\times \hat{S}^1$, $Q_5$ D4-branes on
$\tilde{\Sigma}_2\times S^1\times \hat{S}^1$, $\tilde{K}$ NS5-branes on $K3\times S^1$, $J$ fundamental strings wrapping $\hat{S}^1$ and $n$ units of momentum along $S^1$ in the type IIA frame. We can now lift this configuration to M-theory on $S^1_M$, the M-theory circle. Altogether we have $Q_1$ M5-branes wrapping the 5-cycle $\Sigma_2\times S^1\times \hat{S}^1\times S^1_M$, $Q_5$ M5-branes wrapping $\tilde{\Sigma}_2\times S^1\times \hat{S}^1\times S^1_M$, $\tilde{K}$ M5-branes wrapping $K3\times \hat {S}^1\times S^1_M$, $J$ M2-branes wrapping $\hat{S}^1\times S^1_M$ and $n$ units of momentum along $S^1$.

The presence of spining M2-branes in the background of M5-branes brings additional difficulties in the study of the low energy theory. Therefore we restrict to the case when M2-branes are absent by setting $J=0$.

Our goal will be to compute the index associated with this black hole from a bulk
perspective and then compare it with the index $B_6$ that we determined in
section \S2, in the limit when we take $n$ very large keeping the other charges
finite. As explained before, in the limit of large radius $S^1$, the $AdS_2\times S^2$ near horizon of the four dimensional black hole in M-theory combines with the circle $S^1$ to form a locally $AdS_3\times S^2$ factor \cite{Sen:2007qy,Dabholkar:2010rm}. Moreover as we
send the asymptotic radius $S^1$ to infinity keeping the other moduli fixed the
solution develops an intermediate $AdS_3$ region where the black hole
solution can be seen as a BTZ black hole. The entropy can be computed using the Cardy formula
\begin{equation}
 S_{BH}\approx e^{2\pi\sqrt{nc_L/6}}
\end{equation}where $c_L$ is the left central charge of the dual
$CFT_2$. In the limit when all the charges are very large, we can use formula
(\ref{cextremization}) to compute the leading contribution to the central charge $c_L=6Q_1Q_5\tilde{K}$ which comes, basically, from the Einstein-Hilbert term. This
agrees with the Beckenstein-Hawking entropy $S_{BH}\approx 2\pi\sqrt{Q_1Q_5\tilde{K}n}$,
\cite{Cvetic:1995uj,Cvetic:1995bj}.
 
As explained in the previous section the central charge $c_L$ can be computed
exactly given the Chern-Simons coefficients in the bulk theory. 

M-theory contains, already in eleven dimensions, higher derivative corrections.
One of particular importance in this problem is a eighth derivative Chern-Simons term \cite{Vafa:1995fj,Duff:1995wd}
\begin{equation}\label{eighth derivative}
 \sim \int C^3\wedge I_8(X)
\end{equation}where $C^3$ is the three form gauge field of M theory. The eighth 
form is defined as
\begin{equation}
 I_8(X)=\frac{1}{48}\left(p_2(X)-\frac{1}{4}p_1(X)^2\right)
\end{equation}with $X$ being the eleven dimensional space and $p_n$ the nth
Pontryagin class. Once we reduce the theory on $K3\times \tilde{S}^1\times
S^1_M$ a five dimensional gravitational Chern-Simons on $AdS_3\times S^2$ is generated
\cite{Kraus:2005vz,Dabholkar:2010rm},
\begin{equation}
 S_{CS}=\frac{1}{32\pi^2}\int_{AdS_3\times S^2}\Omega_3(\Gamma)\wedge F
\end{equation}where $F$ is the KK monopole gauge field strength and
$\Omega_3(\Gamma)$ is the gravitational Chern-Simons in $AdS_3$. Using the equation of
motion $\int_{S^2} F=4\pi \tilde{K}$ we get
\begin{equation}
 S_{CS}=\frac{1}{8\pi}\int_{AdS_3}\Omega_3(\Gamma).
\end{equation}From the coefficient of the gravitational Chern-Simons we can determine the difference between the central charges. That is,
\begin{equation}
 c_L-c_R=12\tilde{K}.
\end{equation}Concerning the gauge Chern-Simons it can have two possible origins. First, there is already in eleven dimensions a gauge Chern-Simons term of the form
\begin{equation}
 \int C^3\wedge F^4\wedge F^4
\end{equation}with $C^3$ the M-theory three form and $F^4$ its field strength. After reducing down to five dimensions it originates a five dimensional Chern-Simons on $AdS_3\times S^2$ of the form
\begin{equation}\label{chern simons five dimensions}
 \int C^1\wedge F^2\wedge F^2 
\end{equation}with $F^2=dC^1$ and $C^1$ is the reduction of $C^3$ on $\tilde{S}^1\times
S^1_M$. The reduction of a term of this type down to $AdS_3$ was analyzed in \cite{Hansen:2006wu}. The authors do a careful treatment of the $SU(2)$ gauge fields which result from gauging the $S^2$ isometries. Roughly, they consider a solution where $S^2$ is fibered over $AdS_3$ with the
fibers being the $SU(2)$ connections,
\begin{equation}
 ds^2=ds^2_{AdS_3}+\sum_{i=1}^3(dy^i-A^i_j(x)dx^j)^2,\;\sum_{i=1}^3(dy^i)^2=1.
\end{equation}where $A^i_j$ is the $SO(3)$ R-connection. At infinity, that is, near the boundary $A^i_j$ vanishes and we recover $AdS_3\times S^2$. A careful treatment of the fluxes over a gauged $S^2$ has to be considered in this case. They found that the gauge field $C^1$ that carries magnetic charge on $S^2$ will not be invariant under an $SU(2)$ gauge transformation of the $S^2$ fibers, inducing a $SU(2)$ gauge Chern-Simons via the term (\ref{chern simons five dimensions}). This term will be responsible for an anomalous conservation of the $SU(2)$ right-moving R-current on the boundary theory. Moreover, using $(0,4)$ supersymmetry we determine $c_R=6k_r$ where $k_R$ is the R-current anomaly. Since the Chern-Simons term in (\ref{chern simons five dimensions}) is a two derivative term, the contribution to the central charge will correspond to the leading supergravity approximation, 
\begin{equation}\label{leading 2 derivative c_R}
 c_R=6Q_1Q_5K+\ldots
\end{equation}

The other possible contribution comes from the eighth derivative Chern-Simons term in eleven dimensions (\ref{eighth derivative}). Because this term is higher order in derivatives it will give subleading corrections to $c_R$. The gauging of $S^2$ induces an additional contribution in the tangent space connection $\Gamma$ of $AdS_3\times S^2$. The total connection is the direct sum of the $SO(2,1)$ connection $\Gamma_{AdS_3}$ of $AdS_3$ and the $SO(3)$ connection $A$ associated with the sphere:
$\Gamma=\Gamma_{AdS_3}\oplus A$. The five dimensional gravitational
Chern-Simons term  decomposes as \cite{Dabholkar:2010rm,Freed:1998tg,Harvey:1998bx}
\begin{equation}
 \Omega_3(\Gamma_{AdS_3\times S^2})=\Omega_3(\Gamma_{AdS_3})+\Omega_3(A).
\end{equation}Therefore reducing (\ref{eighth derivative}) on $K3\times\hat{S}^1\times S^1_M\times S^2$ originates an additional Chern-Simons term
\begin{equation}
 \frac{\tilde{K}}{2\pi}\int_{AdS_3}\Omega_3(A_R)
\end{equation}where the gauge field $A_R$ is in the adjoint of $SU(2)$\footnote{In rewriting the $SO(3)$ connection $A=\frac{1}{2}A^{ij}J^{ij}$, where $J^{ij}$ are the $SO(3)$ generators, in terms
of $SU(2)$ generators $J^a$ as $A=A^aJ^a$ there is an additional factor of four in the Chern-Simons term due to the fact that $\textrm{Tr}(J^{ij})^2=4\textrm{Tr}(J^a)^2$ }. This term generates an additional contribution of $12\tilde{K}$ to the central charge $c_R$. Together with the leading contribution (\ref{leading 2 derivative c_R}) we get
\begin{equation}
 c_R=6Q_1Q_5\tilde{K}+12\tilde{K}.
\end{equation}Now using the fact that $c_L-c_R=12\tilde{K}$ we conclude that the
central charge $c_L$ is given by
\begin{equation}
 c_L=6Q_1Q_5\tilde{K}+24\tilde{K}.
\end{equation}

If we consider the most general case of an M5-brane wrapping a five-cycle $P\times
S^1$, where $P$ is a divisor of $M=K3\times S^1_M\times \tilde{S}^1$, we find \cite{Dabholkar:2010rm,Kraus:2005vz,Maldacena:1997de}
\begin{equation}
 c_R=\int_M \tilde{P}\wedge \tilde{P} \wedge \tilde{P}+\frac{1}{2}
\tilde{P}\wedge c_2(M),\; c_L=\int_M \tilde{P}\wedge \tilde{P} \wedge \tilde{P}+
\tilde{P}\wedge c_2(M)
\end{equation}where $\tilde{P}$ is the 2-cycle dual to $P$ in $M$ and $c_2(M)$
is the second chern class of $M$. In terms of a basis of four cycles $\sigma^a$
of $M$ and the charge vector $q_a$, the divisor $P$ is represented as $P=q_a\sigma^a$. In this case we have $P=K\sigma(K3)+Q_1\sigma(\Sigma^2\times S^1\times\tilde{S}^1)+Q_5\sigma(\tilde{\Sigma}^2\times S^1\times \tilde{S}^1)$. 

We could wonder if there are any other terms in the eleven dimensional theory that
could possibly contribute with additional gravitational or gauge Chern-Simons terms.

To study this possibility we use the scaling argument developed in \cite{Sen:2009bm}. This is a useful way to track at which order in perturbation theory potential higher derivative terms can arise. Say we take an extremal black hole carrying NS-NS electric $q^{ele}_{NS-NS}$ and magnetic
$q^{mag}_{NS-NS}$ charges, and RR charges $q_{RR}$. The tree level IIA/B string
theory action has a scaling symmetry under which the dilaton $\phi$ gets
shifted by a constant $-\ln\lambda$, the NS-NS fields remain invariant and the
RR fields are multiplied by $\lambda$. The effect of this scaling is to
multiply the action by $\lambda^2$. In terms of the charges, this corresponds
to multiply by $\lambda^2$ the NS-NS electric charges and by $\lambda$ the RR
charges while leaving invariant the magnetic charges. The formula
(\ref{cextremization}) relates the central charge to the
string theory action computed in the near horizon background. Hence, the central
charge as a function of the charges will have the same scaling symmetry as the tree level
action, that is,
\begin{equation}
 c^0(\lambda^2 q^{ele}_{NS-NS},q^{mag}_{NS-NS},\lambda q^{ele}_{RR})=\lambda^2
c^0(q^{ele}_{NS-NS},q^{mag}_{NS-NS},q^{ele}_{RR})
\end{equation}where $c^0$ denotes the tree level contribution to the central
charge. In the same spirit we can keep track of additional l-loop contributions
through
\begin{equation}
 c^l(\lambda^2 q^{ele}_{NS-NS},q^{mag}_{NS-NS},\lambda
q^{ele}_{RR})=\lambda^{2-2l}
c^l(q^{ele}_{NS-NS},q^{mag}_{NS-NS},q^{ele}_{RR})
\end{equation}where $c^{l}$ is the l-loop contribution to the central charge.
In the case of the D1-D5-KK system we have $Q_1$ and $Q_5$ RR charges and $\tilde{K}$
NS-NS magnetic charge. Since the central charge is an integer we can write it as a polynomial in the charges. Terms linear in $Q_1$ or $Q_5$ are not
allowed because they would correspond to a $1/2$-loop contribution which is absent in closed string theory. Moreover the dependence on the RR charges must
come in the form $Q_1Q_5$ if we want to respect duality. The
contribution $6Q_1Q_5\tilde{K}$ to the central charge scales as $\lambda^2$ since it comes from the tree level action. The remaining $12K$ corresponds to one-loop contribution because it doesn't scale. Indeed the term (\ref{eighth derivative}) is generated at one-loop in string theory \cite{Duff:1995wd,Vafa:1995fj}. Though all this analysis,  we can still ask whether additional one-loop contributions to the gravitational or gauge Chern-Simons can arise after compactification on
$M\times S^2$. A priori we can not rule out such possibility. Equations (\ref{Z_string}) and (\ref{variation Z_string}) show that additional one-loop Chern-Simons terms in $AdS_3$ can be generated. For example in $AdS_5$ there is a  one-loop generated $SU(4)$ gauge Chern-Simons term that is responsible for a constant $-1$ correction to the leading $N^2$ $SU(4)$ R-symmetry anomaly in $SU(N)$ SYM \cite{Bilal:1999ty}.  If we denote by $a_L$ and $a_R$ possible one-loop contributions to $c_L$ and $c_R$ respectively, we have 
\begin{equation}\label{central charges D1D5KK}
 c_L=6Q_1Q_5\tilde{K}+24\tilde{K}+a_L,\;c_R=6Q_1Q_5\tilde{K}+12\tilde{K}+a_R
\end{equation}

To determine the full contribution to the black hole index we have to combine
the bulk results with the exterior modes contribution. The total central charge
which controls the asymptotic growth of the index is
\begin{equation}\label{c^macro}
 c^{macro}_L=6Q_1Q_5\tilde{K}+24\tilde{K}+a_L+c^{eff}_L
\end{equation}In the next section we will see that the effect of $c^{eff}_L$ is to cancel the constant one-loop contribution $a_L$.

\subsection{Exterior modes contribution} 

The central charge of the left Virasoro algebra can be determined in terms of the
coefficients of the gravitational and $SU(2)$ gauge Chern-Simons present in the
bulk action of the intermediate $AdS_3$ geometry. The left central charge is given by a simple formula
\begin{equation}\label{c=c_L+6k_R}
 c_L=c^{grav}_{bulk}+6k_R
\end{equation}where $c^{grav}_{bulk}=c_L-c_R$ is proportional to the
coefficient of the gravitational Chern-Simons and $k_R$ is the R-current
anomaly which is proportional to the coefficient of the $SU(2)$ gauge Chern-Simons term. Part of the
answer came from integrating Chern-Simons terms already existing in eleven dimensional M-theory on $K3\times\tilde{T}^2\times S^2$, down to five dimensions. We also argued that additional one-loop constant contributions generated after compactification were possible.

Imagine that instead of doing a reduction on $K3\times\tilde{T}^2\times
S^2$ we do this in the asymptotic region where the eleven dimensional geometry looks like
$K3\times\tilde{T}^2\times \mbbr^5$. In a vast region of space-time, namely
for $1\ll r_1\ll r\ll r_2$, the space looks like $K3\times\tilde{T}^2\times
S^2\times\mbbr^3$, where $\mbbr^3$ contains the time coordinate, the radius $r$
and the coordinate $y$ corresponding to the circle $S^1$. We can now compactify the
eleven dimensional theory on $K3\times\tilde{T}^2\times S^2$ and compute the coefficients of the gravitational and gauge Chern-Simons with support on $\mbbr^3$. The calculation will be identical to what we have done in the previous section except that in this case the one-loop corrections are not generated after compactification. 

These coefficients will be identical to (\ref{central charges
D1D5KK}) except for the constant shifts. On $\mbbr^3$ we denote
by $c^{asymp}_{grav}$ and $k^{asymp}_R$ the gravitational and gauge Chern-Simons coefficients. For the D1-D5-KK system we have
\begin{equation}
 c^{asymp}_{grav}=12\tilde{K},\;k^{asymp}_R=Q_1Q_5\tilde{K}+2\tilde{K}
\end{equation}

The difference between the anomaly coefficients computed at asymptotic infinity
and those computed at the intermediate $AdS_3$ must be accounted by the
exterior modes which live in between those two regions. That is,
\begin{equation}\label{c^asymp+k^asymp}
 c^{asymp}_{grav}=c^{bulk}_{grav}+c^{exterior}_{grav},\;k^{asymp}_R=k^{bulk}
_R+k^ {exterior}_R
\end{equation}where $c^{exterior}_{grav}$ and $k^ {exterior}_R$ denote the
contributions to the gravitational and $SU(2)$ current anomalies from the exterior modes. Since both $c^{asymp}_{grav}$ and $k^{asymp}_R$ are free from constant shifts, the effect of $c^{exterior}_{grav}$ and $k^{exterior}_R$ is to cancel the one-loop
contributions $a_L$ and $a_R$ in (\ref{central charges D1D5KK}), that is,
\begin{equation}
 6k_R^{exterior}+a_R=0,\;c^{exterior}_{grav}+a_L=0
\end{equation}

Ultimately we are interested in the total index which is controlled by an effective central charge which we denote by $c^{macro}_L$ (\ref{c^macro}). This effective central charge
is the sum of the bulk central charge $c_L$ and an effective central charge
$c^{eff}_L$ that controls the growth of the index for the exterior modes. For the bulk degrees of freedom we used holography to determine the left central charge in terms of the Chern-Simons, that is, $c_L=c^{bulk}_{grav}+6k^{bulk}_R$, while for the exterior modes, a priori there's no relation between $c^{eff}_L$, $c^{exterior}_{grav}$ and
$k^{exterior}_R$ because we don't have a dual description of the theory. Once we find the relation between these coefficients we can relate $c^{macro}_L$ to the known quantities $c^{asymp}_{grav}$ and $k^{asymp}_R$.

In the next we show, based on certain assumptions, that the following relation holds
\begin{equation}\label{c_eff=c_ext+6k_ext}
 c^{eff}_L=c^{exterior}_{grav}+6k^{exterior}_R.
\end{equation}Surprisingly it is identical to the relation (\ref{c=c_L+6k_R})
valid for the bulk theory. In general these exterior modes are associated with the center of mass degrees of freedom of a brane system like the singleton in $AdS_5$ which describes the $U(1)$ factor of the $U(N)$ SYM.  For the exterior modes we are not able to identify the R-symmetry of the two dimensional theory with the rotation group of $S^2$ since the scalars that describe transverse motion are not chiral under rotations. For example in the case of the D1-D5 system they describe motion in the transverse $\mbbr^4$. The $SO(4)$ R-symmetry of the dual $CFT_2$ then acts non-chirally on the bosons as $SO(4)$ rotations. Since the R-symmetry is a chiral symmetry, $SO(4)$ cannot be the R-symmetry of the superconformal $\mbbr^4$ sigma model \cite{Witten:1997yu}. If we were able to identify the R-symmetry of the exterior superconformal theory with the rotation group of the transverse space then $k^{exterior}_R$ would correspond to the R-symmetry current anomaly $k_R$. In addition we could use $(0,4)$
supersymmetry to compute the right super-Virasoro central charge via $c_R=6k_R$ and then conclude $c^{exterior}_{L}=c^{exterior}_{grav}+6k^{exterior}_R$. 

In attempting to derive the formula (\ref{c_eff=c_ext+6k_ext}) we make the assumptions
\begin{enumerate}\label{assumptions}
 \item The exterior modes consist of free massless scalars and fermions
belonging to singlet and/or spinor representations of $SU(2)_L\times SU(2)_R$.

\item The scalar modes that transform in the vector representation of $SO(4)$
are non-chiral. Under this assumption the contribution of the scalars to the
$SU(2)_L$ or $SU(2)_R$ current anomalies always vanish.
\end{enumerate} This is basically the content of the $\mbbr^4$ sigma model we
described in section \S2 (\ref{microsc})  when studying the microscopic degrees of freedom of the D1-D5-KK associated with the center of mass motion of the D1-D5 system.

Note that we are considering the most general case by allowing the fields to be
charged under an additional $SU(2)_L$. This would be relevant for the case when the black hole has a $S^3$ factor in the near horizon geometry, in which case the R-symmetry is $SU(2)_L\times SU(2)_R$.

In addition the $1+1$ dimensional CFT describing the exterior mode part has $(0,4)$ supersymmetry. This follows from the supersymmetry of the solution outside the $AdS_3$ region.

Both the anomaly coefficients $k^{exterior}_R$ and $c^{exterior}_{grav}=c^{exterior}_L-c^{exterior}_R$ can be extracted by reading the quantum numbers of the fields. For example a complex right-moving fermion charged under $SU(2)_R$ will contribute with $1/2$ to the right-moving current anomaly $k_R$ while a left-moving fermion charged under the same group
contributes with $-1/2$ and vice-versa for $SU(2)_L$. As usual, the central charge is
given by the number of bosons plus half the number of fermions. To compute $c_L^{eff}$ we have to read the asymptotic behaviour of the index $B_{6\,exterior}=-1/6!\textrm{Tr}(-1)^{2J_R}(2J_R)^6$ in the limit of large $n$. We use instead $B_{6\,exterior}=\textrm{Tr}(-1)^{2J_R}$ where integration over fermion zero modes has been carried out. Since the theory has $(0,4)$ supersymmetry, BPS condition forces the right-moving excitations to be in the ground state.

Consider the example of left-moving $N_b$ scalars and $N_f$ fermions
uncharged under $SU(2)_R$. The partition function can be easily computed to give
\begin{equation}\no
 \textrm{Tr}(-1)^{2J_R}q^{L_0}=\sum
d(n)q^n=\frac{1}{q^{(N_b-N_f)/24}}\frac{\prod_m(1+q^m)^{N_f}}{\prod_m
(1-q^m)^{N_b}},\;q=e^{2\pi i \tau}.
\end{equation}For large $n$ the index grows as
$d(n)\approx e^{2\pi\sqrt{nc_L/6}}$ with $c_L=N_b+1/2N_f$. In this case $c^{eff}_L=c_L$.
Now consider the same system but with the fermions charged under $SU(2)_R$. The partition function is now
\begin{equation}\no
\sum d(n)q^n=\frac{1}{q^{(N_b-N_f)/24} \prod_m
(1-q^m)^{N_b-N_f}},\;q=e^{2\pi i \tau}.
\end{equation}In this case $d(n)\approx e^{2\pi\sqrt{nc^{eff}_L/6}}$, for large $n$, 
with $c^{eff}_L=N_b-N_f\neq c_L$. The left-moving fermions all together contribute with $-N_f/4$ to the $SU(2)_R$ current anomaly. On the supersymmetric side we have $N$ bosons and $N$ fermions both charged under $SU(2)_R$. Their contribution is $c_R=3N/2$ for the central charge and $N/4$ for the right R-current anomaly. The total anomaly becomes $k_R=N/4-N_f/4$. Note that in this case $c_R\neq 6k_R$ consistent with our assumptions. A straightforward calculation shows  that $c^{eff}_L=c_{grav}+6k_R$. Note that in this example $c^{eff}_L$ can be negative provided $N_f>N_b$ which is possible because the left-moving sector is not supersymmetric.

We could repeat this analysis for many other examples though we would arrive at the
same conclusion
\begin{equation}\label{effective central charge}
 c^{eff}_{L}=c_{grav}+6k_R.
\end{equation}Unfortunately we lack a physical understanding of this result.

We are now in a position to compute the total index in terms of the anomaly coefficients measured at asymptotic infinity. 

Using equations (\ref{effective central charge}) and (\ref{c^asymp+k^asymp}) we compute the total index 
\begin{equation}
 c^{macro}_L=c^{bulk}_L+c^{eff}_L=c^{bulk}_{grav}+c^{exterior}_{grav}+6k^{bulk}
_R+6k^{exterior}_R=c^{asymp}_{grav}+6k^{asymp}_R
\end{equation}from which we conclude that \emph{the coefficient $c^{macro}_L$ that controls the growth of the total index is given in terms of the coefficients of the gravitational and gauge Chern-Simons computed at asymptotic infinity}. For the case of the D1-D5-KK this translates to
\begin{equation}\label{c_macro}
 c^{macro}_L=6Q_1Q_5\tilde{K}+24\tilde{K}
\end{equation}

We end this section by making some remarks on equation (\ref{c^asymp+k^asymp})
based on anomaly inflow \cite{Freed:1998tg,Harvey:1998bx}.

It was pointed out some time ago \cite{Callan:1984sa} that the anomalous conservation of the charge current in a string, due to the presence of chiral fermion zero modes, should be cancelled by an inflow of charge current from the exterior. This mechanism is called anomaly inflow. It was extremely useful in explaining both the tangent and normal bundle anomalies of an M5-brane \cite{Harvey:1998bx,Freed:1998tg}. In the even dimensional world volume theory of the M5-brane there are chiral fields which render both the $SO(5,1)$ and $SO(5)$ ,respectively, tangent and normal diffeomorphisms,  anomalous. The associated charges will therefore be anomalously conserved.  This means that the total charge in the world volume theory will vary with time unless there is an inflow of a charge current from the exterior bulk. This was consistent with the fact that there is already in the eleven dimensional supergravity a eighth derivative Chern-Simons coupling of the form 
\begin{equation}
 \sim \int_{11} C_3\wedge I_8(X) \nonumber
\end{equation}where $C_3$ is the three form gauge field of M-theory and $I_8(X)$ is a eight form constructed out of the Riemann tensor. We had made use of this term to derive the Chern-Simons terms in the five dimensional theory of black holes in M-theory (\ref{eighth derivative}). In this context the Chern-Simons terms are computed at asymptotic infinity and therefore the anomaly coefficients are free from constant one-loop corrections. Equations (\ref{c^asymp+k^asymp}) are equivalent to anomaly inflow.

\subsection{Microscopic results}

We borrow the microscopic results from section \S2 (\ref{microsc}). 

For a primitive dyon, that is, a dyon for which
$\textrm{gcd}(Q\wedge P)=1$ we saw that the index was given by the
fourier coefficient of the inverse of the  Siegel  modular form $\Phi_{10}$, that is,
\begin{equation}
 B_6(Q^2,P^2,Q.P)=(-1)^{Q.P+1}\int_{\mathcal{C}} d\rho d\sigma dv \frac{e^{-i\pi
\rho Q^2-i\pi \sigma P^2 -2\pi i vQ.P }}{\Phi_{10}(\rho,\sigma,v)}
\end{equation}
While in the more general case with $I\geq 1$, for dyons $(Q,P)=(IQ_0,P_0)$ with $\textrm{gcd}(Q_0\wedge P_0)=1$, the index was given by  
\begin{equation}
 B_6(Q,P)=\sum_{s|I}s\,d_1\left(\frac{Q^2}{s^2},P^2,\frac{Q.P}{s}\right),
\end{equation}where $d_1\left(\frac{Q^2}{s^2},P^2,\frac{Q.P}{s}\right)$ is
computed from the primitive answer. 

We are interested in the behaviour of $B_6$ in the limit of large $Q^2=2nK$,
finite $P^2=2Q_1Q_5$ and $Q.P=0$. We use an asymptotic expansion of the primitive answer derived in \cite{David:2006yn} which has the form (\ref{entropy function from Siegel})
\begin{equation}
 B_6\simeq \int \frac{d^2\tau}{\tau_2^2}e^{-F(\tau_1,\tau_2)}
\end{equation} with
\begin{eqnarray}\no
 F(\tau_1,\tau_2)=&-&\ln
\left(26+\frac{\pi}{\tau_2}(Q^2+P^2|\tau|^2)\right)-\frac{\pi}{2\tau_2}
(Q^2+P^2|\tau|^2)\nonumber\\
&+&24\ln\eta(\tau)+24\ln\eta(-\bar{\tau})+12\ln(\tau_2).
\end{eqnarray}We compute this integral using a saddle point approximation. Due to the symmetry $\tau_1\rightarrow -\tau_1$ of the free energy $F$ we can set
$\tau_1=0$ at the saddle point. At the same time we use the ansatz that
$\tau_2$ becomes very large at the saddle point to simplify $F(\tau_1,\tau_2)$
\begin{equation}
 F(0,\tau_2)\simeq-\frac{\pi}{2\tau_2}
(Q^2+P^2{\tau_2}^2)-4\pi\tau_2.
\end{equation}From this expression we determine the value of $\tau_2$ at the saddle point 
\begin{equation}
 \tau_2^*=\sqrt{\frac{Q^2}{P^2+8}}
\end{equation}In the limit considered $\tau_2^*$ becomes very large which justifies our assumption. We can now estimate the asymptotic growth of the index
\begin{equation}
 \ln B_6\simeq \pi\sqrt{Q^2(8+P^2)}
\end{equation}For the charge configuration of our problem we have $Q^2=2n\tilde{K}$ and $P^2=2Q_1Q_5$ which gives
\begin{equation}
 \ln B_6\simeq 2\pi\sqrt{nK(4+Q_1Q_5)}
\end{equation}This is in perfect agreement with the macroscopic derivation (\ref{c_macro}).

\subsubsection{MSW string}
Here we are interested on the derivation of the microscopic index using directly the data from the 1+1 low energy theory of the M5-brane on the divisor $P$.

As explained before the D1-D5-KK system can be mapped to a M5-brane wrapping a five cycle $P\times S^1$ with $P$ a divisor in $M=K3\times \hat{S}^1\times S^1_M$. If the size of the circle $S^1$ is much larger than the typical size of $M$, then theory which describes the low energy fluctuations of a M5-brane on $P\times S^1$ is a $1+1$ dimensional $(0,4)$ SCFT.

The BPS states in this theory involve the left-moving excitations. The growth of the degeneracy of these states is given by a Cardy formula determined in terms of the central charge $c^{micro}_L$ of the left Virasoro algebra. For a theory with $N_b$ left-moving bosons and $N_f$ left-moving fermions the central charge is $c^{micro}_L=N_b+N_f/2$. Since our interest is the index $B_6$ instead of the degeneracy, the computation goes differently. After integrating over the fermionic zero modes, $B_6$ reduces to the Witten index $\textrm{Tr}(-1)^F$. While $(-1)^F$ does not affect the contribution from a bosonic oscillator it can change that of a fermion. The growth of the index is now controlled by an effective central charge \cite{Vafa:1997gr} given by
\begin{equation}
 c^{micro}_{L,eff}=N_b-N_f.
\end{equation}Note that if $N_f=0$ then $c^{micro}_{L,eff}=c^{micro}_{L}$.

According to the analysis of \cite{Maldacena:1997de}, the number of bosons in the low energy theory is
\begin{eqnarray}\label{N^R_b}
 &&N^L_b=d_p(P)+b_2^{-}+3,\no\\
  &&N^R_b=d_p(P)+b_2^{+}+3
\end{eqnarray}The upper indices $L,R$ denote left and right sectors, $d_p(P)$ is the dimension of the moduli space of deformations of the divisor $P$ inside $M$, 3 accounts for the center of mass translations and $b_2^{-},b_2^{+}$ denote the number of self and anti-self dual two forms of $P$. These scalars originate from the reduction of the two form living on the world-volume of the M5-brane. For the fermions we have
\begin{eqnarray}\label{N^R_f}
 &&N^L_f=4h_{1,0}(P),\no\\
  &&N^R_f=4h_{2,0}(P)+4
\end{eqnarray}

For an ample divisor $P$ the authors in \cite{Maldacena:1997de} gave an expression for $d_p(P)$ when $M$ is a manifold without one-cycles. In this case $M=K3\times S^1_M\times \tilde{S}^1$ contains two one-cycles. Therefore we proceed differently following \cite{Lambert:2007is}.  On a Kahler manifold we have the following relations
\begin{equation}
 b_2=b_2^++b_2^{-}=2h_{2,0}+h_{1,1},\;b_2^{-}=h_{1,1}-1
\end{equation}Using these results in equations (\ref{N^R_b}) and (\ref{N^R_f}) together with the fact that for the right-moving sector supersymmetry implies $N^{R}_b=N^{R}_f$, we find
\begin{equation}
 d_p(P)=2h_{2,0}.
\end{equation}For the left-moving fields we get
\begin{equation}
 N^L_b=2h_{2,0}(P)+h_{1,1}+2=b_{even}(P),\;N^L_f=4h_{1,0}=b_{odd}(P).
\end{equation}The effective central charge $c^{micro}_{L,eff}=N^L_b-N^L_f$ is just the Euler character of $P$, that is, $c^{micro}_{L,eff}=b_{even}-b_{odd}=\chi(P)$. 

This has a simple expression in terms of the two-cycle $\tilde{P}$ dual to $P$ in $M$
\begin{equation}
 c^{micro}_{L,eff}=\chi(P)=\int_M \tilde{P}\wedge\tilde{P}\wedge\tilde{P}+\tilde{P}\wedge c_2(M)
\end{equation}where $c_2(M)$ is the second chern class of $M$. In the particular example of the D1-D5-KK the divisor is $P=\tilde{K}\sigma(K3)+Q_1\sigma(\Sigma^2\times S^1\times\tilde{S}^1)+Q_5\sigma(\tilde{\Sigma}^2\times S^1\times \tilde{S}^1)$ where $\sigma$ denotes a four cycle in $M$.  Substituting in the formula above we get
\begin{equation}
 c^{micro}_{L,eff}=6Q_1Q_5\tilde{K}+24\tilde{K}.
\end{equation}in perfect agreement with both the microscopic result computed from $\Phi_{10}^{-1}$ and the macroscopic one computed from the coefficients of the Chern-Simons terms.
If we computed the physical left central charge that controls the degeneracy instead of the index we would find
\begin{equation}\label{c^micro_L}
 c^{micro}_L=N^L_b+N^L_f/2=6Q_1Q_5\tilde{K}+24\tilde{K}+6
\end{equation}where we used $N^L_f=4h_{1,0}(P)=4h_{1,0}(M)=4$. As observed in \cite{Lambert:2007is,LopesCardoso:1999ur}, (\ref{c^micro_L}) fails to agree with the macroscopic result (\ref{c_macro}). Hence we see that the apparent puzzle arose from comparing the microscopic degeneracy with the microscopic index.

\subsection{Five dimensional black holes}

The analysis goes  more or less in the same way as in the four dimensional case. The main difference resides on the fact that the five dimensional black hole can carry angular momentum without breaking supersymmetry.

The spatial rotation group in five dimensions is $SU(2)_L\times SU(2)_R$. We denote by $J_L$ and $J_R$ the  $U(1)$ generators of both factors. Among all the supersymmetry generators of the theory half belong to the $(1_L,2_R)$ and the other half to $(2_L,1_R)$ representations of $SU(2)_L\times SU(2)_R$. We choose the convention that for a state preserving four supersymmetries the unbroken generators are in the $(1_L,2_R)$ representation. Moreover $k$ broken supersymmetries give rise to $k$ fermion zero modes. For those which are charged under $SU(2)_R$, we insert a power of $(2J_R)$ for each pair of fermion zero modes to render the index non vanishing.

For a five dimensional black hole in $\mathcal{N}=4$ string theory, like in the case of the D1-D5-P in type IIB on $K3\times S^1$ \cite{Breckenridge:1996is}, the solution carries two complex fermion zero modes in the $(1_L,2_R)$ representation. If we consider the same brane system but in type IIB on $(T^4\times S^1)$ \cite{Maldacena:1999bp}, we have six complex fermion zero modes instead of two. 

To capture the BPS states we use the spacetime index
\begin{equation}
 B_2=-\frac{1}{2}\textrm{Tr}(-1)^{2J_R-2J_L}(2J_R)^2
\end{equation}where we sum over $J_R$ and fix the angular momentum $J_L$ and the remaining charges. For simplicity we shall use the index defined as
\begin{equation}
 C_2=-\frac{1}{2}\textrm{Tr}(-1)^{2J_R}(2J_R)^2
\end{equation}which is related to $B_2$ by a simple operation $B_2=(-1)^{2J_L}C_2$. 

The near horizon geometry of these black holes has locally an $AdS_3\times S^3$ factor. We proceed similarly as in section \S6.2. By sending the attractor radius of the circle $S^1$ to  infinity keeping the other charges finite, we combine the circle $S^1$ with $AdS_2$ to form a locally $AdS_3$. At the same we take the asymptotic radius to infinity and keep the other moduli fixed to get an intermediate $AdS_3$ region where the solution can be embedded as an extremal BTZ black hole. Since we keep fixed the angular momentum $J$ while taking the asymptotic radius $R$ of the circle to infinity the physical angular momentum $J/R$ vanishes restoring the $S^3$ symmetry as seen from an asymptotic observer.

As before we consider the contribution to the index from the bulk and exterior degrees of freedom. The index $C_2$ becomes
\begin{eqnarray}
 C_2 &=&-\frac{1}{2}\textrm{Tr}(-1)^{2J^{bulk}_R+2J_R^{exterior}}(2J^{bulk}_R+2J_R^{exterior})^2\\
    &=&-\frac{1}{2}\textrm{Tr}(-1)^{2J^{bulk}_R}\textrm{Tr}(-1)^{2J_R^{exterior}}(2J_R^{exterior})^2\\
    &=&\sum_{\substack{q_1+q_2=q\\ J_1+J_2=J_L}} C_{bulk}(q_1,J_1)C_{2\,exterior}(q_2,J_2)
\end{eqnarray}where we have defined $C_{bulk}=\textrm{Tr}(-1)^{2J^{bulk}_R}$ and $C_{2\,exterior}=-\frac{1}{2}\textrm{Tr}(-1)^{2J_R^{exterior}}(2J_R^{exterior})^2$.
   We proceed by constructing the partition function
\begin{eqnarray}
 \tilde{C}_2(\tau,z)&=&\sum_{n,J_L}C_{2}(n,J_L)e^{2\pi i\tau+2\pi iJ_Lz}\\
    &=&\tilde{C}_{bulk}(\tau,z)\tilde{C}_{2\,exterior}(\tau,z)
\end{eqnarray}In the $CFT_2$ dual to the bulk $AdS_3$ we can identify the $SU(2)L$ and $SU(2)_R$ rotation symmetries with the left and right R-symmetries. In the same spirit of (\ref{small tau behaviour}), $\tilde{C}_{bulk}(\tau)$ as the small $\tau$ behavior
\begin{equation}\label{C_bulk small tau}
 \tilde{C}_{bulk}(\tau)\approx e^{\frac{\pi ic_L}{12\tau}-2\pi i\frac{k_Lz^2}{\tau}}
\end{equation}where $c_L$ is the central charge of the left Virasoro algebra and $k_L$ is the $SU(2)_L$ R-current anomaly. 

Using a saddle point approximation, the entropy of the rotating black hole is given by
\begin{equation}
 \ln C_{bulk}\approx 2\pi\sqrt{c_L/6\left(n-\frac{J_1^2}{4k_L}\right)}
\end{equation}in agreement with supergravity computations \cite{Breckenridge:1996is}. We shall also argue that for the exterior modes the corresponding partition function has a similar behaviour for small $\tau$, that is,
\begin{equation}\label{C_exterior small tau}
 \tilde{C}_{2\,exterior}(\tau)\approx e^{\frac{\pi ic^{eff}_L}{12\tau}-2\pi i\frac{k^{eff}_Lz^2}{\tau}}
\end{equation}where $c^{eff}_L$ and $k^{eff}_L$ shouldn't be confused neither with the left central charge nor with the current anomaly. These are coefficients that control the asymptotic growth and should be computed case by case. Their physical origin is not known for the moment. As in section \S6.1 these coefficients must be determined directly from the index. Joining both bulk (\ref{C_bulk small tau}) and exterior (\ref{C_exterior small tau}) contributions, the partition function $\tilde{C}_2(\tau,z)$ has small $\tau$ behaviour
\begin{equation}
 \tilde{C}_{2}(|\tau|\ll 1)\approx e^{\frac{\pi i(c_L+c^{eff}_L)}{12\tau}-2\pi i\frac{(k_L+k^{eff}_L)z^2}{\tau}}.
\end{equation}Performing a saddle point approximation, the growth of the total index is given by
\begin{equation}
 \ln C_2\approx 2\pi\sqrt{c^{macro}_L/6\left(n-\frac{J_1^2}{4k^{macro}_L}\right)}
\end{equation}where we have defined $c^{macro}_L=c_L+c^{eff}_L$ and $k^{macro}_L=k_L+k^{eff}_L$.

So far we have analysed the dependence on the index in terms of microscopic quantities like the central charges or the R-current anomalies. Since we don't know the details of the dual $CFT_2$ we use the same technology of section \S6.2 to determine $c_L$ and $k_L$ using the coefficients of the Chern-Simons terms in the bulk of $AdS_3$. In this case we need to consider in addition the $SU(2)_L$ gauge Chern-Simons term to determine $k_L$.

For the exterior modes we have to compute $c^{eff}_L$ and $k^{eff}_L$ directly from the index. The analysis goes in the same manner as in the four dimensional case. We study case by case two dimensional field theories which obeying the assumptions (\ref{assumptions}) may correspond to an exterior contribution. We found the following relations 
\begin{eqnarray}
 &&c^{eff}_L=c^{exterior}_{grav}+6k^{exterior}_R\\
  &&k^{eff}_L=k^{exterior}_L
\end{eqnarray}The coefficient $c^{eff}_L$ is the sum of $c^{exterior}_{grav}=c^{exterior}_L-c^{exterior}_R$ and the $SU(2)_R$ current anomaly $k^{exterior}_R$ while $k^{eff}_L$ equals the $SU(2)_L$ current anomaly.

The anomalous contributions from the bulk and the exterior modes must be combined to give the anomaly coefficients measured at asymptotic infinity
\begin{eqnarray}
 &&c^{asymp}_{grav}=c^{bulk}_{grav}+c^{exterior}_{grav}\\
  &&k^{asymp}_{R}=k^{bulk}_{R}+k^{exterior}_{R}\\
&&k^{asymp}_{L}=k^{bulk}_{L}+k^{exterior}_{L}
\end{eqnarray}These together with the previous results give
\begin{eqnarray}
 &&c^{macro}_L=c_L+c^{eff}_{L}=c^{asymp}_{grav}+6k^{asymp}_R\\
  &&k^{macro}_L=k_L+k^{eff}_L=k^{asymp}_L
\end{eqnarray}Since both $c^{asymp}_{grav}$ and $k^{asymp}_R$ don't receive one-loop constant corrections we arrive at the same conclusion that the effect of the exterior mode contribution is to cancel the one-loop contributions in the bulk anomaly coefficients which appear after compactification. The coefficients $c^{macro}_L$ and $k^{macro}_L$ are the same as $c_L$ and $k_L$ except for the constants shifts. 

The asymptotic growth of the total index, as in the case of four dimensional black holes, \emph{is determined in terms of the coefficients of the gravitational and gauge Chern-Simons computed at asymptotic infinity}. We find perfect agreement with the microscopic answer for the BMPV black hole \cite{Dabholkar:2010rm} computed from the low energy dynamics of the D1-D5 system on $K3$ \cite{Maldacena:1999bp,David:2006yn,Dabholkar:2008zy}. 

\subsubsection{Microscopic derivation}

For the D1-D5 black hole in type IIB on $K3\times S^1$ with $Q_1$ D1-branes wrapping $S^1$, $Q_5$ D5-branes wrapping $K3\times S^1$ and $n$ units of momentum on $S^1$, the microscopic index is given by the formula \cite{Maldacena:1999bp,Dabholkar:2010rm,Banerjee:2009uk}
\begin{equation}\label{micro partition 5D}
 C_2(n,Q_1Q_5,J)=(-1)^{J+1}\int_{0}^{1}d\rho\int_{0}^{1}d\sigma \int_{0}^{1}dv (e^{\pi iv}-e^{-\pi iv})^4\frac{\eta(\rho)^{24}}{\Phi_{10}(\rho,\sigma,v)}e^{-2\pi i(n\rho+\sigma Q_1Q_5+Jv)}
\end{equation}which is obtained from $\Phi_10$ using 4D-5D lift. Once we go from five to four dimensions there is an additional contribution coming from the KK monopole excitations \cite{David:2006yn}. This explains the factor $\eta(\rho)^{24}$ in the formula above.

In the limit of large charges \footnote{This limit is particularly different from the limit used in \cite{David:2006yn} in the sense that only one of the charges is taken to be very large. This is an important difference to consider when performing the asymptotic expansion. We refer the reader to the appendix of \cite{Dabholkar:2010rm} where this analysis has been carried out carefully} we can rewrite the index as an entropy function \cite{Dabholkar:2010rm}
\begin{equation}
 C_2(n,Q_1Q_5,J)\simeq \int \frac{d^2\tau}{\tau_{2}^2}e^{-F(\tau_1,\tau_2)}
\end{equation}with the free energy $F(\tau_1,\tau_2)$ given by
\begin{eqnarray}\label{five dimensional entropy function}
 F(\tau_1,\tau_2)&=&-\frac{\pi}{\tau_2}(n+Q_1Q_5|\tau|^2-\tau_1J)+24\ln \eta(\tau)+24\ln \eta(-\bar{\tau})+12\ln 2\tau_2 \no\\
&+&-24\ln \eta(i/2\tau_2)-4\ln\{2\cosh(\pi\tau_1/2\tau_2)\}\no \\
&-&\ln \left[\frac{1}{4\pi}\left\{26+\frac{2\pi}{\tau_2}(n+Q_1Q_5|\tau|^2-\tau_1J)+i\frac{24}{\tau_2}\frac{\eta'(i/2\tau_2)}{\eta(i/2\tau_2)}+4\pi\frac{\tau_1}{\tau_2}\tanh\frac{\pi \tau_1}{2\tau_2}\right\}\right]\no\\
\end{eqnarray}
In the limit of large $n$ and finite $Q_1Q_5$ and $J$ we use the ansatz that  at the saddle point $\tau_2$ becomes very large. Therefore, neglecting some terms in (\ref{five dimensional entropy function}), we keep the first square bracket term and approximate $24\ln\eta(\tau)+24\ln\eta(-\bar{\tau})\simeq -4\pi \tau_2$ and $24\ln \eta(i/2\tau_2)\simeq -4\pi \tau_2$. The saddle point is at
\begin{equation}
 \tau_1^*=\frac{J}{2Q_1Q_5},\;\tau_2^*=\sqrt{\frac{4nQ_1Q_5-J^2}{4Q_1Q_5}}
\end{equation}which in the limit considered justifies our ansatz. The entropy function evaluated at the saddle point gives the Beckenstein-Hawking entropy of the BMPV black hole
\begin{equation}
 \ln C_2\approx \pi\sqrt{4nQ_1Q_5-J^2}
\end{equation}This limit is also known as Cardy-limit.

Another interesting limit is to consider the case when the number of D1-branes becomes very large while keeping the other charges finite. This case is easier to understand from the type IIA perspective. Using ten dimensional S-duality followed by a T-duality along $S^1$ we map the D1-D5-P system to a system with $Q_5$ NS5-branes, $n$ fundamental strings wrapped along $\tilde{S}^1$, the dual circle, and momenta $Q_1$ along the same circle. The limit $Q_1$ very large corresponds to the Cardy limit of the low energy theory of the F-NS5 system. This example is interesting due to the presence of NS5-branes. Even though we don't know the microscopic theory, the anomaly coefficients can be computed from the bulk using the technique already described. Moreover since the index is invariant under duality, we can use the answer (\ref{micro partition 5D}) obtained in the type IIB frame and compare it with the macroscopic answer in the IIA frame.

In this case we assume the ansatz that $\tau_2$ is small at the saddle point. Defining a new variable $\sigma_1+i\sigma_2=\tau^{-1}$ and using modular properties, we rewrite (\ref{five dimensional entropy function}) in terms of this variable. The limit of small $\tau_2$ corresponds to large $\sigma_2$. The saddle point is at
\begin{equation}
 \sigma_1=-\frac{J}{2(n-1)},\;\sigma_2=\sqrt{\left(Q_1Q_5-\frac{J^2}{4(n-1)}\right)/(n+3)}
\end{equation}which in the limit considered justifies our ansatz. The Beckenstein-Hawking entropy becomes
\begin{equation}
 \ln C_{2}\simeq2\pi \sqrt{(n+3)\left(Q_1Q_5-\frac{J^2}{4(n-1)}\right)}
\end{equation}

In summary,
\begin{enumerate}
 \item Cardy limit: $n\rightarrow \infty$, fixed $Q_1, Q_5$ charges and angular momenta $J$,
\begin{equation}
 \ln B_{2\,micro}\simeq\ln B_{2\,macro}\simeq2\pi \sqrt{nQ_1Q_5-J^2/4}
\end{equation}where $\simeq$ means equality up to corrections suppressed by powers of $n$.
\item Anti-Cardy limit: $Q_1\rightarrow \infty$, fixed $n,Q_5$ charges and angular momenta $J$,
\begin{equation}
 \ln B_{2\,micro}\simeq\ln B_{2\,macro}\simeq2\pi \sqrt{(n+3)\left(Q_1Q_5-\frac{J^2}{4(n-1)}\right)}
\end{equation}where $\simeq$ means equality up to corrections suppressed by powers of $Q_1$.
\end{enumerate}
\subsubsection{Macroscopic derivation}
The analysis of the Chern-Simons in five dimensions requires a bit more work than in four dimensions. The main difference is the additional $SU(2)_L$ R-symmetry current dual to the bulk $SU(2)_L$ gauge field. Because the black hole has a $S^3$ factor in its near horizon geometry, after reducing down to $AdS_3$ a $SU(2)_L$ gauge Chern-Simons will be generated. The Cardy formula which depends explicitly on the anomaly coefficient $k_L$, is given by
\begin{equation}\label{five dimensional cardy formula}
 \ln C_{2,\,macro}\simeq 2\pi\sqrt{c^{macro}_L/6\left(n-\frac{J_1^2}{4k^{macro}_L}\right)}
\end{equation}

We give the results for both the Cardy and Anti-Cardy limits. For additional details on the computation we refer the reader to \cite{Dabholkar:2010rm}.
\begin{enumerate}
 \item Cardy limit:
\begin{eqnarray}
 c^{macro}_L=6Q_1Q_5,\;k^{macro}_L=Q_1Q_5 \no\\
c^{macro}_R=6Q_1Q_5,\;k^{macro}_R=Q_1Q_5\no
\end{eqnarray}

\begin{equation}
 \ln C_{2,\,macro}\approx \pi\sqrt{4nQ_1Q_5-J^2}\no
\end{equation}

\item Anti-Cardy limit:
\begin{eqnarray}
 c^{macro}_L=6Q_5(n+3),\;k^{macro}_L=Q_5(n-1)\no \\
c^{macro}_R=6Q_5(n+1),\;k^{macro}_R=Q_5(n+1)\no
\end{eqnarray}
\begin{equation}
 \ln C_{2,\,macro}\approx 2\pi \sqrt{(n+3)\left(Q_1Q_5-\frac{J^2}{4(n-1)}\right)}\no
\end{equation}

\end{enumerate}
  
Both the Cardy and Anti-Cardy limits of the macroscopic index are in perfect agreement with the microscopic results.

\section{Discussion, conclusions and outlook\label{Open}}

 It is remarkable that a functional integral of string theory in  $AdS_{2}$  precisely reproduces the first term
 in the Rademacher expansion that already captures all power-law suppressed corrections to the Bekenstein-Hawking-Wald formula as described in \S\ref{Evalu}. As we have seen in \S\ref{Nonpert}, the functional integral has all the ingredients to reproduce even the subleading nonperturbative corrections in  the  Rademacher expansion. It would be insteresting to see how string theory functional integral reproduces the detailed number theoretic details of the Kloosterman sum.  Since $d(\Delta)$ is an integer $W(\Delta)$ would also have to be an integer. This suggests an underlying integral structure in  quantum gravity at a deeper level.
 
Our  computation suggests that the bulk $AdS$ string theory is every bit as fundamental as the boundary $CFT$. Even though one sometimes refers to the $AdS$ computation as macroscopic and thermodynamic,  quantum gravity in $AdS_{2}$ does  not appear to be an emergent, coarse-grained description of the more microscopic boundary theory. Each theory has its own   rules of computation. It seems  more natural to regard $AdS/CFT$ holography as an exact strong-weak coupling duality.
 
So far we have used holography in its original sense to mean a complete accounting of the degrees of freedom associated with the  $AdS_{2}$ black hole horizon in terms of  the states of a $CFT_{1}$ in one lower dimension.  The  $AdS_{2}/CFT_{1}$ correspondence actually extends this idea further to apply correlation functions as well. The boundary $CFT_{1}$  has a  $GL(d)$ symmetry that acts upon $d(q, p)$ zero energy states.  The observables of the theory are thus simply $d \times d$ matrices $\{M_{i}\}$.  A precise state-operator correspondence has been suggested \cite{Sen:2011cn} that allows one to define, at least formally, the corresponding correlation functions for some of the observables in the bulk theory.
 In the boundary theory it is easy to define correlation functions of 
observables as traces of strings of operators such as 
\begin{equation}
\textrm{Tr} (M_{1}M_{2}\ldots M_{k}) \ .
\end{equation}
We have seen that localization techniques can be successfully applied for computing the partition function to compute the integer $d$. A natural question is if localization can be useful for computing the  correlation functions such as above. Such a computation would allow us to  recover the discrete information  about the microstates of a black hole from observables living in the bulk near the horizon. This of course goes to the heart of the problem of information retrieval from black holes. It is likely that one would need to extend the localization analysis beyond the massless fields to higher string modes to access this information.

The content of the boundary $CFT_{1}$ is essentially completely determined by the integer $d$. The  bulk theory  has an elaborate field content and action that depends on the compactification $K$ and the charges of the black hole. Imagine two different bulk theories $AdS_{2}\times K$ and $AdS_{2} \times K'$  but with the same black hole degeneracy  $d$. This would suggest 
that the two string theories near the horizon of two  very different black holes in  very different compactifications are dual to the same $CFT_{1}$. 
 By transitivity of duality, this would imply that the two string theories themselves are dual to each other. This  conclusion seems inescapable  from the perspective of the $CFT_{1}$.  Note that it is not easy to arrange the situation when the degeneracies of two different black holes are given by the same integer.  For example, if the degeneracy is given by the Fourier coefficients of some modular form, it would be  rare, but not impossible,  that two such Fourier coefficients are precisely equal.

Our analysis uses an $\CN=2$ reduction of the full $\CN=8$ theory by dropping six gravitini multiplets of $\CN=2$ and the hypermultiplets. This was motivated by the fact that hypermultiplets are flat directions of the classical entropy function and the black hole is not charged under the gauge fields in the gravitini multiplets. We have also ignored  D-terms. This is partially justified by the fact that the black hole horizon is supersymmetric and a large class of D-terms are known not to contribute to the Wald entropy as a consequence of this supersymmetry \cite{deWit:2010za}. 
Our final answer strongly suggests that these assumptions are justified and our reduced theory fully captures the physics. A technical obstacle in justifying these assumptions stems from the fact that the incorporation of  the hypermultiplets and the gravitini multiplets  would require infinite number of auxiliary fields if all $\CN=8$ supersymmetries are realized off-shell. It may be possible to make progress in this direction  perhaps by using a formulation where only the Q-supersymmetry used for localization is realized off-shell but on all fields of $\CN=8$ supergravity. Alternatively, it may be possible to repeat the localization analysis in a different off-shell formalism such as the harmonic superspace \cite{Galperin:2001uw}
where all $\CN=8$ supersymmetries are realized off-shell with infinite number of auxiliary fields; but perhaps only a small number of auxiliary fields get excited for the localizing solution.

We think that the application of localization techniques in $AdS_3$ could be very interesting as a means to understand the elliptic genus from a gravity perspective. We have already given some exact results in this context. In the last section we saw that in a particular charge regime, namely when only one of the charges is taken to be very large, the asymptotic growth of the index is controlled by the coefficients of the Chern-Simons terms. Even if this seems to follow naturally in a theory which has an holographic description, in this case it is surprising because we had to take into account an exterior contribution for which we don't have an holographic dual. 

In establishing an exact $AdS_2/CFT_1$ holography it is necessary not only to just compute exactly the quantum entropy, for which we give an important contribution, but also to have in hands precise microscopic answers. Duality plays a very important role in this matter. It is therefore important that the results are consistent with the duality symmetries of the theory. Much of this work has been accomplished here for quarter-BPS dyons in $\mathcal{N}=4$ string theory. In section \S2 we proposed a two dimensional supersymmetric sigma model whose index captures dyons with non-trivial values of $I$. Part of the microscopic answer, namely the divisor structure, has already been understood from the bulk perspective in \cite{Sen:2009vz}. Inclusion of orbifold geometries is necessary to explain the divisor structure in the microscopic answer (\ref{non primitive degeneracy}). Although there are still some caveats, mainly concerning the symmetrization of fermion zero modes, the index obtained passes many physical checks and is consistent with duality. The perturbative analysis of a set of two charge configurations presented in \S\ref{perturbative test} gives an additional and non-trivial important check. We think this work is worth to be explored in $\mathcal{N}=4$ CHL models or in $\mathcal{N}=8$ string theory.

\subsection*{Acknowledgments}
I would like to thank first Atish Dabholkar, Sameer Murthy and Ashoke Sen. Much of this review is based on the work done in various collaborations with them. I would like also to thank Boris Pioline, Nadav Drukker, Bernard de Wit, Gabriel Lopes Cardoso, Miguel Paulos and Ayan Mukhopadhyay for many useful discussions.  
The work of J.~G. was supported in part by Fundac\~{a}o para Ci\^{e}ncia e
Tecnologia (FCT).  J.~G. would like to acknowledge the hospitality of the LPTHE, HRI and TIFR where part of this work was completed.

\appendix
\section{Killing spinors in the attractor geometry \label{Killing}}

To apply localization arguments, it is necessary to identify the supercharge $Q$ that squares to the compact bosonic generator $L - J$. For this purpose, it is useful to know first the explicit form of the Killing spinors in the on-shell attractor geometry. 

Recall that in the superconformal formalism, there are fermionic variations corresponding to $Q$ as well as $S$, 
which we label by $\ve$ and $\eta$ respectively \cite{Mohaupt:2000mj}. One can only impose $Q$-invariance up to a uniform 
$S$-supertranslation. This corresponds to the fact that the physical supersymmetries in the Poincar\'e theory are 
found after the gauge fixing procedure to be a linear combination of these two variations. 
In general, this combination has a complicated dependence on the other fields as well as the choice of prepotential. 
The method of \cite{LopesCardoso:2000qm} is to surpass this problem by finding spinor fields whose variation under $S$ vanishes. 
One can then simply use the $Q$-invariance conditions for these spinor fields, which by construction is gauge 
independent. This construction was very useful in \cite{LopesCardoso:2000qm} to find the half-BPS solution in asymptotically flat space.

However, these gauge-independent supersymmetry transformations then depend on the choice of prepotential and hence the choice of the Lagrangian. This  is not well-suited for our purposes since we are really interested in the  off-shell localizing solutions that are determined direcly by the off-shell supersymmetry transformation without any reference to the prepotential. Moreover, we are only interested in the near horizon geometry which is much simpler to analyze than the full black hole solution including the asymptotic infinity. For the near horizon supersymmetries,  we make the simple observation that a choice of the bosonic fields corresponding to the near horizon attractor 
background leads to a particularly simple choice of gauge for the physical theory, namely $\eta=0$. 
This choice then permits us to work with the simpler supersymmetry transformations of the superconformal theory. 

To see this, we begin by imposing the vanishing of the variations of fermionic fields of the Weyl mutiplet:
\begin{eqnarray} \label{Weylvar1}
   0 =   \delta \psi^i_{\mu} & = & 2D_{\mu} \epsilon^i-\frac{1}{8}\gamma_a\gamma_b T^{abij}\gamma_{\mu} \epsilon_j+
      \gamma_{\mu}\eta^i \ , \\
 \label{Weylvar2} 
    0 =  \delta\chi^i & = & -\frac{1}{12}\gamma_a\gamma_b\displaystyle{\not}DT^{abij} \epsilon_j+D \epsilon^i
      +\frac{1}{12}T^{ij}_{ab}\gamma^a\gamma^b\eta_j \ , \\
  \label{Weylvar3}   
   0 =  \delta\phi^i_{\mu} & = & -2f^a_{\mu}\gamma_a \epsilon^i-\frac{1}{4}\displaystyle{\not}DT^{ij}_{cd}\sigma^{cd}
      +2D_{\mu}\eta^i \ .
\end{eqnarray}
At the attractor values, we have 
\be \label{attval}
v=\frac{16}{\omega\bar{\omega}} \ , \qquad  T_{rt}^-= v \omega \ ,
\ee
and the above variations simplify to 
\begin{eqnarray} \label{Weylvar22}
 \delta \psi^i_{\mu} & =& 2D_{\mu} \epsilon^i-\frac{1}{8}\gamma_a\gamma_b T^{abij}\gamma_{\mu} \epsilon_j+\gamma_{\mu}\eta^i \ , \\
      \delta\chi^i & =& \frac{1}{12}T^{ij}_{ab}\gamma^a\gamma^b\eta_j \ , \\
      \delta\phi^i_{\mu} & =& 2D_{\mu}\eta^i \ .
\end{eqnarray}
From here, we deduce the $AdS_2\times S^2$ Killing spinors equations
\begin{eqnarray} \label{killing eq}
 D_{\mu} \epsilon^i & =& \frac{1}{16}\gamma_a\gamma_b T^{abij}\gamma_{\mu} \epsilon_j  \ , \nonumber\\
  D_{\mu} \epsilon_i & =& \frac{1}{16}\gamma_a\gamma_b T^{ab}_{\quad ij}\gamma_{\mu} \epsilon^j\\
 \eta_i=\eta^{i} & = & 0 \nonumber . 
\end{eqnarray}
We thus see that $\eta^{i}=0$ as promised. Before solving the equation for $ \epsilon^{i},\epsilon_i$, 
note that in the Euclidean theory in four dimensions, the spinors should have a symplectic-Majorana condition imposed on them, while in Minkowski spacetime they can be majorana or symplectic-Majorana \cite{Cortes:2003zd}. In addition, the Weyl projection is not compatible with the majorana condition in the Minkowski case and therefore the left and right-handed spinors are complex conjugate to each other. On the contrary, in the Euclidean case, we can have symplectic Majorana-Weyl spinors but not majorana
\begin{equation} \label{Majsymp}
(\zeta^{i}_{\pm})^{*}=-i \ve_{ij} \, (\sigma_1\times \sigma_2) \,\zeta^{j}_{\pm},
\end{equation}
where the indices $i,j$ are $SU(2)'$ quantum numbers and $\ve_{ij}$ is the antisymmetric tensor of $SU(2)$. In the literature \cite{Mohaupt:2000mj} the spinors used obeyed a majorana condition in Minkowski space. They used the convention that positive/negative chirality is correlated with upper/down $SU(2)'$ indice due to complex conjugation. Since the killing spinor equations \ref{killing eq} were derived from the Lorentzian theory, we shall use an ansatz which reproduces the killing spinor equations in Euclidean $AdS_2\times S^2$. The ansatz is the following
\begin{eqnarray}\label{ansatz}
 \epsilon_i=i\ve_{ij}\xi^j_{-}\\
  \epsilon^i=\xi^i_{+}\\ 
\nonumber
\end{eqnarray}Note that we explicitly show the chirality of the spinor. We should therefore solve the Killing spinor condition for an unconstrained Dirac spinor $\xi^i=\xi^i_{+}+\xi^i_{-}$, double the space and 
then impose the above constraint \eqref{Majsymp}. 
We represent the Dirac spinor $ \xi$ as a direct product $ \xi= \xi_{AdS_2}\otimes \xi_{S^2}$ 
where $ \xi_{AdS_{2}}$ and $ \xi_{S^{2}}$ are two component spinors,  
and use the following gamma matrix representation 
\begin{equation}
\gamma_{\theta}= \sqrt{v} \, \sinh\eta\,\sigma_1\otimes 1\ , \quad \gamma_{\eta}=  \sqrt{v} \, \sigma_2\otimes
1\ , \quad \gamma_{\phi}=  \sqrt{v} \, \sin\psi\, \sigma_3\otimes \sigma_1\ , \quad 
\gamma_{\psi}=  \sqrt{v} \, \sigma_3\otimes \sigma_2 \ ,
\end{equation}
where $v \equiv v_{1} (=v_{2})$ is the classical size of the $AdS_{2}$ (and the $S^{2}$). 

Equations \eqref{killing eq} simplify to the diagonal form 
\begin{eqnarray}
 D_{\mu} \xi^i_{AdS_2} & = & \frac{\omega}{|\omega|}\frac{i}{2}(\sigma_3\times 1) \, \gamma_{\mu} \,  \xi^i_{AdS_2} \ , \\
 D_{j} \xi^i_{S^2} & = & \frac{\omega}{|\omega|}\frac{i}{2}(\sigma_3\times 1) \, \gamma_{j} \,  \xi^i_{S^2} \ . \\
\end{eqnarray}
which are easily solved \cite{Lu:1998nu}. In the bispinor basis
\begin{eqnarray}
  \xi&=& a_1\left(\begin{array}{c}
            1\\ 
	    0
           \end{array}\right)\times\left(\begin{array}{c}
            1\\ 
	    0
           \end{array}\right)+a_2\left(\begin{array}{c}
            0\\ 
	    1
           \end{array}\right)\times\left(\begin{array}{c}
            1\\ 
	    0
           \end{array}\right)\nonumber\\ \nonumber\\
	    &&+a_3\left(\begin{array}{c}
            1\\ 
	    0
           \end{array}\right)\times\left(\begin{array}{c}
            0\\ 
	    1
           \end{array}\right)+a_4\left(\begin{array}{c}
            0\\ 
	    1
           \end{array}\right)\times\left(\begin{array}{c}
            0\\ 
	    1
           \end{array}\right)\nonumber\\ \nonumber\\
&\equiv&\left(\begin{array}{c}
            a_1\\
	    a_2\\	
	    a_3\\	
	    a_4
          \end{array}\right)\nonumber \\
\end{eqnarray}the solutions are (this is assuming that $w \in \IR^{+}$, and we have fixed a certain normalization for the spinors):
\begin{eqnarray} \label{Killspin1}
 \xi^i_{--}= 2 \, e^{-\frac{i}{2}(\theta+\phi)}\left(\begin{array}{c}
\cosh\frac{\eta}{2}\cos\frac{\psi}{2}\\
                                                         \sinh\frac{\eta}{2}
\cos\frac{\psi}{2}\\
                                                        
-\cosh\frac{\eta}{2}\sin\frac{\psi}{2}\\
                                                        
-\sinh\frac{\eta}{2}\sin\frac{\psi}{2}\end{array}\right)  & \ , \qquad & 
 \xi^i_{-+} = 2 \, e^{-\frac{i}{2}(\theta-\phi)}\left(\begin{array}{c}
\cosh\frac{\eta}{2}\sin\frac{\psi}{2}\\
                                                         \sinh\frac{\eta}{2}
\sin\frac{\psi}{2}\\
                                                        
\cosh\frac{\eta}{2}\cos\frac{\psi}{2}\\
                                                        
\sinh\frac{\eta}{2}\cos\frac{\psi}{2}\end{array}\right) \ , \nonumber \\
%\label{Killspin2}
 \xi^i_{+-} = 2 \, e^{\frac{i}{2}(\theta-\phi)}\left(\begin{array}{c}
\sinh\frac{\eta}{2}\cos\frac{\psi}{2}\\
                                                         \cosh\frac{\eta}{2}
\cos\frac{\psi}{2}\\
                                                        
-\sinh\frac{\eta}{2}\sin\frac{\psi}{2}\\
                                                        
-\cosh\frac{\eta}{2}\sin\frac{\psi}{2}\end{array}\right) & \ , \qquad &  
 \xi^i_{++} = 2 \, e^{\frac{i}{2}(\theta+\phi)}\left(\begin{array}{c}
\sinh\frac{\eta}{2}\sin\frac{\psi}{2}\\
                                                         \cosh\frac{\eta}{2}
\sin\frac{\psi}{2}\\
                                                        
\sinh\frac{\eta}{2}\cos\frac{\psi}{2}\\
                                                        
\cosh\frac{\eta}{2}\cos\frac{\psi}{2}\end{array}\right) \nonumber \\ . 
\end{eqnarray}As explained above, we should impose a symplectic-Majorana conditon on the spinors.
In the above basis, equation \eqref{Majsymp} implies: 
\begin{eqnarray}
 &&\xi^{+}_{++}=(\xi^{-}_{--})^* \nonumber\\
 &&\xi^{+}_{-+}=(\xi^{-}_{+-})^*\nonumber\\
 &&\xi^{-}_{++}=(-\xi^{+}_{--})^*\nonumber\\
 &&\xi^{-}_{-+}=(-\xi^{+}_{+-})^*\nonumber
\end{eqnarray}
%CHECK ZZZ
 %\be\label{epsplminrel}
 %\epsilon^{+}_{\a,\b} = ( \epsilon^{-}_{-\a, -\b})^{*} \ . 
%( \epsilon^{i}_{\a,\b})^{*} = -i(\sigma_1\times \sigma_2) \, \epsilon_{ij} \,  \epsilon^{j}_{-\a, -\b} \ . 
%\ee 
where the star is not the ordinary complex conjugation but the complex
conjugation condition as defined by the symplectic-majorana condition.

One can now identify the spinors $ \epsilon^{i}_{ra}$ as the generators of  $G^{ia}_{r}$, the supercharges 
of the near horizon $\CN=4$ superalgebra \S\ref{Supersymmetries}. 
The real combinations $Q_{\mu}, \wt Q_{\mu}, \m = 1,\dots , 4$ are generated by the combinations:
\begin{eqnarray} \label{defzeta}
\begin{array}{l}
\zeta_1= \xi^+_{++}+ \xi^-_{--},\\
\zeta_2=-i\left( \xi^+_{++}- \xi^-_{--}\right),\\
\zeta_3=-i\left( \xi^-_{++}+ \xi^+_{--}\right),\\
\zeta_4= \xi^-_{++}- \xi^+_{--},\end{array}
\,\,
\begin{array}{l}
\tilde{\zeta}_1= \xi^+_{-+}+ \xi^-_{+-},\\
\tilde{\zeta}_2=-i\left( \xi^+_{-+}- \xi^-_{+-}\right),\\
\tilde{\zeta}_3=-i\left( \xi^-_{-+}+ \xi^+_{+-}\right),\\
\tilde{\zeta}_4= \xi^-_{-+}- \xi^+_{+-},\end{array}
\end{eqnarray}We can easily see that these killing spinors are real under the
complex conjugation condition defined by \eqref{Majsymp}. As an instructive
exercise take for example $\zeta^1$. The $SU(2)'$ components are
$\zeta^{1+}=\xi^{+}_{++}$ and $\zeta^{1-}=\xi^{-}_{--}$. Both are
complex conjugate to each other
\begin{eqnarray}
 (\zeta^{1+})^*=-i \ve_{+-} \, (\sigma_1\times \sigma_2) \, \zeta^{1-}
\nonumber\\
  (\zeta^{1-})^*=-i \ve_{-+} \, (\sigma_1\times \sigma_2) \, \zeta^{1+}\nonumber
\end{eqnarray}

\subsection{Supersymmetry variations}

Recall that the supersymmetry variations for fermions and scalars of the vector multiplets in Minkowski theory are \cite{Mohaupt:2000mj}
\begin{eqnarray}
 &&\delta X^I=\bar{\epsilon}^i\Omega^I_i \nonumber\\
  &&\delta \bar{X}^I=\bar{\epsilon}_i\Omega^{Ii} \nonumber\\
&& \delta \Omega^I_i=2\displaystyle{\not} \partial X^I\epsilon_i+\frac{1}{2}\ve_{ij}\mathcal{F}^{I\mu\nu-}\gamma_{\mu}\gamma_{\nu}\epsilon^j+Y^{I}_{ij}\epsilon^j+2X^I\eta_i \nonumber\\
&& \delta \Omega^{Ii}=2\displaystyle{\not} \partial \bar{X}^I\epsilon^i+\frac{1}{2}\ve^{ij}\mathcal{F}^{I\mu\nu+}\gamma_{\mu}\gamma_{\nu}\epsilon_j+Y^{Iij}\epsilon_j+2\bar{X}^I\eta^i\nonumber
\end{eqnarray} where  $\Omega_i$ has positive chirality while $\Omega^i$ has negative chirality. Changing basis from the $\epsilon$ spinors to the $\zeta$ spinors using (\ref{ansatz}),we can reexpress the susy variations  as
\begin{eqnarray}\label{susy variantions}
 &&\delta X^I=-(\zeta^i_{+})^{\dagger}\lambda^{Ii}_{+} \nonumber\\
  &&\delta \bar{X}^I= -(\zeta^i_{-})^{\dagger}\lambda^{Ii}_{-}\nonumber\\
&& \delta \lambda^{Ii}_{+}=\frac{1}{2}(F_{\mu\nu}^{I-}-\frac{1}{4}\bar{X}^{I} \, T^{-}_{\mu\nu}) \, 
 \gamma^{\mu} \, \gamma^{\nu} \, \zeta^{i}_+ +2i \displaystyle{\not}\partial X^{I} \, \zeta^i_-+Y^{Ii}_j \, \zeta^j_+ \nonumber\\
&& \delta \lambda^{Ii}_{-}=\frac{1}{2}(F_{\mu\nu}^{I+}-\frac{1}{4}X^{I} \, T^{+}_{\mu\nu}) \, 
 \gamma^{\mu} \, \gamma^{\nu} \, \zeta^{i}_- +2 i \displaystyle{\not}\partial \bar{X}^{I} \, \zeta^i_+ +Y^{Ii}_j \, \zeta^j_- \,
\end{eqnarray}
where $\lambda$ are related to $\Omega$ spinors by 
\begin{equation}
\Omega_{i } = \varepsilon_{ij} \lambda^{j}_{-} \qquad \Omega^{i} = -i \lambda^{i}_{+} \, .
\end{equation}

Under a transformation generated by $\zeta_i$ or $\tilde{\zeta}_i$, given in (\ref{defzeta}), we can show that the action of $\delta^2$ is $L-J$ or $L+J$ respectively
\begin{eqnarray}
 \delta^2X^I=-(\zeta^i_{+})^{\dagger}\delta \lambda^{Ii}_{+}=2i(\zeta^i_{+})^{\dagger} \displaystyle{\not}\partial X^{I} \, \zeta^i_-\\
\delta^2\bar{X}^I=-(\zeta^i_{-})^{\dagger}\delta\lambda^{Ii}_{-}=2i(\zeta^i_{-})^{\dagger}\displaystyle{\not}\partial \bar{X}^{I} \, \zeta^i_+
\end{eqnarray}where the remaining contractions vanish identically for the spinors chosen. After a straightforward computation we find 
 \begin{eqnarray}
 \delta^2X^I=-2i (\partial_{\theta}-\partial_{\phi}) X^{I} =2(L-J)X^I\\
\delta^2\bar{X}^I=-2i (\partial_{\theta}-\partial_{\phi}) \bar{X}^{I}=2(L- J)\bar{X}^I \, .
\end{eqnarray}

\section{Some aspects of the superconformal multiplet calculus \label{susyvar}}

In this appendix, we shall summarize some aspects of the superconformal multiplet calculus 
which we briefly presented in \S\ref{off-shell}. We shall first present the supersymmetry 
variation of the various multiplets. We shall then present the invariant Lagrangian density formula 
for a chiral multiplet. We shall then present the rule which defines the various
components of a scalar function of chiral superfields {\it e.g.} the prepotential superfield 
$\bf F (\bf X^{I})$. These are the basic ingredients that go into building the superconformal action. 
We shall borrow the presentation of the recent \cite{deWit:2010za} wherein a lot of these 
facts (and more) have been collected, this can be referred to for more details. 

The invariance of the bulk Lagrangian under the superconformal transformations are 
well established, we provide these details for the sake of completeness. Using the same transformations, 
in another appendix, we shall sketch the supersymmetry invariance of 
our conjectured boundary action. This, as far as we know, is new, and there is scope to develop it further. 

As in the text, $\epsilon_{i}$ and $\eta_{i}$ denote the parameters of the $Q$ and $S$ supersymmetry 
transformations. The transformation rules for a chiral multiplet of Weyl weight $w$ are:
\begin{eqnarray}
  \label{eq:conformal-chiral}
  \delta A =&\,\bar\epsilon^i\Psi_i\,, \nonumber\\[.2ex]
  \delta \Psi_i =&\,2\,\Slash{D} A\epsilon_i + B_{ij}\,\epsilon^j +
  \tfrac12   \gamma^{ab} F_{ab}^- \,\varepsilon_{ij} \epsilon^j + 2\,w
  A\,\eta_i\,,  \nonumber\\[.2ex]   
  \delta B_{ij} =&\,2\,\bar\epsilon_{(i} \Slash{D} \Psi_{j)} -2\,
  \bar\epsilon^k \Lambda_{(i} \,\varepsilon_{j)k} + 2(1-w)\,\bar\eta_{(i}
  \Psi_{j)} \,, \nonumber\\[.2ex] 
  \delta F_{ab}^- =&\,\tfrac12
  \varepsilon^{ij}\,\bar\epsilon_i\Slash{D}\gamma_{ab} \Psi_j+
  \tfrac12 \bar\epsilon^i\gamma_{ab}\Lambda_i
  -\tfrac12(1+w)\,\varepsilon^{ij} \bar\eta_i\gamma_{ab} \Psi_j \,,
  \nonumber\\[.2ex]   
  \delta \Lambda_i =&\,-\tfrac12\gamma^{ab}\Slash{D}F_{ab}^-
   \epsilon_i  -\Slash{D}B_{ij}\varepsilon^{jk} \epsilon_k +
  C\varepsilon_{ij}\,\epsilon^j 
  +\tfrac14\big(\Slash{D}A\,\gamma^{ab}T_{abij}
  +w\,A\,\Slash{D}\gamma^{ab} T_{abij}\big)\varepsilon^{jk}\epsilon_k
  \nonumber\\ 
  &\, -3\, \gamma_a\varepsilon^{jk}
  \epsilon_k\, \bar \chi_{[i} \gamma^a\Psi_{j]} -(1+w)\,B_{ij}
  \varepsilon^{jk}\,\eta_k + \tfrac12 (1-w)\,\gamma^{ab}\, F_{ab}^-
    \eta_i \,, \nonumber\\[.2ex]
    \delta C =&\,-2\,\varepsilon^{ij} \bar\epsilon_i\Slash{D}\Lambda_j
  -6\, \bar\epsilon_i\chi_j\;\varepsilon^{ik}
    \varepsilon^{jl} B_{kl}   \nonumber\\ 
  &\, -\tfrac14\varepsilon^{ij}\varepsilon^{kl} \big((w-1)
  \,\bar\epsilon_i \gamma^{ab} {\Slash{D}} T_{abjk}
    \Psi_l + \bar\epsilon_i\gamma^{ab}
    T_{abjk} \Slash{D} \Psi_l \big) + 2\,w \varepsilon^{ij}
    \bar\eta_i\Lambda_j \,. 
\end{eqnarray}

The independent fields of the Weyl multiplet transform
as follows,
\begin{eqnarray}
  \label{eq:weyl-multiplet}
    \delta e_\mu{}^a & =& \bar{\epsilon}^i \, \gamma^a \psi_{ \mu i} +
  \bar{\epsilon}_i \, \gamma^a \psi_{ \mu}{}^i \, , \nonumber\\ 
  \delta \psi_{\mu}{}^{i} & =& 2 \,\mathcal{D}_\mu \epsilon^i - \tfrac{1}{8}
  T_{ab}{}^{ij} \gamma^{ab}\gamma_\mu \epsilon_j - \gamma_\mu \eta^i
  \, \nonumber \\  
  \delta b_\mu & =& \tfrac{1}{2} \bar{\epsilon}^i \phi_{\mu i} -
  \tfrac{3}{4} \bar{\epsilon}^i \gamma_\mu \chi_i - \tfrac{1}{2}
  \bar{\eta}^i \psi_{\mu i} + \mbox{h.c.} + \Lambda^a_K e_{\mu a} \, ,
  \nonumber \\ 
  \delta A_{\mu} & =& \tfrac{1}{2} \mathrm{i} \bar{\epsilon}^i \phi_{\mu i} +
  \tfrac{3}{4} \mathrm{i} \bar{\epsilon}^i \gamma_\mu \, \chi_i +
  \tfrac{1}{2} \mathrm{i} 
  \bar{\eta}^i \psi_{\mu i} + \mbox{h.c.} \, , \nonumber\\  
  \delta \mathcal{V}_\mu{}^{i}{}_j &=& 2\, \bar{\epsilon}_j
  \phi_\mu{}^i - 3 
  \bar{\epsilon}_j \gamma_\mu \, \chi^i + 2 \bar{\eta}_j \, \psi_{\mu}{}^i
  - (\mbox{h.c. ; traceless}) \, , \nonumber \\   
  \delta T_{ab}{}^{ij} &=& 8 \,\bar{\epsilon}^{[i} R(Q)_{ab}{}^{j]} \,
  , \nonumber \\ 
  \delta \chi^i & =& - \tfrac{1}{12} \gamma^{ab} \, \Slash{D} T_{ab}{}^{ij}
  \, \epsilon_j + \tfrac{1}{6} R(\mathcal{V})_{\mu\nu}{}^i{}_j
  \gamma^{\mu\nu} \epsilon^j -
  \tfrac{1}{3} \mathrm{i} R_{\mu\nu}(A) \gamma^{\mu\nu} \epsilon^i + D
  \epsilon^i + 
  \tfrac{1}{12} \gamma_{ab} T^{ab ij} \eta_j \, , \nonumber \\ 
  \delta D & =& \bar{\epsilon}^i \,  \Slash{D} \chi_i +
  \bar{\epsilon}_i \,\Slash{D}\chi^i \, , 
\end{eqnarray}
where 
\be
  R(Q)_{\mu \nu}{}^i =  \, 2 \, \mathcal{D}_{[\mu} \psi_{\nu]}{}^i -
  \gamma_{[\mu}   \phi_{\nu]}{}^i - \tfrac{1}{8} \, T^{abij} \,
  \gamma_{ab} \, \gamma_{[\mu} \psi_{\nu]j} \ . 
\ee

Based on these two multiplets, one can write down a Lagrangian density 
for the chiral multiplet which is invariant under the superconformal transformations:
\begin{align}
  \label{eq:chiral-density}
  e^{-1}\mathcal{L} =&\, C - \varepsilon^{ij}\, \bar\psi_{\mu i} \gamma^\mu
  \Lambda_j-\tfrac18\bar \psi_{\mu i} T_{ab\,jk}\gamma^{ab}\gamma^\mu
  \Psi_l \,\varepsilon^{ij}\varepsilon^{kl}
   -\tfrac1{16}A( T_{ab\,ij} \varepsilon^{ij})^2 \nonumber\\
  &\, 
  -\tfrac12\bar\psi_{\mu i}\gamma^{\mu\nu}\psi_{\nu j}\,
  B_{kl}\,\,\varepsilon^{ik}\varepsilon^{jl} 
    + \varepsilon^{ij} \bar \psi_{\mu i}\psi_{\nu j}(F^{-\mu\nu}
    -\tfrac12 A\, T^{\mu\nu}{}_{kl}\,\varepsilon^{kl} )\nonumber\\
  &\,
  -\tfrac12 \varepsilon^{ij}\varepsilon^{kl} e^{-1}
  \varepsilon^{\mu\nu\rho\sigma} \bar\psi_{\mu i}\psi_{\nu j}
  (\bar\psi_{\rho k}\gamma_\sigma\Psi_{l} +\bar\psi_{\rho k}
  \psi_{\sigma j}\, A)\,.
\end{align}
This density is built such that the variation of the Lagrangian is equal to a total derivative
in spacetime. 

The Lagrangian for vector multiplets is based on first viewing the gauge invariant quantities 
of the vector multiplet as a reduced chiral multiplet with weight $w=1$. The components are:
\begin{align}
  \label{eq:vect-mult}
  A\vert_{\text{vector}}=&\,X\,,\nonumber\\
  \Psi_i\vert_{\text{vector}}=&\, \Omega_i\,,\nonumber\\
  B_{ij}\vert_{\text{vector}}=&\, Y_{ij}
  =\varepsilon_{ik}\varepsilon_{jl}Y^{kl}\,,\nonumber\\
  F_{ab}^-\vert_{\text{vector}}=&   \big(\delta_{ab}{}^{cd} -\tfrac12
    \varepsilon_{ab}{}^{cd}\big) e_c{}^\mu e_d{}^\nu \,\partial_{[\mu}
    A_{\nu]}\nonumber\\ 
    &\, 
  +\tfrac14\big[\bar{\psi}_{\rho}{}^i\gamma_{ab} \gamma^\rho\Omega^{j}
  + \bar{X}\,\bar{\psi}_\rho{}^i\gamma^{\rho\sigma}\gamma_{ab}
  \psi_\sigma{}^j
  - \bar{X}\, T_{ab}{}^{ij}\big]\varepsilon_{ij}  \,,\nonumber\\
  \Lambda_i\vert_{\text{vector}}
  =&\,-\varepsilon_{ij}\Slash{D}\Omega^j \,,\nonumber\\
  C\vert_{\text{vector}}= &\,-2\, \Box_\mathrm{c}  \bar X  -\tfrac14  F_{ab}^+\,
   T^{ab}{}_{ij} \varepsilon^{ij} - 3\,\bar\chi_i \Omega^i\,.
\end{align}
The transformations of the vector multiplet are:
\begin{align}
  \label{eq:variations-vect-mult}
  \delta X =&\, \bar{\epsilon}^i\Omega_i \,,\nonumber\\
  \delta\Omega_i =&\, 2 \Slash{D} X\epsilon_i
     +\half \varepsilon_{ij}  F_{\mu\nu}
   \gamma^{\mu\nu}\epsilon^j +Y_{ij} \epsilon^j
     +2X\eta_i\,,\nonumber\\
  \delta A_{\mu} = &\, \varepsilon^{ij} \bar{\epsilon}_i
  (\gamma_{\mu} \Omega_j+2\,\psi_{\mu j} X)
  + \varepsilon_{ij}
  \bar{\epsilon}^i (\gamma_{\mu} \Omega^{j} +2\,\psi_\mu{}^j
  \bar X)\,,\nonumber\\
\delta Y_{ij}  = &\, 2\, \bar{\epsilon}_{(i}
  \Slash{D}\Omega_{j)} + 2\, \varepsilon_{ik}
  \varepsilon_{jl}\, \bar{\epsilon}^{(k} \Slash{D}\Omega^{l)
  } \ . 
\end{align}

One then has to choose a meromorphic homogeneous function $F$ of weight $2$ and 
build the multiplet $\bf F(\bf X^{I})$ with lowest component $F(X^{I})$. 
The components of this is given in terms of the components of the vector multiplet as follows: 
\begin{align}
  \label{eq:chiral-mult-exp}
  A\vert_F =&\, F(A) \,,\nonumber\\
  \Psi_i\vert_F =&\, F(A)_I \,\Psi_i{}^I
  \,,\nonumber\\ 
  B_{ij}\vert_F =&\, F(A)_I\, B_{ij}{}^I -\tfrac12
  F(A)_{IJ} \,\bar \Psi_{(i}{}^I 
  \Psi_{j)}{}^J \,,\nonumber\\ 
  F_{ab}^-\vert_F =&\, F(A)_I \,F_{ab}^-{}^I -\tfrac18
  F(A)_{IJ}\, \varepsilon^{ij} \bar 
  \Psi_{i}{}^I \gamma_{ab} \Psi_{j}{}^J \,,\nonumber\\ 
  \Lambda_{i}\vert_F =&\, F(A)_I \,\Lambda_{i}{}^I
  -\tfrac12 
  F(A)_{IJ}\big[B_{ij}{}^I   \varepsilon^{jk} \Psi_{k}{}^J  
   +\tfrac12 F^{-}_{ab}{}^I\gamma^{ab} \Psi_{k}{}^J\big] \nonumber\\
   &\, 
   + \tfrac1{48}  F(A)_{IJK}\,\gamma^{ab} \Psi_i{}^I \,
   \varepsilon^{jk} \bar 
   \Psi_{j}{}^J \gamma_{ab}  \Psi_{k}{}^K \,,\nonumber\\ 
   C\vert_F =&\, F(A)_I\, C^I  -\tfrac14
   F(A)_{IJ}\big[ B_{ij}{}^I B_{kl}{}^J\, 
   \varepsilon^{ik} \varepsilon^{jl} 
   -2\, F^{-}_{ab}{}^I F^{-abJ} +4\,\varepsilon^{ik} \bar
   \Lambda_i{}^I \Psi_j{}^J\big]  \,,\nonumber\\ 
        &\,   +\tfrac14 F(A)_{IJK} \big[ \varepsilon^{ik}
        \varepsilon^{jl} 
        B_{ij}{}^I \Psi_{k}{}^J \Psi_{l}{}^K  -\tfrac12
        \varepsilon^{kl} \bar\Psi_{k}{}^I F^{-}_{ab}{}^J\gamma^{ab}
        \Psi_{l}{}^K\big] \nonumber\\ 
        &\,
        + \tfrac1{192} F(A)_{IJKL} \,\varepsilon^{ij}  \bar 
        \Psi_{i}{}^I \gamma_{ab} \Psi_{j}{}^J \,\varepsilon^{kl}  \bar
        \Psi_{k}{}^K \gamma_{ab} \Psi_{l}{}^L\,. 
\end{align}

\section{Boundary terms and supersymmetry of the renormalized action \label{Supersymmetry}}

In \S\ref{RenAction}, we conjectured the boundary action \eqref{Sbdry} 
\be
%\CS_{\rm bdry} = \pi r_{0} \left( \frac{q_{I} \, e^{I}_{*}}{2} + 2 i \, \big(F(X_{*}^{I}) -  \bar F(X_{*}^{I}) \big)  \right) \ . 
\CS_{\rm bdry} =  - 2 \pi r_{0} \left(  \frac{q_{I} \, e^{I}_{*}}{2} + i \, \big(F(X_{*}^{I}) -  \bar F(X_{*}^{I}) \big)  \right) \ . 
\ee
%so that $\CS_{\rm ren} =  \CS_{\rm bulk} + \CS_{\rm bdry} + i \frac{q}{2} \oint A$ is finite. We also mentioned that 
%this action is supersymmetric. In this appendix, we shall discuss the supersymmetry of the action $S_{\rm ren}$. 
so that $\CS_{\rm ren}$ is finite. We also mentioned that 
this action is supersymmetric. In this appendix, we shall discuss the action $S_{\rm ren}$, and 
show that it is supersymmetric.

To motivate this, we note that we can rewrite $\CS_{\rm ren}$  as the sum of two pieces 
\bea\label{Srentwopieces}
\CS_{\rm ren} & = & \CS_{\rm bulk} + \CS_{\rm bdry} + i \frac{q}{2} \oint A \cr
& = & \left( \CS_{\rm bulk} +  \CS_{\rm bdry}^{1} \right) +  \left( \frac{i}{2}  q_I   \int_{0}^{2\pi} A^I_{\theta} \, d\theta +\CS^{2}_{\rm bdry} \right) \ , 
\eea
where we have split the boundary action \eqref{Sbdry} into a sum of two pieces:
\bea\label{Sbdrypieces}
\CS_{\rm bdry} & = & \CS_{\rm bdry}^{1} + \CS_{\rm bdry}^{2} \ , \\
%= - \int_{0}^{2\pi} \big( \CL_{bdry}^{1} +  \CL_{bdry}^{2} \big) \, e^{\hat \theta} \, d\theta \ , 
\CS_{\rm bdry}^{1} & = & - \int_{0}^{2\pi}   i \, \Big[ F(X) - \bar{F (X)} \Big]_{\rm bdry} \, e^{\theta}_{\theta} \, d\theta \ ,  \\ 
\CS_{\rm bdry}^{2} & = & - \int_{0}^{2\pi}   \frac{q_I}{2} \,   \Big[ X^{I} + \bar X^{I} \Big]_{\rm bdry} \, e^{\theta}_{\theta} \, d\theta \ . 
\eea
%with 
%\bea\label{L12bdry}
%\mathcal{L}^{1}_{\rm bdry} & \equiv &    i \, (F(X) - \bar{F (X)})  \ , \\
%\mathcal{L}^{2}_{\rm bdry} & \equiv  &  \frac{q_I}{2} \,   (X^{I} + \bar X^{I})  \ .
%\ee
Here,  $e^{\hat \theta} = \sinh{\eta_{0}}$ is the induced vielbein on the boundary. To verfiy
\eqref{Sbdrypieces}, we use the same algebra used in \eqref{Sbulk1}, namely, an expansion 
of the field $X^{I}$ into its fixed part $X^{I}_{*}$ and varying part which is $\CO(1/r_{0})$, followed by a Taylor 
expansion and the use of  attractor equations. 

With such a split of the action, the two pieces in \eqref{Srentwopieces} 
have a very natural interpretation as we discuss below. 
We will show further that each of them is finite and supersymmetric, implying the same for $\CS_{\rm ren}$. 

Recall that the bulk action \eqref{Sbulk} evaluated on the solution 
can be written as the difference of two pieces 
%\be
%\CS_{\rm bulk}  = 2 \pi i r_{0} \Big [F\big(X_{*}^{I} + i \frac{C^{I}}{r_{0}}\big) 
%- \bar  F\big(X_{*}^{I} + i \frac{C^{I}}{r_{0}}\big) \Big]  
% - 2 \pi i \Big[F(X_{*}^{I} + i C^{I}) -  \bar F(X_{*}^{I} + i C^{I}) \Big],   
%\ee
\be
\CS_{\rm bulk}  = 2 \pi i r_{0} \Big [F\big(X^{I}\big) 
- \bar{F\big(X^{I}\big)} \Big]_{\rm bdry}  
 - 2 \pi i\Big [F\big(X^{I}\big) - \bar{F\big(X^{I}\big)} \Big]_{\rm origin} \ . 
\ee
We see that $\CS_{\rm bulk}$ + $\CS^{1}_{\rm bdry}$ is manifestly finite. 
Thus, $\CS^{1}_{\rm bdry}$ has the natural interpretation of a canonical boundary term 
which cancels the boundary part of the bulk action, so that any variation of 
$\CS_{\rm bulk}$ + $\CS^{1}_{\rm bdry}$ will be finite and not contain boundary terms.

The second piece of the boundary action combines with the Wilson line to give the operator 
%\bea\label{Swilsonsugra}
%\exp \big[ - \frac{i}{2}  q_I   \int_{0}^{2\pi} A^I_{\theta} \, d\theta - \CS^{2}_{\rm bdry} \big]
% & = &  \exp \big[  -\frac{i}{2} \, q_I \int_{0}^{2\pi}  \left(A^I_{\theta} + i e^{\theta}_{\theta} \, (X^{I} + \bar X^{I}) \right)  
%d\theta   \big]  \\ 
%& = &   \exp \big[ - \pi \, q_I \, e_{*}^I \, (r_{0} -1)
% + \pi  \, q_{I} \, r_{0}\Big( X^I_{*}+i\frac{C^I}{r_{0}} + \bar X^I_{*}+i\frac{C^I}{r_{0}} \Big)   
% + \CO(1/r_{0})  \big] \nonumber \\
%%-i \, q_i \int_{0}^{2\pi}  \left(A^i_{\theta} - \sinh{\eta_{0}} \, (X^{i} + \bar X^{i}) \right)  d\theta \Big]  
%&= & \exp \big[ -  \pi \, q_I \, e^I_* \, (r_{0}-1) + \pi \, q_{I} \, r_{0} \Big( e_{I}  + 2 i\frac{C^I}{r_{0}}  \Big) +  \CO(1/r_{0})  \big] \nonumber \\  
%& = & \exp \big[   \pi \, q_I \, e^I_* + 2\pi i \, q_{I} \, C^I  + \CO(1/r_{0})  \big] \label{Swilsoneval} \ 
%\eea
\be\label{Swilsonsugra}
\exp \big[ - \frac{i}{2}  q_I   \int_{0}^{2\pi} A^I_{\theta} \, d\theta - \CS^{2}_{\rm bdry} \big]
 =   \exp \big[  -\frac{i}{2} \, q_I \int_{0}^{2\pi}  \left(A^I_{\theta} + i e^{\theta}_{\theta} \, (X^{I} + \bar X^{I}) \right)  
d\theta   \big] 
\ee
This operator has the natural interpretation as the supersymmetric Wilson line of 
gauge theory \cite{Maldacena:1998im, Rey:1998ik}. Recalling the boundary behavior of the fields 
\bea
- \frac{i}{2}  q_I   \int_{0}^{2\pi} A^I_{\theta} \, d\theta & = &  - \pi \, q_I \, e_{*}^I \, r_{0}(1 + \CO(1/r_{0})) \ ,  \\
- \frac{i}{2} \, q_I \int_{0}^{2\pi}  i e^{\theta}_{\theta} \, (X^{I} + \bar X^{I})  d\theta & = & 
 \pi  \, q_{I} \, r_{0}\Big( X^I_{*} + \bar X^I_{*} + \CO(1/r_{0}) \Big)   \ ,  \\
& = & \pi \, q_I \, e_{*}^I \, r_{0}(1 + \CO(1/r_{0})) \ , 
%-i \, q_i \int_{0}^{2\pi}  \left(A^i_{\theta} - \sinh{\eta_{0}} \, (X^{i} + \bar X^{i}) \right)  d\theta \Big]  
%&= & \exp \big[ -  \pi \, q_I \, e^I_* \, (r_{0}-1) + \pi \, q_{I} \, r_{0} \Big( e_{*}^{I}  + 2 i\frac{C^I}{r_{0}}  \Big) +  \CO(1/r_{0})  \big] \nonumber \\  
%& = & \exp \big[   \pi \, q_I \, e^I_* + 2\pi i \, q_{I} \, C^I  + \CO(1/r_{0})  \big] \label{Swilsoneval} \ 
\eea
it is easy to see that this operator is manifestly finite. 

Evaluated on the solutions $A^{I}_{\theta} = - i e^{I}_{*}(r_{0}-1)$, $X^{I} = X^{I}_{*} + \frac{C^I}{r_{0}}$, $\bar X^{I} = \bar X^{I}_{*} + \frac{C^I}{r_{0}} $ that we consider in \S\ref{Solution}, we see that the two pieces of the 
renormalized action \eqref{Srentwopieces} above give the two pieces of the final renormalized action 
\eqref{Srenfinal} which we found in \S\ref{RenAction}, as indeed should happen.

In the rest of the appendix, we shall sketch the proof of supersymmetry of these two operators.
The supersymmetry of the operator  \eqref{Swilsonsugra} above follows from the transformation 
rules of $X^{I}$ and $A_{\mu}^{I}$ of the vector multiplet \eqref{eq:variations-vect-mult}. 
We use the fact that the Killing spinors obey 
\be\label{killingsprel}
\zeta^{i} = \varepsilon_{ij} \gamma^{0} \zeta^{j} \ . 
\ee
The extra term in the variation of the vector field which is proportional to the gravitino 
is cancelled by the variation of the vielbein in the definition of the super Wilson line. 
This is the new ingredient in the super Wilson line of a gravitational theory compared to that of gauge 
theory. 

Now we come to the supersymmetry of the combination $\CS_{\rm bulk} + \CS^{1}_{\rm bdry}$.
The statement that $\CS_{\rm bulk}$ is supersymmetric \cite{deWit:1979ug, deWit:1984px, deWit:1980tn} 
really means that the 
variation of $\CS_{\rm bulk}$ is a boundary term which can be ignored in certain circumstances. 
In our situation, there is a non-trivial boundary, and therefore what we need to check is that the 
variation of the bulk Lagrangian is indeed equal to the derivative of the boundary Lagrangian. 

To investigate this, we need to understand the structure of the Lagrangian built using the 
chiral superfield \cite{deRoo:1980mm}. 
In the case of rigid supersymmetry, the variation of the top component of the chiral superfield 
is a total derivative in spacetime, and therefore the top component (picked by a chiral superspace integral) 
is an invariant Lagrangian. 
For chiral superfields coupled to superconformal gravity, the transformation rules undergo 
a modification and the derivatives become covariant derivatives, and there are additional 
terms in the variation \eqref{eq:conformal-chiral} of the top component $C$. 
%At the same time, the Lagrangian is an integral over chiral superspace of the chiral superfield times the 
%superspace density, and so the Lagrangian is not equal to the top component of the chiral superfield. 
The invariant  Lagrangian density  \eqref{eq:chiral-density} contains 
new terms whose variation cancel the additional non-derivative terms present in $\delta C$. 

The net result of this procedure is that the variation of the invariant 
Lagrangian is equal to the total derivative terms that are present in the variation of the top component 
$C$ of the chiral multiplet, {\it i.e.} essentially one can drop the extra terms which arise due to 
the covariantization. 
As an example,  the term proportional to the auxiliary field $B_{ij}$ in $\delta C$ contains $\chi_{i}$
which is an auxiliary field of the superconformal multiplet constrained to be proportional 
to $R(Q)^i$. This term is cancelled by the term proportional to $B_{ij}$ in the higher corrections 
to the Lagrangian density \eqref{eq:chiral-density} after solving for the auxiliary field $\chi$ in terms of 
the gravitini. 

%
%Looking at the $Q$ variation \eqref{eq:conformal-chiral} of a chiral multiplet of weight $w=2$, 
%we see that the variation of $C$ contains two total derivative pieces 
%\be\label{totalder1}
%-2 \varepsilon^{ij} \bar\epsilon_i\slash{\p}\Lambda_j \ , 
%\ee 
%and 
%\be\label{totalder2}
% -\tfrac14\varepsilon^{ij}\varepsilon^{k \ell} \big(\bar\epsilon_i \gamma^{ab} {\slash{\p}} T_{abjk}
%\Psi_l + \bar\epsilon_i\gamma^{ab} T_{abjk} \slash{\p} \Psi_\ell \big)  
%=  -\tfrac14\varepsilon^{ij}\varepsilon^{k \ell} \bar\epsilon_i \gamma^{ab} {\slash{\p}} 
%(T_{abjk} \Psi_\ell \big)  \ . 
%\ee  
%The rest of the terms in the variation $\delta C$ are either terms 
%which are cancelled by the higher terms in the Lagrangian, or terms with more than two fermions. 
%Our interest is in the bosonic boundary terms whose variation is cancelled by the variation 
%of the bulk action, and so we can focus only on terms with two or less fermions. 

Looking at the $Q$ variation \eqref{eq:conformal-chiral} of a chiral multiplet of weight $w=2$, 
we see that the variation of $C$ contains two total derivative pieces 
\be\label{totalder1}
-2 \varepsilon^{ij} \slash{\p} (\bar\epsilon_i \, \Lambda_j ) \ , 
\ee 
and 
\be\label{totalder2}
 -\tfrac14\varepsilon^{ij}\varepsilon^{k \ell} \big( ({\slash{\p}}  \bar\epsilon_i \, \gamma^{ab} \, T_{abjk})
\Psi_l + \gamma^{ab} \, T_{abjk} \, \slash{\p} (\bar\epsilon_i \, \Psi_\ell )\big)  
=  -\tfrac14\varepsilon^{ij}\varepsilon^{k \ell} \, {\slash{\p}} \,
( \bar\epsilon_i \, \gamma^{ab} \, T_{abjk} \Psi_\ell \big)  \ . 
\ee  

In our problem where we have a bunch of vector multiplets, the way to build a Lagrangian 
is by using the homogeneous function $F(X^{I})$. One 
first builds a chiral multiplet $\bf F (\bf X^{I})$ whose bottom component is $F(X^{I})$,
and then uses the invariant Lagrangian described above for this chiral multiplet. 
The variation of our Lagrangian is therefore equal to the total derivative terms that appear 
in the variation of the top component of the chiral multiplet $\bf F (\bf X^{I})$.  
Looking at the form of the components of this superfield \eqref{eq:chiral-mult-exp}, and then 
substituting for the components of the reduced chiral multiplet \eqref{eq:vect-mult}, 
we find that the first type of total derivative term from integration of \eqref{totalder1} is 
\bea
&-& 2 \varepsilon^{ji} \int_{\rm bulk} \slash{\p} (\bar\epsilon_j \Lambda_i )\vert_{F}\cr
 &=& 
-2 \varepsilon^{ji}  \int_{\rm bulk}  \slash{\p} \bigg( - \bar\epsilon_j  F(X)_I \, \varepsilon_{ik}\Slash{D}\Omega^{kI}
  -\tfrac12 
  \bar\epsilon_j F(X)_{IJ}\big[B_{ij}{}^I   \varepsilon^{jk} \Omega_{k}{}^J  
   + \tfrac12 \bar\epsilon_j F^{-}_{ab}{}^I\gamma^{ab} \Omega_{k}{}^J\big]     \nonumber\\
   && + \tfrac1{48}  \bar\epsilon_j \, F(X)_{IJK}\,\gamma^{ab} \, \Omega_i{}^I \,
   \varepsilon^{k\ell} \, \bar 
   \Omega_{k}{}^J  \gamma_{ab} \, \Omega_{\ell}{}^K \bigg) \ . 
\eea

We are interested in the bosonic part of the boundary counterterm Lagrangian. 
The third term on the right hand side contains three fermions and so cannot 
appear from the variation of a pure bosonic term, so we shall ignore that term here.
The second term on the RHS proportional to $F_{IJ}$ is equal to the variation of 
$F_{IJ} \Omega^{I} \Omega^{J}$ minus a total derivative term on the boundary. We 
can therefore drop this term since it is fermionic. Using the variation 
$\delta X^{I} = \bar{\epsilon}^i\Omega_i^{I}$, the first term on the RHS is 
proportional to the variation of the derivative of $F_{I}$, which integrates to zero 
 on the closed boundary, and therefore does not produce any boundary counterterms.

This leaves us with the second term \eqref{totalder2} which gives rise to a boundary term 
\be\label{deltaCfin}
-\tfrac14\varepsilon^{ij}\varepsilon^{k \ell} \int_{\rm bulk}  \slash{\p} \, 
 \big( \bar\epsilon_i \, \gamma^{ab} \, T_{abjk} \, \Psi_\ell \vert_{F} \big)  
= -\tfrac14\varepsilon^{ij}\varepsilon^{k \ell} \int_{\rm bulk}  \slash{\p} \, 
\big(  \bar\epsilon_i \, \gamma^{ab} \, T_{abjk} \, F(X)_I \,\Omega_\ell{}^I \big) \ . 
\ee 
Now, the variation of $T_{abjk}$   \eqref{eq:weyl-multiplet} is proportional to the 
curvature $R(Q)_{ab}$ which integrates to zero on the boundary. Therefore, $T_{abjk}$ can be 
treated as a constant on the boundary for the purpose of supersymmetry 
variations. Plugging in the attractor value for $T_{abjk}$, and 
using $\delta X^{I} = \bar{\epsilon}^i\Omega_i^{I}$ again, and the Killing spinor relation \eqref{killingsprel}, 
we see that the remaining term \eqref{deltaCfin}
is equal and opposite to the variation of the boundary term $\CS^{1}_{\rm bdry}$, thus showing that 
the supersymmetry variation of $\CS_{\rm bulk} + \CS^{1}_{\rm bdry}$ vanishes.

\bibliographystyle{JHEP}
\bibliography{review}

\end{document}